\documentclass[onecolumn,notitlepage,prx,superscriptaddress]{revtex4-1}
\usepackage{latexsym}
\usepackage{amsthm}
\usepackage{amssymb}
\usepackage{amsmath}
\usepackage[all]{xy}
\usepackage[pdftex]{graphicx}
\usepackage{dcolumn}
\usepackage{bm}
\usepackage{color}
\usepackage{multirow}
\usepackage{comment}
\usepackage{epic}
\usepackage{subcaption}
\usepackage{float}
\usepackage[justification=raggedright, singlelinecheck=off]{caption}
\usepackage{tikz}
\usepackage{hyperref}
\usetikzlibrary{shapes,arrows,decorations.markings}

\SelectTips{eu}{11}

\newtheorem{theorem}{Theorem}[section]
\newtheorem{remark}[theorem]{Remark}

\newcommand{\ket}[1]{|#1\rangle}

\begin{document}

\title {Hamiltonian gadgets with reduced resource requirements}
\author {Yudong Cao}\email{cao23@purdue.edu}
\address{Department of Computer Science, Purdue University, 601 Purdue Mall, West Lafayette, IN 47907, USA}
\address{Qatar Energy and Environment Research Institute (QEERI), Ar-Rayy\={a}n, P.O Box 5825, Doha, Qatar} 
\author {Ryan Babbush}\email{babbush@fas.harvard.edu}
\address{Department of Chemistry and Chemical Biology, Harvard University, Cambridge, MA 02138, USA}
\author{Jacob Biamonte}\email{jacob.biamonte@qubit.org}
\address{ISI Foundation, Via Alassio 11/c, 10126, Torino, Italy}
\author{Sabre Kais}\email{kais@purdue.edu}
\address{Department of Computer Science, Purdue University, 601 Purdue Mall, West Lafayette, IN 47907, USA}
\address{Qatar Energy and Environment Research Institute (QEERI), Ar-Rayy\={a}n, P.O Box 5825, Doha, Qatar} 
\address{Department of Chemistry, Physics and Birck Nanotechnology Center, Purdue University, 601 Purdue Mall, West Lafayette, IN 47907, USA} 
\address{Santa Fe Institute, 1399 Hyde Park Rd, Santa Fe, NM 87501, USA}
\date{\today }

%\yudong{See the bulletins $\blacksquare$ for the changes made.}

\begin{abstract}
{Application of the adiabatic model of quantum computation requires efficient encoding of the solution to computational problems into the lowest eigenstate of a 
Hamiltonian that supports universal adiabatic quantum computation}.
%\yudong{``programmable Hamiltonian" changed to "Hamiltonian that supports universal adiabatic quantum computation"} 
Experimental systems are typically limited to restricted forms of 2-body interactions.  Therefore, {universal adiabatic quantum computation requires a method for approximating} quantum many-body Hamiltonians up to arbitrary spectral error using at most 2-body interactions.  Hamiltonian gadgets, introduced around a decade ago, offer the only current means to address this requirement.  Although the applications of Hamiltonian gadgets have steadily grown since their introduction, little progress has been made in overcoming the limitations of the gadgets themselves.  In this experimentally motivated theoretical study, we {introduce} several gadgets which {require significantly more realistic control parameters than similar gadgets in the literature}.  We employ analytical techniques which result in a reduction of the resource scaling as a function of spectral error for the commonly used subdivision, 3- to 2-body and $k$-body gadgets. Accordingly, our improvements reduce the resource requirements of all proofs and experimental proposals making use of these common gadgets. Next, we numerically optimize these new gadgets {to illustrate the tightness of our analytical bounds}. 
%\yudong{Removed the sentence ``Our analysis points to regimes and trade-offs where various $k$-body gadget constructions are shown to afford advantages."}
Finally, we introduce a new gadget that simulates a $YY$ interaction term using Hamiltonians containing only $\{X,Z,XX,ZZ\}$ terms. Apart from possible implications in a theoretical context, this work could also be useful for a first experimental implementation of these key building blocks  by requiring less control precision without introducing extra ancillary qubits. 

\end{abstract}

\maketitle

Although adiabatic quantum computation is known to be a universal model of quantum computation \cite{2004quant.ph..5098A, 2007PhRvL..99g0502M, OT06, BL07, CL08} and hence, in principle equivalent to the circuit model, the mappings between an adiabatic process and an arbitrary quantum circuit require significant overhead. 
%\yudong{Added clarifications after ``mappings"}
Currently the approaches to universal adiabatic quantum computation require implementing multiple higher order and non-commuting interactions by means of perturbative gadgets \cite{BL07}. 
%\yudong{Added citation to Biamonte and Love 2008.}
Such gadgets arose in early work on quantum complexity theory and the resources required for their implementation are the subject of this study.  

Early work by Kitaev \emph{et al}.\ \cite{KSV02} established that an otherwise arbitrary Hamiltonian restricted to have at most 5-body interactions has a ground state energy problem which is complete for the quantum analog of the complexity class \textsc{NP} (\textsc{QMA-complete}).  Reducing the locality of the Hamiltonians from 5-body down to 2-body remained an open problem for a number of years.  In their 2004 proof that \textsc{2-local Hamiltonian} is \textsc{QMA-Complete}, Kempe, Kitaev and Regev formalized the idea of a perturbative gadget, which finally accomplished this task \cite{KKR06}. Oliveira and Terhal further reduced the problem, showing completeness when otherwise arbitrary 2-body Hamiltonians were restricted to act on a square lattice \cite{OT06}. The form of the simplest \textsc{QMA-complete} Hamiltonian is reduced to physically relevant models in \cite{BL07} (see also \cite{CM13}), e.g. 
\begin{equation}
H = \sum_i h_i Z_i + \sum_{i<j} J_{ij}Z_iZ_j + \sum_{i<j}K_{ij}X_iX_j.  
\end{equation}

Although this model contains only physically accessible terms, programming problems into a universal adiabatic quantum computer \cite{BL07} or an adiabatic quantum simulator \cite{sim11,2014arXiv1401.3186V} involves several types of $k$-body interactions (for bounded $k$).
%\yudong{changed ``requires several types..." to ``involves several types..."} 
To reduce from $k$-body interactions to 2-body is accomplished through the application of gadgets. Hamiltonian gadgets were introduced as theorem-proving tools in the context of quantum complexity theory yet their experimental realization currently offers the only path towards universal adiabatic quantum computation. In terms of experimental constraints, an important parameter in the construction of these gadgets is a large spectral gap introduced into the ancilla space as part of a penalty Hamiltonian. This large spectral gap often requires control precision well beyond current experimental capabilities and must be improved for practical physical realizations. 

A perturbative gadget consists of an ancilla system acted on by Hamiltonian $H$, characterized by the spectral gap $\Delta$ between its ground state subspace and excited state subspace, and a perturbation $V$ which acts on both the ancilla and the system.
%\yudong{Added clarification on where $V$ acts on.}
$V$ perturbs the ground state subspace of $H$ such that the perturbed low-lying spectrum of the gadget Hamiltonian $\widetilde H=H+V$ captures the spectrum of the target Hamiltonian, $H_\text{targ}$, up to error $\epsilon$. The purpose of a gadget is dependent on the form of the target Hamiltonian $H_\text{targ}$. For example, if the target Hamiltonian is $k$-local with $k\ge 3$ while the gadget Hamiltonian is 2-local, the gadget serves as a tool for reducing locality. Also if the target Hamiltonian involves interactions that are hard to implement experimentally and the gadget Hamiltonian contains only interactions that are physically accessible, the gadget becomes a generator of physically inaccesible terms from accessible ones. For example the gadget which we introduce in Sec.\ \ref{sec:yy} might fall into this category.
Apart from the physical relevance to quantum computation, gadgets have been central to many results in quantum complexity theory \cite{BDLT08, BL07, BDOT06, CM13}.  Hamiltonian gadgets were also used to characterize the complexity of density functional theory \cite{Schuch09} and are required components in current proposals related to error correction on an adiabatic quantum computer \cite{Ganti2013} and the adiabatic and ground state quantum simulator \cite{sim11,2014arXiv1401.3186V}.  Since these works employ known gadgets which we provide improved constructions of here, our results hence imply a reduction of the resources required in these past works.  

The first use of perturbative gadgets \cite{KKR06} relied on a 2-body gadget Hamiltonian to simulate a 3-body Hamiltonian of the form $H_\text{targ} = H_\text{else}+\alpha \cdot A\otimes B \otimes C$ with three auxiliary spins in the ancilla space. Here $H_\text{else}$ is an arbitrary Hamiltonian that does not operate on the auxiliary spins. Further, $A$, $B$ and $C$ are unit-norm operators and $\alpha$ is the desired coupling. For such a system, it is shown that it suffices to construct $V$ with $\|V\|<\Delta/2$ to guarantee that the perturbative self-energy expansion approximates $H_\text{targ}$ up to error $\epsilon$ \cite{OT06,KKR06,BDLT08}. Because the gadget Hamiltonian is constructed such that in the perturbative expansion (with respect to the low energy subspace), only virtual excitations that flip all 3 ancilla bits would have non-trivial contributions in the $1^\text{st}$ through $3^\text{rd}$ order terms. 
%We briefly discuss its $k$-body generalization \cite{JF08} in Sec.\ \ref{sec:alt_kbody} and present more details in Appendix B.
In \cite{JF08} Jordan and Farhi generalized the construction in \cite{KKR06} to a general $k$-body to 2-body reduction using a perturbative expansion due to Bloch \cite{bloch58}. They showed that one can approximate the low-energy subspace of a Hamiltonian containing $r$ distinct $k$-local terms using a 2-local Hamiltonian. Two important gadgets were introduced by Oliveira and Terhal \cite{OT06} in their proof that \textsc{2-local Hamiltonian on square lattice} is \textsc{QMA-Complete}. In particular, they introduced  an alternative 3- to 2-body gadget which uses only one additional spin for each 3-body term as well as a ``subdivision gadget'' that reduces a $k$-body term to a $(\lceil k/2 \rceil+1)$-body term using only one additional spin \cite{OT06}.  These gadgets, which we improve in this work, find their use as the de facto standard whenever the use of gadgets is necessitated. For instance, the gadgets from \cite{OT06} were used by Bravyi, DiVincenzo, Loss and Terhal \cite{BDLT08} to show that one can combine the use of subdivision and 3- to 2-body gadgets to recursively reduce a $k$-body Hamiltonian to $2$-body, which is useful for simulating quantum many-body Hamiltonians. We note that these gadgets solve a different problem than the type of many-body operator simulations considered previously \cite{cory99,cory00} for gate model quantum computation, where the techniques developed therein are not directly applicable to our situation. 

While recent progress in the experimental implementation of adiabatic quantum processors \cite{2006cond.mat..8253H,Boixo2012,BCM+13,Lidar2014} suggests the ability to {perform} sophisticated adiabatic quantum computing experiments, the perturbative gadgets require very large values of $\Delta$. This places high demands on experimental control precision by requiring that devices enforce very large couplings between ancilla qubits while still being able to resolve couplings from the original problem --- even though those fields may be orders of magnitude smaller than $\Delta$. Accordingly, if perturbative gadgets are to be used, it is necessary to find gadgets which can efficiently approximate their target Hamiltonians with significantly lower values of $\Delta$.
$\quad$\\
$\quad$\\
\noindent{\bf Results summary and manuscript structure}. 
Previous works in the literature \cite{KKR06,OT06,BDOT06,BL07,BDLT08} choose $\Delta$ to be a polynomial function of $\epsilon^{-1}$ which is sufficient for yielding a spectral error $O(\epsilon)$ between the gadget and the target Hamiltonian. Experimental realizations however, will require a recipe for assigning the minimum $\Delta$ that guarantees error within specified $\epsilon$, which we consider here. This recipe will need to depend on three parameters: (i) the desired coupling, $\alpha$; (ii) the magnitude of the non-problematic part of the Hamiltonian, $\|H_\text{else}\|$; and (iii) the specified error tolerance, $\epsilon$.  For simulating a target Hamiltonian up to error $\epsilon$, previous constructions \cite{OT06,BDOT06,BDLT08} use $\Delta=\Theta(\epsilon^{-2})$ for the subdivision gadget and $\Delta=\Theta(\epsilon^{-3})$ for the 3- to 2-body gadget. We will provide analytical results and numerics which indicate that $\Delta=\Theta(\epsilon^{-1})$ is sufficient for the subdivision gadget (Sec.\ \ref{sec:sub} and \ref{sec:par_sub}) and $\Delta=\Theta(\epsilon^{-2})$ for the 3- to 2-body gadget (Sec.\ \ref{sec:3body} and Appendix \ref{sec:3body_par}), showing that the physical resources required to realize the gadgets are less than previously assumed elsewhere in the literature.

In our derivation of the $\Delta$ scalings, we use an analytical approach that involves bounding the infinite series in the perturbative expansion. For the 3- to 2-body reduction, in Appendix \ref{sec:3body_par} we show that complications arise when there are multiple 3-body terms in the target Hamiltonian that are to be reduced concurrently and bounding the infinite series in the multiple-bit perturbative expansion requires separate treatments of odd and even order terms. Furthermore, in the case where $\Delta=\Theta(\epsilon^{-2})$ is used, additional terms which are dependent on the commutation relationship among the 3-body target terms are added to the gadget in order to compensate for the perturbative error due to cross-gadget contributions (Appendix \ref{appendix:4local}). 

The next result of this paper, described in Sec.\ \ref{sec:5th_32}, is a 3- to 2-body gadget construction that uses a 2-body Ising Hamiltonian with a local transverse field. This opens the door to use existing flux-qubit hardware \cite{2006cond.mat..8253H} to simulate $H_\text{targ}=H_\text{else}+\alpha Z_iZ_jZ_k$ where $H_\text{else}$ is not necessarily diagonal. One drawback of this construction is that it requires $\Delta=\Theta(\epsilon^{-5})$, rendering it challenging to realize in practice. For cases where the target Hamiltonian is diagonal, there are non-perturbative gadgets \cite{B08,WFB12,BOA13} that can reduce a $k$-body Hamiltonian to 2-body. In this work, however, we focus on perturbative gadgets.
%These gadgets will be considered along with their limitations and interrelations with perturbative gadgets. Non-perturbative gadget constructions place significantly less stringent requirements on the gap $\Delta$ needed and thus have significant advantages over the perturbative gadgets. However, in general, non-perturbative gadgets cannot replace perturbative gadgets for arbitrary $k$- to 2-body reduction. 

The final result of this paper in Sec.\ \ref{sec:yy} is to propose a gadget which is capable of reducing arbitrary real-valued Hamiltonians to a Hamiltonian with only XX and {ZZ} couplings. In order to accomplish this, we go to fourth-order in perturbation theory to find an XXZZ Hamiltonian which serves as an effective Hamiltonian {dominated by} YY coupling terms. Because YY terms are especially difficult to realize {in some experimental architectures}, this result is useful for those wishing to encode arbitrary \textsc{QMA-Hard} problems on existing hardware.  This gadget in fact now opens the door to solve electronic structure problems on an adiabatic quantum computer.  

To achieve both fast readability and completeness in presentation, each section from Sec.\ \ref{sec:sub} to Sec.\ \ref{sec:yy} consists of a {\bf Summary} subsection and an {\bf Analysis} subsection. The former is mainly intended to provide a high-level synopsis of the main results in the corresponding section. Readers could only refer to the {\bf Summary} sections on their own for an introduction to the results of the paper. The {\bf Analysis} subsections contain detailed derivations of the results in the {\bf Summary}.
%\yudong{This paragraph is added to clarify what Summary and Analysis subsections do.}

\section{Perturbation theory}\label{sec:perturbation}
In our notation the spin-1/2 Pauli operators will be represented as $\{X,Y,Z\}$ with subscript indicating which spin-1/2 particle (qubit) it acts on. For example $X_2$ is a Pauli operator $X=|0\rangle\langle{1}|+|1\rangle\langle{0}|$ acting on the qubit labelled as $2$.

In the literature there are different formulations of the perturbation theory that are adopted when constructing and analyzing the gadgets. This adds to the challenge faced in comparing the physical resources required among the various proposed constructions. 
For example, Jordan and Farhi \cite{JF08} use a formulation due to Bloch, while Bravyi et al.\ use a formulation based on the Schrieffer-Wolff transformation \cite{BDLT08}. Here we employ the formulation used in \cite{KKR06,OT06}. For a review on various formulations of perturbation theory, refer to \cite{BDL11}. 

A gadget Hamiltonian $\tilde{H}=H+V$ consists of a penalty Hamiltonian $H$, which applies an energy gap onto an ancilla space, and a perturbation $V$. To explain in further detail how the low-lying sector of the gadget Hamiltonian $\tilde{H}$ approximates the entire spectrum of a certain target Hamiltonian $H_\text{targ}$ with  error $\epsilon$, we set up the following notations: let $\lambda_j$ and $|\psi_j\rangle$ be the $j^\text{th}$ eigenvalue and eigenvector of $H$ and similarly define $\tilde\lambda_j$ and $|\tilde\psi_j\rangle$  as those of $\tilde{H}$, assuming all the eigenvalues are labelled in a weakly increasing order ($\lambda_1\le\lambda_2\le\cdots$, same for $\tilde{\lambda}_j$). Using a cutoff value $\lambda_*$, let $\mathcal{L}_-=\text{span}\{|\psi_j\rangle|\forall j:\lambda_j\le\lambda_*\}$ be the low energy subspace and $\mathcal{L}_+=\text{span}\{|\psi_j\rangle|\forall j:\lambda_j>\lambda_*\}$ be the high energy subspace. Let ${\Pi_-}$ and ${\Pi_+}$ be the orthogonal projectors onto the subspaces $\mathcal{L}_-$ and $\mathcal{L}_+$ respectively. For an operator $O$ we define the partitions of $O$ into the subspaces as $O_-={\Pi_-}O{\Pi_-}$, $O_+={\Pi_+}O{\Pi_+}$, $O_{-+}={\Pi_-}O{\Pi_+}$ and $O_{+-}={\Pi_+}O{\Pi_-}$.  

With the definitions above, one can turn to perturbation theory to approximate $\tilde{H}_-$ using $H$ and $V$. We now consider the operator-valued resolvent $\tilde{G}(z)=(z\openone-\tilde H)^{-1}$. Similarly one would define $G(z)=(z\openone-H)^{-1}$. Note that $\tilde{G}^{-1}(z)-G^{-1}(z)=-V$ so that this allows an expansion in powers of $V$ as \begin{equation}\label{eq:G_expand}
\tilde{G}=(G^{-1}-V)^{-1}=G(\openone-VG)^{-1}=G+GVG+GVGVG+GVGVGVG+\cdots.
\end{equation}
\noindent{}It is then standard to define the self-energy $\Sigma_-(z)=z\openone-({\tilde G}_-(z))^{-1}$. The self-energy is important because the spectrum of $\Sigma_-(z)$ gives an approximation to the spectrum of $\tilde{H}_-$ since by definition $\tilde{H}_-=z\openone-{\Pi_-}(\tilde{G}^{-1}(z)){\Pi_-}$ while $\Sigma_-(z)=z\openone-({\Pi_-}\tilde G(z){\Pi_-})^{-1}$. As is explained by Oliveira and Terhal \cite{OT06}, loosely speaking, if $\Sigma_-(z)$ is roughly constant in some range of $z$ (defined below in Theorem \ref{th:perturbation}) then $\Sigma_-(z)$ is playing the role of $\tilde{H}_-$. This was formalized in \cite{KKR06} and improved in \cite{OT06} where the following theorem is proven (as in \cite{OT06} we state the case where $H$ has zero as its lowest eigenvalue and a spectral gap of $\Delta$. We use operator norm $\|\cdot\|$ which is defined as $\|M\|\equiv\max_{|\psi\rangle\in\mathcal{M}}|\langle\psi|M|\psi\rangle|$ for an operator $M$ acting on a Hilbert space $\mathcal{M}$):

\begin{theorem}[Gadget Theorem \cite{KKR06,OT06}]\label{th:perturbation}
 Let $\|V\|\le\Delta/2$ where $\Delta$ is the spectral gap of $H$ and let the low and high spectrum of $H$ be separated by a cutoff $\lambda_*=\Delta/2$. Now let there be an effective Hamiltonian $H_\text{eff}$ with a spectrum contained in $[a,b]$. If for some real constant $\epsilon>0$ and $\forall z\in[a-\epsilon,b+\epsilon]$ with $a<b<\Delta/2-\epsilon$, the self-energy $\Sigma_-(z)$ has the property that $\|\Sigma_-(z)-H_\text{eff}\|\le\epsilon$, then each eigenvalue $\tilde\lambda_j$ of $\tilde{H}_-$ differs to the $j^\text{th}$ eigenvalue of $H_\text{eff}$, $\lambda_j$, by at most $\epsilon$. In other words $|\tilde{\lambda}_j - \lambda_j|\le\epsilon$, $\forall j$.
\end{theorem}

To apply Theorem \ref{th:perturbation}, a series expansion for $\Sigma_-(z)$ is truncated at low order for which $H_\text{eff}$ is approximated. The 2-body terms in $H$ and $V$ by construction can give rise to higher order terms in $H_\text{eff}$. For this reason it is possible to engineer $H_\text{eff}$ from $\Sigma_-(z)$ to approximate $H_\text{targ}$ up to error $\epsilon$ in the range of $z$ considered in Theorem \ref{th:perturbation} by introducing auxiliary spins and a suitable selection of 2-body $H$ and $V$. Using the series expansion of $\tilde{G}$ in Eq.\ \ref{eq:G_expand}, the self-energy $\Sigma_-(z)=z\openone-\tilde{G}_-^{-1}(z)$ can be expanded as (for further details see \cite{KKR06})
\begin{equation}\label{eq:selfenergy}
\Sigma_-(z)=H_-+V_-+V_{-+}G_+(z)V_{+-}+V_{-+}G_+(z)V_+G_+(z)V_{+-}+\cdots.
\end{equation}
The terms of $2^\text{nd}$ order and higher in this expansion give rise to the effective many-body interactions.

\begin{figure}
\makebox[1.6cm][l]{ }\includegraphics[scale=0.15]{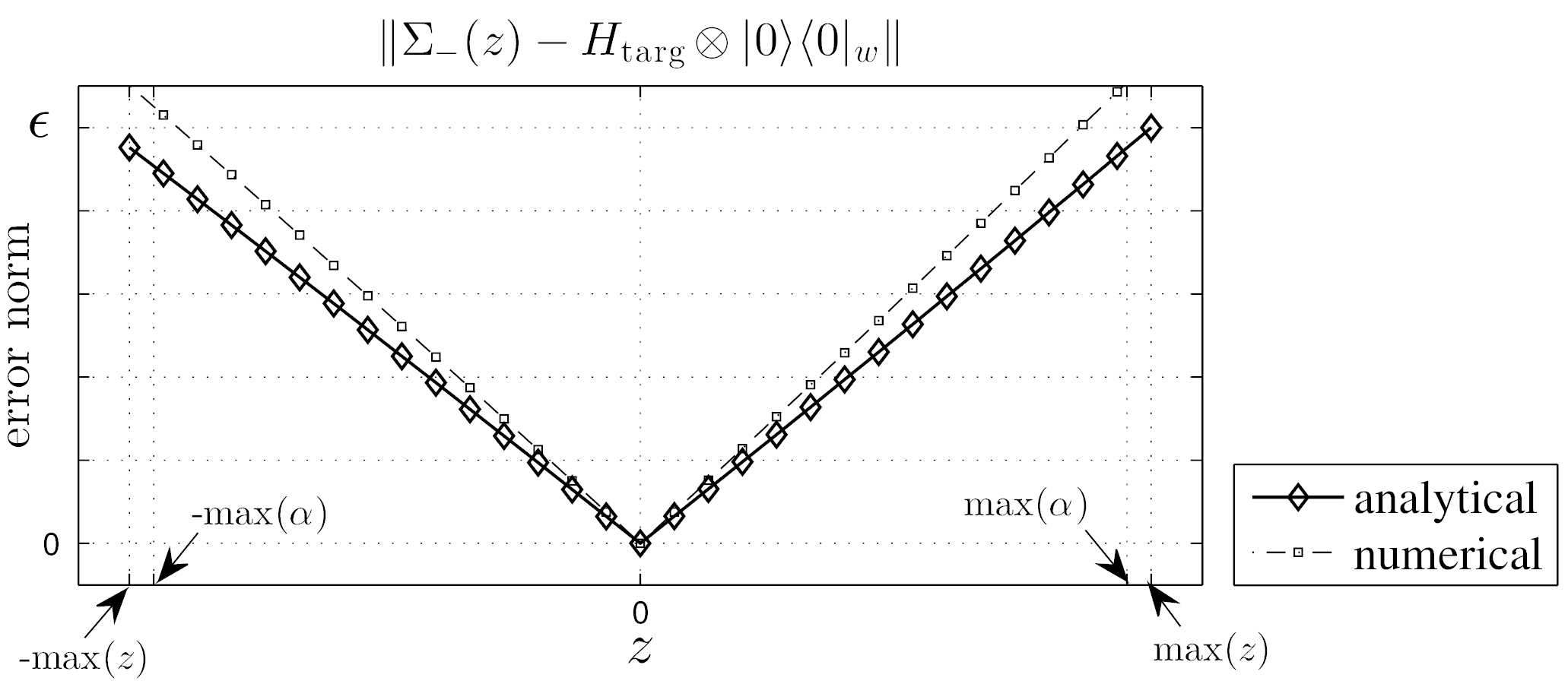}
\centerline{\makebox[-0.2cm][l]{ }(a)}
\makebox[1.6cm][l]{ }\includegraphics[scale=0.15]{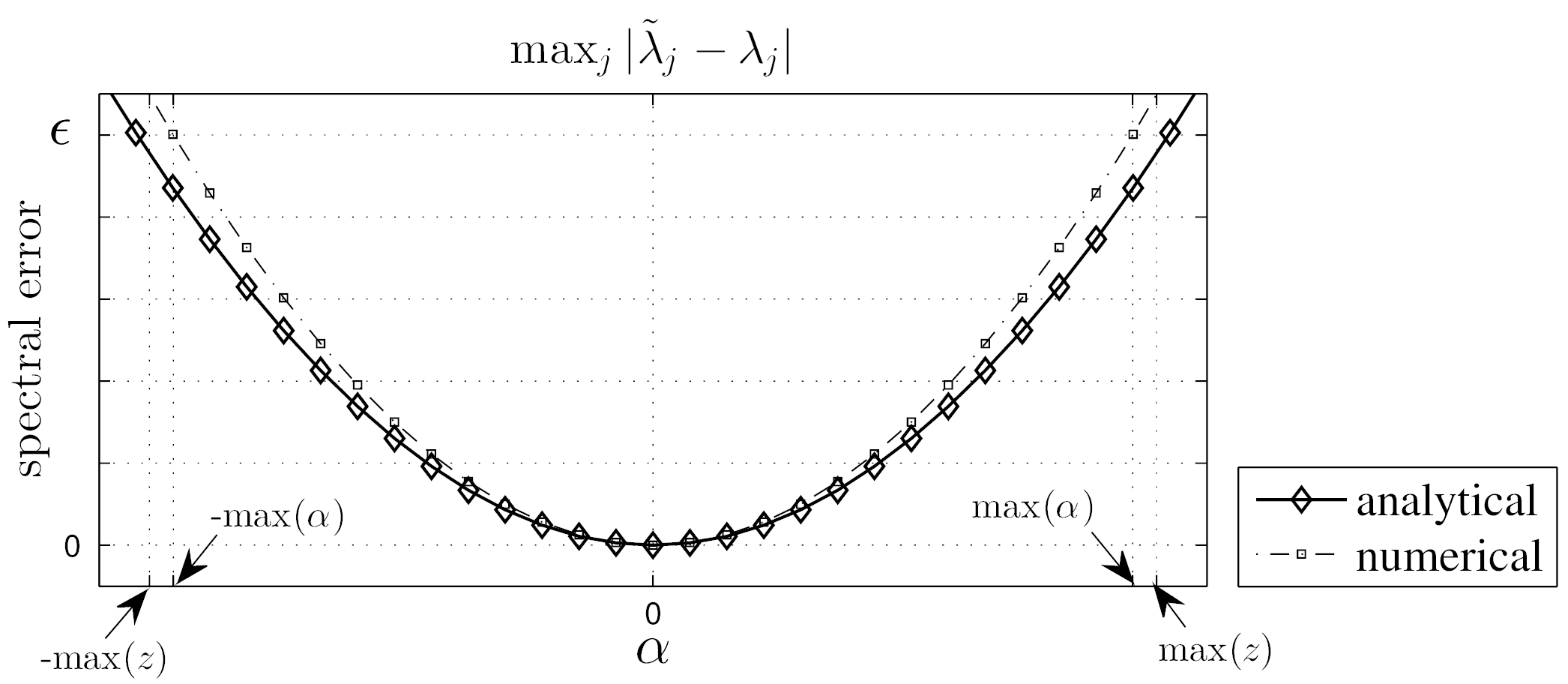}
\centerline{\makebox[-0.2cm][l]{ }(b)}
\caption{\normalsize Numerical illustration of gadget theorem using a subdivision gadget. Here we use a subdivision gadget to approximate $H_\text{targ}=H_\text{else}+\alpha Z_1Z_2$ with $\|H_\text{else}\|=0$ and $\alpha\in[-1,1]$. $\epsilon=0.05$. ``analytical'' stands for the case where the value of $\Delta$ is calculated using Eq.\ \ref{eq:2bodyDelta} when $|\alpha|=1$. ``numerical'' represents the case where $\Delta$ takes the value that yield the spectral error to be $\epsilon$. In (a) we let $\alpha=1$. $z\in[-\max z,\max z]$ with $\max z=\|H_\text{else}\|+\max\alpha+\epsilon$. The operator $\Sigma_-(z)$ is computed up to the $3^\text{rd}$ order. Subplot (b) shows for every value of $\alpha$ in its range, the maximum difference between the eigenvalues $\tilde\lambda_j$ in the low-lying spectrum of $\tilde{H}$ and the corresponding eigenvalues $\lambda_j$ in the spectrum of $H_\text{targ}\otimes|0\rangle\langle{0}|_w$.}
\label{fig:2bodygadget}
\end{figure}

\section{Improved Oliveira and Terhal subdivision gadget}\label{sec:sub}

\noindent{\bf Summary.} The subdivision gadget is introduced by Oliveira and Terhal \cite{OT06} in their proof that \textsc{2-local Hamiltonian on square lattice} is \textsc{QMA-Complete}. Here we show an improved lower bound for the spectral gap $\Delta$ needed on the ancilla of the gadget.
%\yudong{A sentence is added to clarify our improvement up front.}
A subdivision gadget simulates a many-body target Hamiltonian $H_\text{targ}=H_\text{else}+\alpha \cdot A\otimes B$ ($H_\text{else}$ is a Hamiltonian of arbitrary norm, $\|A\|=1$ and $\|B\|=1$) by introducing an ancilla spin $w$ and applying onto it a penalty Hamiltonian $H=\Delta|1\rangle\langle{1}|_w$ so that its ground state subspace $\mathcal{L}_-=\text{span}\{|0\rangle_w\}$ and its excited subspace $\mathcal{L}_+=\text{span}\{|1\rangle_w\}$ are separated by energy gap $\Delta$. In addition to the penalty Hamiltonian $H$, we add a perturbation $V$ of the form
\begin{equation}\label{eq:2body_V}
V = H_\text{else}+|\alpha||0\rangle\langle{0}|_w  + \sqrt{\frac{|\alpha|\Delta}{2}}(\text{sgn}(\alpha)A-B)\otimes X_w.
\end{equation}
Hence if the target term $A\otimes B$ is $k$-local, the gadget Hamiltonian $\tilde{H}=H+V$ is at most $(\lceil{k/2}\rceil+1)$-local, accomplishing the locality reduction. Assume $H_\text{targ}$ acts on $n$ qubits.
Prior work \cite{OT06} shows that $\Delta=\Theta(\epsilon^{-2})$ is a sufficient condition for the lowest $2^n$ levels of the gadget Hamiltonian $\widetilde{H}$ to be $\epsilon$-close to the corresponding spectrum of $H_\text{targ}$. However, by bounding the infinite series of error terms in the perturbative expansion, we are able to obtain a tighter lower bound for $\Delta$ for error $\epsilon$.  Hence we arrive at our first result (details will be presented later in this section), that it suffices to let
\begin{equation}\label{eq:2body_D}
\Delta\ge\left(\frac{2|\alpha|}{\epsilon}+1\right)(2\|H_\text{else}\|+|\alpha|+\epsilon).
\end{equation}

\begin{figure*}
\begin{minipage}{0.4\textwidth}
\setlength{\unitlength}{1cm}
\begin{picture}(10,7.5)
\put(-4.85,0.42){\includegraphics[scale=0.13]{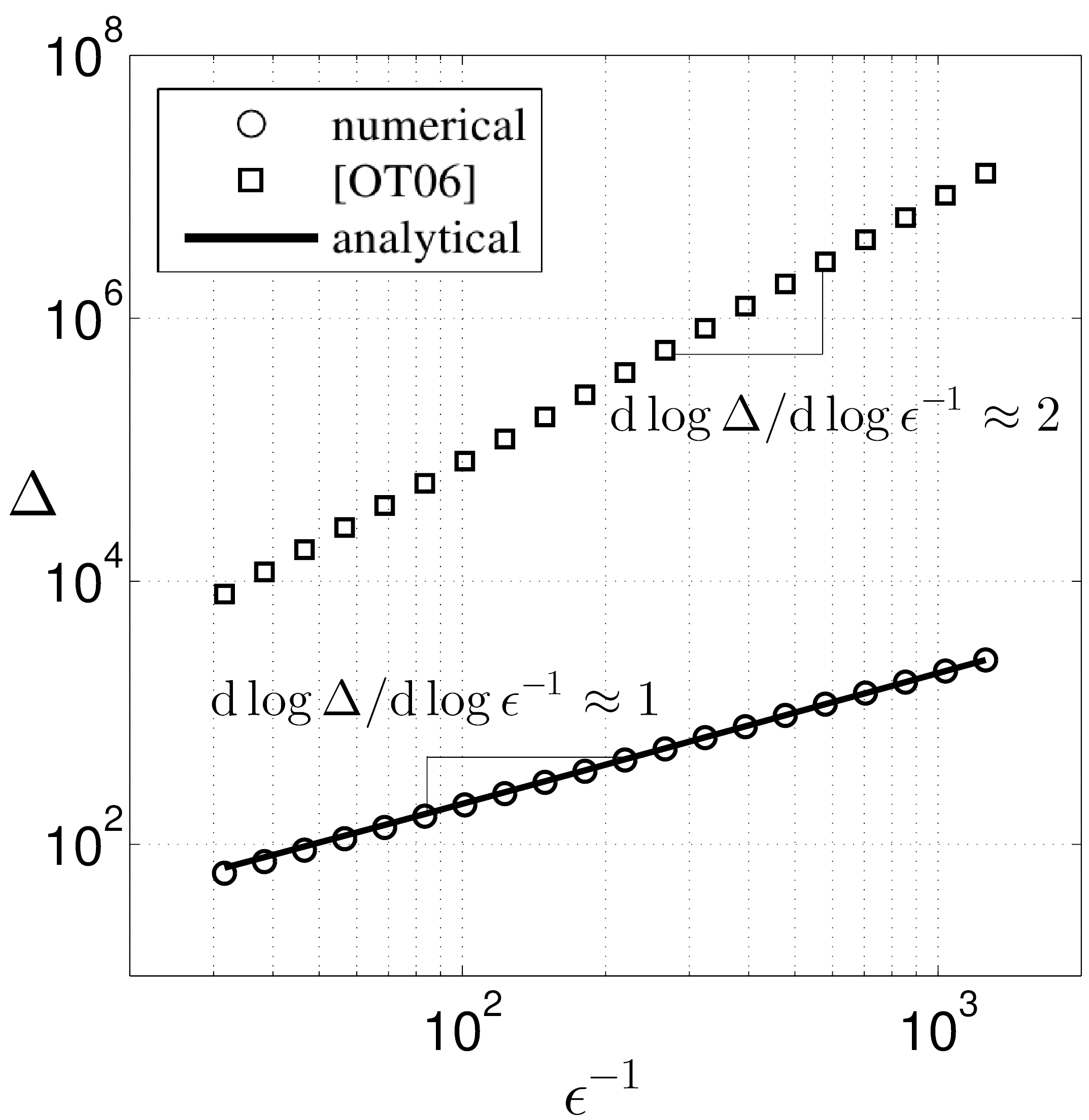}}
\put(2.65,0.28){\includegraphics[scale=0.188]{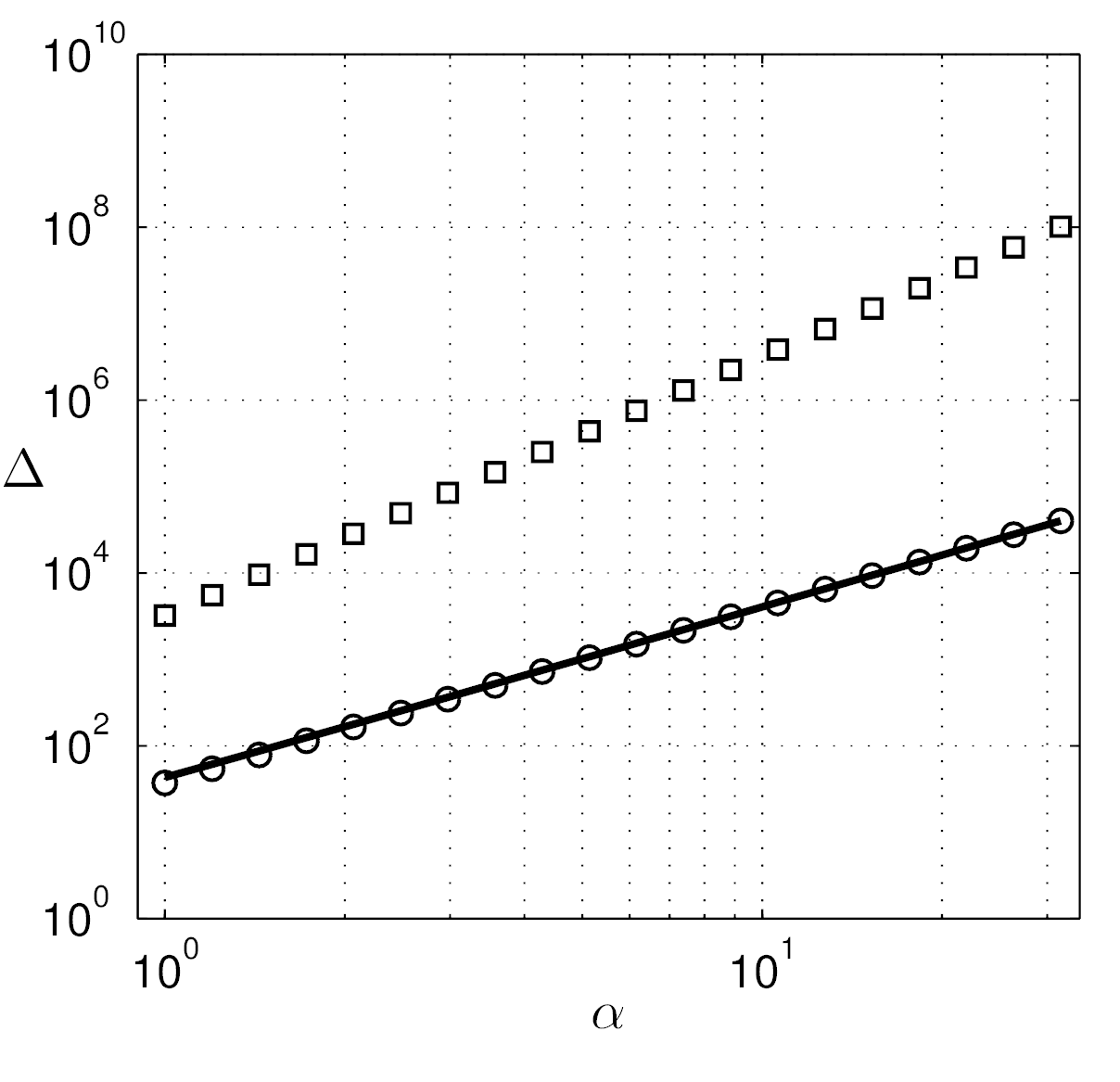}}
\put(-1.1,0){\text{(a)}}
\put(6.8,0){\text{(b)}}
\end{picture}
\end{minipage}
\caption{Comparison between our subdivision gadget with that of Oliveira and Terhal \cite{OT06}. The data labelled as ``numerical'' represent the $\Delta$ values obtained from the numerical search such that the spectral error between $H_\text{targ}$ and $\widetilde{H}_-$ is $\epsilon$. The data obtained from the calculation using Eq.\ \ref{eq:2body_D} are labelled as ``analytical''. ``[OT06]'' refers to values of $\Delta$ calculated according to the assignment by Oliveira and Terhal \cite{OT06}. In this example we consider $H_\text{targ}=H_\text{else}+\alpha Z_1Z_2$. (a) Gap scaling with respect to $\epsilon^{-1}$. Here $\|H_\text{else}\|=0$ and $\alpha=1$. (b) The gap $\Delta$ as a function of the desired coupling $\alpha$. Here $\|H_\text{else}\|=0$,  $\epsilon=0.05$.}
\label{fig:delta_compare_sub}
\end{figure*}

In Fig.\ \ref{fig:delta_compare_sub} we show numerics indicating the minimum $\Delta$ required as a function of $\alpha$ and $\epsilon$. In Fig.\ \ref{fig:delta_compare_sub}a the numerical results and the analytical lower bound in Eq.\ \ref{eq:2body_D} show that for our subdivision gadgets, $\Delta$ can scale as favorably as $\Theta(\epsilon^{-1})$. For the subdivision gadget presented in \cite{OT06},  $\Delta$ scales as $\Theta(\epsilon^{-2})$. Though much less than the original assignment in \cite{OT06}, the lower bound of $\Delta$ in Eq.\ \ref{eq:2body_D}, still satisfies the condition of Theorem \ref{th:perturbation}. In Fig.\ \ref{fig:delta_compare_sub} we numerically find the minimum value of such $\Delta$ that yields a spectral error of exactly $\epsilon$. 

$\quad$\\
\noindent{\bf Analysis.}  The currently known subdivision gadgets in the literature assume that the gap in the penalty Hamiltonian $\Delta$ scales as $\Theta(\epsilon^{-2})$ (see for example \cite{OT06,BDLT08}). Here we employ a method which uses infinite series to find the upper bound to the norm of the high order terms in the perturbative expansion. We find that in fact $\Delta=\Theta(\epsilon^{-1})$ is sufficient for the error to be within $\epsilon$. A variation of this idea will also be used to reduce the gap $\Delta$ needed in the $3$- to 2-body gadget (see Sec. \ref{sec:3body}).

The key aspect of developing the gadget is that given $H=\Delta|1\rangle\langle{1}|_w$, we need to determine a perturbation $V$ to perturb the low energy subspace \[\mathcal{L}_-=\text{span}\{|\psi\rangle\otimes|0\rangle_w,\makebox[0.05cm]{}\text{ $\ket{\psi}$ is any state of the system excluding the ancilla spin $w$}\}\] such that the low energy subspace of the gadget Hamiltonian $\tilde{H}=H+V$ approximates the spectrum of the entire operator $H_\text{targ}\otimes|0\rangle\langle{0}|_w$ up to error $\epsilon$.  Here we will define $V$ and work backwards to show that it satisfies Theorem \ref{th:perturbation}. We let
\begin{equation}\label{eq:V}
V=H_\text{else}+\frac{1}{\Delta}({\kappa}^2A^2+{\lambda}^2B^2)\otimes|0\rangle\langle{0}|_w+({\kappa}A+{\lambda}B)\otimes X_w
\end{equation}
\noindent{}where ${\kappa}$, ${\lambda}$ are constants which will be determined such that the dominant contribution to the perturbative expansion which approximates $\tilde{H}_-$ gives rise to the target Hamiltonian $H_\text{targ}=H_\text{else}+\alpha \cdot A\otimes B$. In Eq.\ \ref{eq:V} and the remainder of the section, by slight abuse of notation, we use $\kappa A+\lambda B$ to represent $\kappa(A\otimes\openone_\mathcal{B})+\lambda(\openone_\mathcal{A}\otimes B)$ for economy. Here $\openone_\mathcal{A}$ and $\openone_\mathcal{B}$ are identity operators acting on the subspaces $\mathcal{A}$ and $\mathcal{B}$ respectively. The partitions of $V$ in the subspaces, as defined in Sec. \ref{sec:perturbation} are
\begin{equation}
\begin{array}{c}
\displaystyle
V_+=H_\text{else}\otimes|1\rangle\langle{1}|_w,\quad V_-=\left(H_\text{else}+\frac{1}{\Delta}({\kappa}^2A^2+{\lambda}^2B^2)\openone\right)\otimes|0\rangle\langle{0}|_w,\\[0.1in] V_{-+}=({\kappa}A+{\lambda}B)\otimes|0\rangle\langle{1}|_w,\quad V_{+-}=({\kappa}A+{\lambda}B)\otimes|1\rangle\langle{0}|_w.
\end{array}
\end{equation}

\noindent{}We would like to approximate the target Hamiltonian $H_\text{targ}$ {and so} expand {the} self-energy in Eq.\ \ref{eq:selfenergy} up to $2^\text{nd}$ order. Note that $H_-=0$ and $G_+(z)=(z-\Delta)^{-1}|1\rangle\langle{1}|_w$. Therefore the self energy $\Sigma_-(z)$ can be expanded as
\begin{equation}\label{eq:Sz_2}
\begin{array}{ccl}
\Sigma_-(z) & = & \displaystyle V_-+\frac{1}{z-\Delta}V_{-+}V_{+-}+\sum_{k=1}^\infty\frac{V_{-+}V_+^kV_{+-}}{(z-\Delta)^{k+1}} \\[0.1in]
 & = & \displaystyle 
\underbrace{\left(H_\text{else}-\frac{2{\kappa}{\lambda}}{\Delta} A\otimes B\right)\otimes|0\rangle\langle{0}|_w}_{H_\text{eff}}+
\underbrace{\frac{z}{\Delta(z-\Delta)}({\kappa}A+{\lambda}B)^2\otimes|0\rangle\langle{0}|_w+\sum_{k=1}^\infty\frac{V_{-+}V_+^kV_{+-}}{(z-\Delta)^{k+1}}}_\text{error term}.
\end{array}
\end{equation}

\noindent{}{By selecting} ${\kappa}=\text{sgn}(\alpha)(|\alpha|\Delta/2)^{1/2}$ and ${\lambda}=-(|\alpha|\Delta/2)^{1/2}$, the leading order term in $\Sigma_-(z)$ becomes $H_\text{eff}=H_\text{targ}\otimes|0\rangle\langle{0}|_w$. {We must now} show that the condition of Theorem \ref{th:perturbation} is satisfied i.e.\ for a small real number $\epsilon>0$, $\|\Sigma_-(z)-H_\text{eff}\|\le\epsilon,\forall z\in[\min z,\max z]$ where $\max z=\|H_\text{else}\|+|\alpha|+\epsilon=-\min z$. {Essentially this amounts to choosing a value of $\Delta$ to cause the error term in Eq.\ \ref{eq:Sz_2} to be $\le\epsilon$}. In order to derive a tighter lower bound for $\Delta$, we bound the norm of the error term in Eq.\ \ref{eq:Sz_2} by letting $z\mapsto\max z$ and {from} the triangle inequality for operator norms:
\begin{equation}\label{eq:Sz_derive}
\begin{array}{rcl}
\displaystyle
\left\|\frac{z}{\Delta(z-\Delta)}(\kappa A+\lambda B)^2\otimes|0\rangle\langle{0}|_w\right\| & \le & \displaystyle \frac{\max{z}}{\Delta(\Delta-\max{z})}\cdot 4\kappa^2=\frac{2|\alpha|\max{z}}{\Delta-\max{z}} \\[0.1in]
\displaystyle
\left\|\sum_{k=1}^\infty\frac{V_{-+}V_+^kV_{+-}}{(z-\Delta)^{k+1}}\right\| & \le & \displaystyle \sum_{k=1}^\infty\frac{\|V_{-+}\|\cdot\|V_+\|^k\cdot\|V_{+-}\|}{(\Delta-\max{z})^{k+1}} \\[0.1in]
& \le & \displaystyle \sum_{k=1}^\infty\frac{2|\kappa|\cdot\|H_\text{else}\|^k\cdot 2|\kappa|}{(\Delta-\max{z})^{k+1}}=\sum_{k=1}^\infty\frac{2|\alpha|\Delta\|H_\text{else}\|^k}{(\Delta-\max{z})^{k+1}}.
\end{array}
\end{equation}
Using $H_\text{eff}=H_\text{targ}\otimes|0\rangle\langle{0}|_w$, from \eqref{eq:Sz_2} we see that
\begin{eqnarray}
\label{eq:5}\begin{array}{ccl}
\|\Sigma_-(z)-H_\text{targ}\otimes|0\rangle\langle{0}|_w\| & \le & \displaystyle \frac{2|\alpha|\max z}{\Delta-\max z}+\sum_{k=1}^\infty\frac{2|\alpha|\Delta\|H_\text{else}\|^k}{(\Delta-\max z)^{k+1}} \end{array} \\
\label{eq:Sz_bound}\begin{array}{ccl}
& = & \displaystyle \frac{2|\alpha|\max z}{\Delta-\max z}+\frac{2|\alpha|\Delta}{\Delta-\max z}\cdot\frac{\|H_\text{else}\|}{\Delta-\max z-\|H_\text{else}\|}.
\end{array}
\end{eqnarray}
\noindent{}Here going from Eq.\ \ref{eq:5} to Eq.\ \ref{eq:Sz_bound} we have assumed the convergence of the infinite series in Eq.\ \ref{eq:5}, which adds the reasonable constraint that $\Delta>|\alpha|+\epsilon+2\|H_\text{else}\|$.  To ensure that $\|\Sigma_-(z)-H_\text{targ}\otimes|0\rangle\langle{0}|_w\|\le\epsilon$ it is sufficient to let expression Eq.\ \ref{eq:Sz_bound} be $\le\epsilon$, which implies that
\begin{equation}\label{eq:2bodyDelta}
\Delta\ge\left(\frac{2|\alpha|}{\epsilon}+1\right)(|\alpha|+\epsilon+2\|H_\text{else}\|)
\end{equation}
\noindent{}which is $\Theta(\epsilon^{-1})$, a tighter bound than $\Theta(\epsilon^{-2})$ in the literature \cite{BDLT08,KKR06,OT06}. This bound is illustrated with a numerical example (Fig.\ \ref{fig:2bodygadget}). From the data labelled as ``analytical'' in Fig.\ \ref{fig:2bodygadget}a we see that the error norm $\|\Sigma_-(z)-H_\text{eff}\|$ is within $\epsilon$ for all $z$ considered in the range, which satisfies the condition of the theorem for the chosen example. In Fig.\ \ref{fig:2bodygadget}b, the data labelled ``analytical'' show that the spectral difference between $\tilde{H}_-$ and $H_\text{eff}=H_\text{targ}\otimes|0\rangle\langle{0}|_w$ is indeed within $\epsilon$ as the theorem promises. Furthermore, note that the condition of Theorem \ref{th:perturbation} is only sufficient, which justifies why in Fig.\ \ref{fig:2bodygadget}b for $\alpha$ values at $\max\alpha$ and $\min\alpha$ the spectral error is strictly below $\epsilon$. This indicates that an even smaller $\Delta$, although below the bound we found in Eq.\ \ref{eq:2bodyDelta} to satisfy the theorem, could still yield the spectral error within $\epsilon$ for all $\alpha$ values in the range. The smallest value $\Delta$ can take would be one such that the spectral error is exactly $\epsilon$ when $\alpha$ is at its extrema. We numerically find this $\Delta$ (up to numerical error which is less than $10^{-5}\epsilon$) and as demonstrated in Fig.\ \ref{fig:2bodygadget}b, the data labelled ``numerical" shows that the spectral error is indeed $\epsilon$ at $\max(\alpha)$ and $\min(\alpha)$, yet in Fig.\ \ref{fig:2bodygadget}a the data labelled ``numerical" shows that for some $z$ in the range the condition of the Theorem \ref{th:perturbation}, $\|\Sigma_-(z)-H_\text{targ}\otimes|0\rangle\langle{0}|_w\|\le\epsilon$, no longer holds. In Fig.\ \ref{fig:2bodygadget} we assume that $\epsilon$ is kept constant. In Fig.\ \ref{fig:delta_compare_sub}a we compute both analytical and numerical $\Delta$ values for different values of $\epsilon$.
$\quad$\\
$\quad$\\
\noindent{\emph{Comparison with Oliveira and Terhal \cite{OT06}.}} We also compare our $\Delta$ assignment with the subdivision gadget by Oliveira and Terhal \cite{OT06}, where given a target Hamiltonian $H_\text{targ}=H_\text{else}+Q\otimes R$ it is assumed that $Q$ and $R$ are operators with finite norm operating on two separate spaces $\mathcal{A}$ and $\mathcal{B}$. 

The construction of the subdivision gadget in \cite{OT06} is the same as the construction presented earlier: introduce an ancillary qubit $w$ with energy gap $\Delta$, then the unperturbed Hamiltonian is $H=\Delta|1\rangle\langle{1}|_w$. In \cite{OT06} they add a perturbation $V$ that takes the form of \cite[Eq.\ 15]{OT06}
\begin{equation}\label{eq:V_terhal}
V=H'_\text{else}+\sqrt{\frac{\Delta}{2}}(-Q+R)\otimes{X_w}
\end{equation}
\noindent{}where $H'_\text{else}=H_\text{else}+Q^2/2+R^2/2$. Comparing the form of Eq.\ \ref{eq:V_terhal} and Eq.\ \ref{eq:V} we can see that if we redefine $Q=\sqrt{|\alpha|}A$ and $R=\sqrt{|\alpha|}B$, the gadget formulation is identical to our subdivision gadget approximating $H_\text{targ}=H_\text{else}+\alpha{A\otimes B}$ with $\alpha>0$. In the original work $\Delta$ is chosen as \cite[Eq.\ 20]{OT06} \[\Delta = \frac{(\|H'_\text{else}\|+C_2r)^6}{\epsilon^2}\] where $C_2\ge\sqrt{2}$ and $r=\max\{\|Q\|,\|R\|\}$. In the context of our subdivision gadget, this choice of $\Delta$ translates to a lower bound
\begin{equation}\label{eq:terhal_sub_gap}
\Delta \ge \frac{(\|H_\text{else}+|\alpha|\openone \|+\sqrt{2|\alpha|})^6}{\epsilon^2}.
\end{equation}
In Fig.\ \ref{fig:delta_compare_sub}a we compare the lower bound in Eq.\ \ref{eq:terhal_sub_gap} with our lower bound in Eq.\ \ref{eq:2bodyDelta} and the numerically optimized $\Delta$ described earlier.

\section{Parallel subdivision and $k$- to $3$-body reduction}\label{sec:par_sub}

\noindent{\bf Summary.} Applying subdivision gadgets iteratively one can reduce a $k$-body Hamiltonian $H_\text{targ}=H_\text{else}+\alpha\bigotimes_{i=1}^k\sigma_i$ to 3-body. Here each $\sigma_i$ is a single spin Pauli operator. Initially, the term $\bigotimes_{i=1}^k\sigma_i$ can be broken down into $A\otimes B$ where $A=\bigotimes_{i=1}^r\sigma_i$ and $B=\bigotimes_{i=r+1}^k\sigma_i$. Let $r=k/2$ for even $k$ and $r=(k+1)/2$ for odd $k$. The gadget Hamiltonian will be $(\lceil{k/2}\rceil+1)$-body, which can be further reduced to a $(\lceil{\lceil{k/2}\rceil+1}\rceil/2+1)$-body Hamiltonian in the same fashion. Iteratively applying this procedure, we can reduce a $k$-body Hamiltonian to $3$-body, with the $i^\text{th}$ iteration introducing the same number of ancilla qubits as that of the many-body term to be subdivided. Applying the previous analysis on the improved subdivision gadget construction, we find that $\Delta_i=\Theta(\epsilon^{-1}\Delta_{i-1}^{3/2})$ is sufficient such that during each iteration the spectral difference between $\widetilde{H}_i$ and $\widetilde{H}_{i-1}$ is within $\epsilon$. From the recurrence relation $\Delta_i=\Theta(\epsilon^{-1}\Delta_{i-1}^{3/2})$, we hence were able to show a quadratic improvement over previous $k$-body constructions \cite{BDLT08}.
$\quad$\\
$\quad$\\
\noindent{\bf Analysis.} The concept of parallel application of gadgets has been introduced in \cite{OT06,KKR06}. The idea of using subdivision gadgets for iteratively reducing a $k$-body Hamiltonian to 3-body has been mentioned in \cite{OT06,BDLT08}. Here we elaborate the idea by a detailed analytical and numerical study. We provide explicit expressions of all parallel subdivision gadget parameters which guarantees that during each reduction the error between the target Hamiltonian and the low-lying sector of the gadget Hamiltonian is within $\epsilon$. For the purpose of presentation, let us define the notions of ``parallel" and ``series" gadgets in the following remarks.

\begin{remark}[Parallel gadgets]\label{rem:par} Parallel application of gadgets refers to using gadgets on multiple terms $H_\text{targ,i}$ in the target Hamiltonian $H_\text{targ}=H_\text{else}+\sum_{i=1}^mH_\text{targ,i}$ concurrently. Here one will introduce $m$ ancilla spins $w_1,\cdots,w_m$ and the parallel gadget Hamiltonian takes the form of $\tilde{H}=\sum_{i=1}^mH_i+V$ where $H_i=\Delta|1\rangle\langle{1}|_{w_i}$ and $V=H_\text{else}+\sum_{i=1}^mV_i$. $V_i$ is the perturbation term of the gadget applied to $H_\text{targ,i}$.
\end{remark}

\begin{remark}[Serial gadgets] Serial application of gadgets refers to using gadgets sequentially. Suppose the target Hamiltonian $H_\text{targ}$ is approximated by a gadget Hamiltonian $\tilde{H}^{(1)}$ such that $\tilde{H}^{(1)}_-$ approximates the spectrum of $H_\text{targ}$ up to error $\epsilon$. If one further applies onto $\tilde{H}^{(1)}$ another gadget and obtains a new Hamiltonian $\tilde{H}^{(2)}$ whose low-lying spectrum captures the spectrum of $\tilde{H}^{(1)}$, we say that the two gadgets are applied in series to reduce $H_\text{targ}$ to $\tilde{H}^{(2)}$.
\end{remark}

Based on Remark \ref{rem:par}, a parallel subdivision gadget deals with the case where $H_\text{targ,i}=\alpha_iA_i\otimes B_i$. $\alpha_i$ is a constant and $A_i$, $B_i$ are unit norm Hermitian operators that act on separate spaces $\mathcal{A}_i$ and $\mathcal{B}_i$. Note that with $H_i=\Delta|1\rangle\langle{1}|_{w_i}$ for every $i\in\{1,2,\cdots,m\}$ we have the total penalty Hamiltonian $H=\sum_{i=1}^mH_i=\sum_{x\in\{0,1\}^m}h(x)\Delta|x\rangle\langle{x}|$ where $h(x)$ is the Hamming weight of the $m$-bit string $x$. This penalty Hamiltonian ensures that the ground state subspace is $\mathcal{L}_-=\text{span}\{|0\rangle^{\otimes{m}}\}$ while all the states in the subspace $\mathcal{L}_+=\text{span}\{|x\rangle|x\in\{0,1\}^m,x\neq 00\cdots 0\}$ receives an energy penalty of at least $\Delta$. The operator-valued resolvent $G$ for the penalty Hamiltonian is (by definition in Sec.\ \ref{sec:perturbation})
\begin{equation}\label{eq:Gz}
G(z)=\sum_{x\in\{0,1\}^m}\frac{1}{z-h(x)\Delta}|x\rangle\langle{x}|.
\end{equation}
\noindent{}The perturbation Hamiltonian $V$ is defined as
\begin{equation}\label{eq:V_sub_par}
V=H_\text{else}+\frac{1}{\Delta}\sum_{i=1}^m({\kappa_i^2}A_i^2+{\lambda_i^2}B_i^2)+\sum_{i=1}^m({\kappa_i}A_i+{\lambda_i}B_i)\otimes X_{u_i}
\end{equation}
\noindent{}where the coefficients ${\kappa_i}$ and ${\lambda_i}$ are defined as ${\kappa_i}=\text{sgn}(\alpha_i)\sqrt{{|\alpha_i|\Delta}/{2}},{\lambda_i}=-\sqrt{{|\alpha_i|\Delta}/{2}}$.
% and the function ``sgn" is defined as
%\[
%\text{sgn}(x)=\left\{
%\begin{array}{cr}
%+1 & x>0 \\
%0 & x=0 \\
%-1 & x<0.
%\end{array}
%\right.
%\]
Define $P_-=|0\rangle^{\otimes m}\langle{0}|^{\otimes m}$ and $P_+=\openone-P_-$. Then if $H_\text{targ}$ acts on the Hilbert space $\mathcal{M}$, $\Pi_-=\openone_\mathcal{M}\otimes P_-$ and $\Pi_+=\openone_\mathcal{M}\otimes P_+$. Comparing Eq.\ \ref{eq:V_sub_par} with Eq.\ \ref{eq:V} we see that the projector to the low-lying subspace $|0\rangle\langle{0}|_w$ in Eq.\ \ref{eq:V} is replaced by an identity $\openone$ in Eq.\ \ref{eq:V_sub_par}. This is because in the case of $m$ parallel gadgets $P_-$ cannot be realized with only 2-body terms when $m\ge 3$. 

The partition of $V$ in the subspaces are
\begin{equation}\label{eq:V+-}
\begin{array}{ll}
\displaystyle V_- = \left(H_\text{else}+\frac{1}{\Delta}\sum_{i=1}^m({\kappa_i^2}A_i^2+{\lambda_i^2}B_i^2)\right)\otimes{P_-}, & \displaystyle V_+ = \left(H_\text{else}+\frac{1}{\Delta}\sum_{i=1}^m({\kappa_i^2}A_i^2+{\lambda_i^2}B_i^2)\right)\otimes{P_+} \\[0.1in]
\displaystyle V_{-+}= \sum_{i=1}^m({\kappa_i}A_i+{\lambda_i}B_i)\otimes{P_-}X_{u_i}{P_+}, & \displaystyle V_{+-} =\sum_{i=1}^m({\kappa_i}A_i+{\lambda_i}B_i)\otimes{P_+}X_{u_i}{P_-}. \\[0.1in]
\end{array}
\end{equation} 
\noindent{}The self-energy expansion in Eq.\ \ref{eq:selfenergy} then becomes
\begin{equation}
\begin{array}{ccl}
\displaystyle \Sigma_-(z) & = & \displaystyle \left(H_\text{else}+\frac{1}{\Delta}\sum_{i=1}^m({\kappa_i^2}A_i^2+{\lambda_i^2}B_i^2)\right)\otimes{P_-}+\frac{1}{z-\Delta}\sum_{i=1}^m({\kappa_i}A_i+{\lambda_i}B_i)^2\otimes{P_-} \\[0.1in]
& + & \displaystyle \sum_{k=1}^\infty V_{-+}(G_+V_+)^kG_+V_{+-}.
\end{array}
\end{equation}
\noindent{}Rearranging the terms we have
\begin{equation}
\begin{array}{ccl}
\displaystyle \Sigma_-(z) & = & 
\displaystyle 
\underbrace{\left(H_\text{else}+\sum_{i=1}^m\left(-\frac{2{\kappa_i}{\lambda_i}}{\Delta}A_i\otimes B_i \right)\right)\otimes {P_-}}_{H_\text{eff}}+
\underbrace{\left(\frac{1}{\Delta}+\frac{1}{z-\Delta}\right)\sum_{i=1}^m({\kappa_i^2}A_i^2+{\lambda_i^2}B_i^2)\otimes {P_-}}_{E_1} \\[0.1in]
& + & \displaystyle 
\underbrace{\left(\frac{1}{\Delta}+\frac{1}{z-\Delta}\right)\sum_{i=1}^m2{\kappa_i}{\lambda_i}A_i\otimes B_i\otimes {P_-}}_{E_2}+
\underbrace{\sum_{k=1}^\infty V_{-+}(G_+V_+)^kG_+V_{+-}}_{E_3} \\[0.1in]
\end{array}
\end{equation}
\noindent{}where the term $H_\text{eff}=H_\text{targ}\otimes{P_-}$ is the effective Hamiltonian that we would like to obtain from the perturbative expansion and $E_1$, $E_2$, and $E_3$ are error terms. Theorem \ref{th:perturbation} states that for $z\in[-\max(z),\max(z)]$, if $\|\Sigma_-(z)-H_\text{targ}\otimes{P_-}\|\le\epsilon$ then $\tilde{H}_-$ approximates the spectrum of $H_\text{targ}\otimes P_-$ by error at most $\epsilon$. Similar to the triangle inequality derivation shown in \eqref{eq:Sz_derive}, to derive a lower bound for $\Delta$, let $z\mapsto\max(z)=\|H_\text{else}\|+\sum_{i=1}^m|\alpha_i|+\epsilon$ and the upper bounds of the error terms $E_1$ and $E_2$ can be found as
\begin{equation}\label{eq:par_sub_E12}
\begin{array}{ccl}
\|E_1\| & \le & \displaystyle \frac{\max(z)}{\Delta-\max(z)}\sum_{i=1}^m|\alpha_i|\le\frac{\max(z)}{\Delta-\max(z)}\left(\sum_{i=1}^m|\alpha_i|^{1/2}\right)^2 \\[0.1in]
\|E_2\| & \le & \displaystyle \frac{\max(z)}{\Delta-\max(z)}\left(\sum_{i=1}^m|\alpha_i|^{1/2}\right)^2.
\end{array}
\end{equation}
\noindent{}From the definition in Eq.\ \ref{eq:Gz} we see that $\|G_+(z)\|\le\frac{1}{\Delta-\max(z)}$. Hence the norm of $E_3$ can be bounded by
\begin{equation}\label{eq:par_sub_E3}
\begin{array}{ccl}
\|E_3\| & \le & \displaystyle \sum_{k=1}^\infty\frac{\|\sum_{i=1}^m({\kappa_i}A_i+{\lambda_i}B_i)\|^2\|H_\text{else}+\frac{1}{\Delta}\sum_{i=1}^m({\kappa_i^2}A_i^2+{\lambda_i^2}B_i^2){\openone}\|^k}{(\Delta-\max(z))^{k+1}} \\[0.1in]
& \le & \displaystyle \sum_{k=1}^\infty\frac{2\Delta(\sum_{i=1}^m|\alpha_i|^{1/2})^2(\|H_\text{else}\|+\sum_{i=1}^m|\alpha_i|)^k}{(\Delta-\max(z))^{k+1}} \\[0.1in]
& = & \displaystyle \frac{2\Delta(\sum_{i=1}^m|\alpha_i|^{1/2})^2}{\Delta-\max(z)}\frac{\|H_\text{else}\|+\sum_{i=1}^m|\alpha_i|}{\Delta-\max(z)-(\|H_\text{else}\|+\sum_{i=1}^m|\alpha_i|)}.
\end{array}
\end{equation}
\noindent{}Similar to the discussion in Sec.\ \ref{sec:sub}, to ensure that $\|\Sigma_-(z)-H_\text{targ}\otimes{P_-}\|\le\epsilon$, which is the condition of Theorem \ref{th:perturbation}, it is sufficient to let $\|E_1\|+\|E_2\|+\|E_3\|\le\epsilon$:
\begin{equation}
\begin{array}{ccl}
\|E_1\|+\|E_2\|+\|E_3\| & \le & \displaystyle \frac{2\max(z)}{\Delta-\max(z)}\left(\sum_{i=1}^m|\alpha_i|^{1/2}\right)^2 \\[0.1in]
& + & \displaystyle \frac{2\Delta(\sum_{i=1}^m|\alpha_i|^{1/2})^2}{\Delta-\max(z)}\cdot\frac{\|H_\text{else}\|+\sum_{i=1}^m|\alpha_i|}{\Delta-\max(z)-(\|H_\text{else}\|+\sum_{i=1}^m|\alpha_i|)} \\[0.1in]
& = & \displaystyle \frac{2(\sum_{i=1}^m|\alpha_i|^{1/2})^2(\max(z)+\|H_\text{else}\|+\sum_{i=1}^m|\alpha_i|)}{\Delta-\max(z)-(\|H_\text{else}\|+\sum_{i=1}^m|\alpha_i|)}\le\epsilon
\end{array}
\end{equation}
\noindent{}where we find the lower bound of $\Delta$ for parallel subdivision gadget
\begin{equation}\label{eq:D_par_sub}
\Delta\ge\left[\frac{2(\sum_{i=1}^m|\alpha_i|^{1/2})^2}{\epsilon}+1\right](2\|H_\text{else}\|+2\sum_{i=1}^m|\alpha_i|+\epsilon).
\end{equation}
\noindent{}Note that if one substitutes $m=1$ into Eq.\ \ref{eq:D_par_sub} the resulting expression is a lower bound that is less tight than that in Eq.\ \ref{eq:2bodyDelta}. This is because of the difference in the perturbation $V$ between Eq.\ \ref{eq:V_sub_par} and Eq.\ \ref{eq:V} which is explained in the text preceding Eq.\ \ref{eq:V+-}. Also we observe that the scaling of this lower bound for $\Delta$ is $O(\text{poly}(m)/\epsilon)$ for $m$ parallel applications of subdivision gadgets, assuming $|\alpha_i|=O(\text{poly}(m))$ for every $i\in\{1,2,\cdots,m\}$. This confirms the statement in \cite{OT06,KKR06,BDLT08} that subdivision gadgets can be applied to multiple terms in parallel and the scaling of the gap $\Delta$ in the case of $m$ parallel subdivision gadgets will only differ to that of a single subdivision gadget by a polynomial in $m$.
$\quad$\\*
$\quad$\\*
\noindent{\emph{Iterative scheme for $k$- to 3-body reduction.}} The following iterative scheme summarizes how to use parallel subdivision gadgets for reducing a $k$-body Ising Hamiltonian to 3-body ({Here we use superscript $^{(i)}$ to represent the $i^\text{th}$ iteration and subscript $_i$ for labelling objects within the same iteration}):
\begin{equation}\label{eq:k3_algo}
\begin{array}{c l}
\tilde{H}^{(0)}= & H_\text{targ};\text{$H_\text{targ}$ acts on the Hilbert space $\mathcal{M}^{(0)}$.} \\
\text{\bf while} & \text{$\tilde{H}^{(i)}$ is more than 3-body} \\
& \text{Step 1: Find all the terms that are no more than 3-body (including $H_\text{else}$ from $\tilde{H}^{(0)}$) in $\tilde{H}^{(i-1)}$} \\
& \text{$\qquad\quad$ and let their sum be $H_\text{else}^{(i)}$.} \\
& \text{Step 2: Partition the rest of the terms in $\tilde{H}^{(i-1)}$ into $\alpha_1^{(i)}A_1^{(i)}\otimes B_1^{(i)}$, } \\ 
& \text{$\qquad\quad$ $\alpha_2^{(i)}A_2^{(i)}\otimes B_2^{(i)}$, $\cdots$, $\alpha_m^{(i)}A_m^{(i)}\otimes B_m^{(i)}$. Here $\alpha_j^{(i)}$ are coefficients.} \\
& \text{Step 3: Introduce $m$ ancilla qubits $w_1^{(i)}$, $w_2^{(i)}$, $\cdots w_m^{(i)}$ and construct $\tilde{H}^{(i)}$ using the} \\
& \text{$\qquad\quad$ parallel subdivision gadget. Let $P^{(i)}_-=|0\cdots 0\rangle\langle 0\cdots 0|_{w_1^{(i)}\cdots w_m^{(i)}}$. Define $\Pi_-^{(i)}=\openone_{\mathcal{M}^{(i)}}\otimes P_-^{(i)}$.} \\
& \text{$\qquad$ 3.1: Apply the penalty Hamiltonian $H^{(i)}=\sum_{x\in\{0,1\}}^mh(x)\Delta^{(i)}|x\rangle\langle x|$.} \\
& \text{$\qquad\qquad$ Here $\Delta^{(i)}$ is calculated by the lower bound in Eq.\ \ref{eq:D_par_sub}.} \\
& \text{$\qquad$ 3.2: Apply the perturbation $V^{(i)}=H_\text{else}^{(i)}+\sum_{j=1}^m\sqrt{\frac{|\alpha_j^{(i)}|\Delta^{(i)}}{2}}(\text{sgn}(\alpha_j^{(i)})A_j^{(i)}-B_j^{(i)})\otimes X_{w_j^{(i)}}$} \\
& \text{$\qquad\qquad+\sum_{j=1}^m|\alpha_j^{(i)}|{\openone}$.} \\
& \text{$\qquad$ 3.3: $\tilde{H}^{(i)}=H^{(i)}+V^{(i)}$ acts on the space $\mathcal{M}^{(i)}$ and the maximum spectral difference} \\
& \text{$\qquad\qquad$ between $\tilde{H}^{(i)}_-=\Pi^{(i)}_-\tilde{H}^{(i)}\Pi^{(i)}_-$ and $\tilde{H}^{(i-1)}\otimes P^{(i)}_-$ is at most $\epsilon$.} \\
& \text{$i\rightarrow{i+1}$} \\
\text{\bf end} & \\
\end{array}
\end{equation}
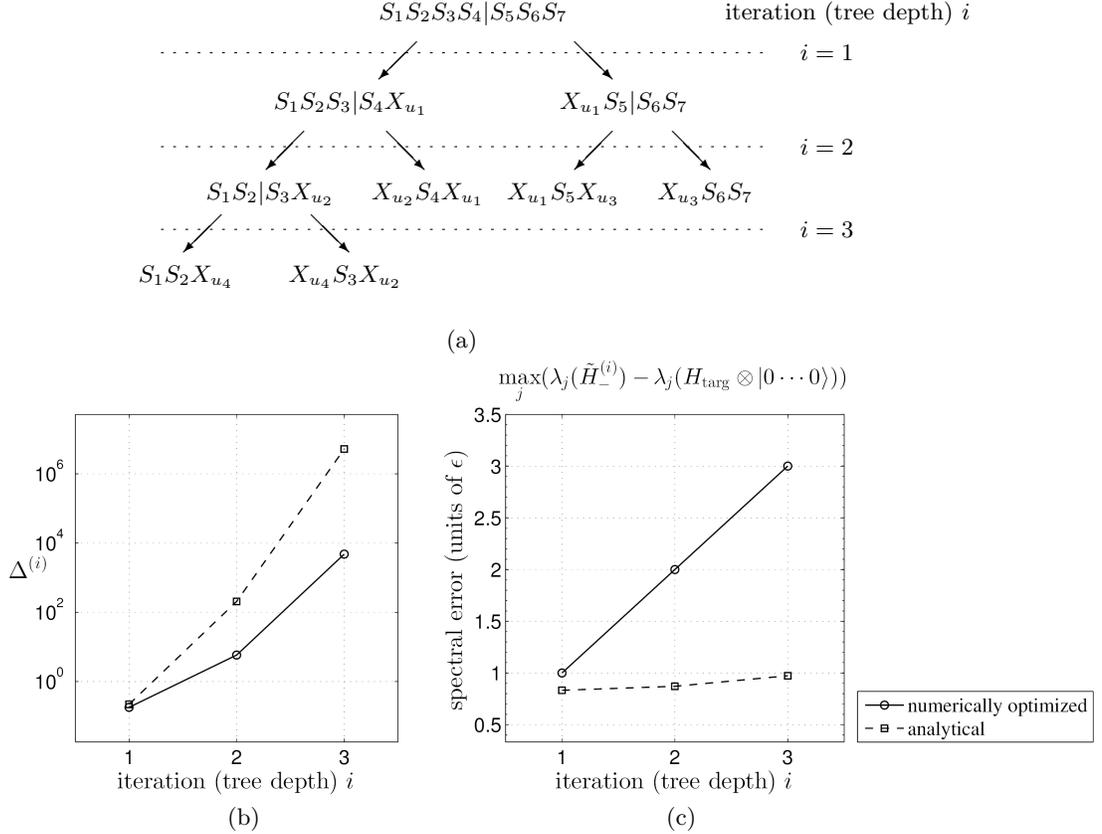
\begin{figure}
\setlength{\unitlength}{1cm}
\centerline{
\begin{picture}(7,5)
\put(2.4,4){\makebox{${S_1}{S_2}{S_3}{S_4}|{S_5}{S_6}{S_7}$}}
\put(7,4){\makebox{iteration (tree depth) $i$}}
\put(2.9,3.7){\vector(-1,-1){0.5}}
{\dashline{0.05}(-0.5,3.55)(7.5,3.55)}
\put(8,3.45){\makebox{$i=1$}}
\put(5,3.7){\vector(1,-1){0.5}}
\put(1,2.8){\makebox{${S_1}{S_2}{S_3}|{S_4}X_{u_1}$}}
\put(4.8,2.8){\makebox{$X_{u_1}{S_5}|{S_6}{S_7}$}}
\put(1.4,2.5){\vector(-1,-1){0.5}}
\put(2.5,2.5){\vector(1,-1){0.5}}
\put(5.5,2.5){\vector(-1,-1){0.5}}
\put(6.3,2.5){\vector(1,-1){0.5}}
{\dashline{0.05}(-0.5,2.3)(7.5,2.3)}
\put(8,2.2){\makebox{$i=2$}}
\put(0.1,1.6){\makebox{${S_1}{S_2}|{S_3}X_{u_2}$}}
\put(2.3,1.6){\makebox{$X_{u_2}{S_4}X_{u_1}$}}
\put(4.1,1.6){\makebox{$X_{u_1}{S_5}{X_{u_3}}$}}
\put(6.1,1.6){\makebox{$X_{u_3}{S_6}{S_7}$}}
\put(0.3,1.4){\vector(-1,-1){0.5}}
\put(1.5,1.4){\vector(1,-1){0.5}}
{\dashline{0.05}(-0.5,1.2)(7.5,1.2)}
\put(8,1.1){\makebox{$i=3$}}
\put(-0.8,0.5){\makebox{${S_1}{S_2}X_{u_4}$}}
\put(1.2,0.5){\makebox{$X_{u_4}{S_3}X_{u_2}$}}
\end{picture}
}
\centerline{(a)}
\makebox[2.2cm][l]{ }\includegraphics[scale=0.12]{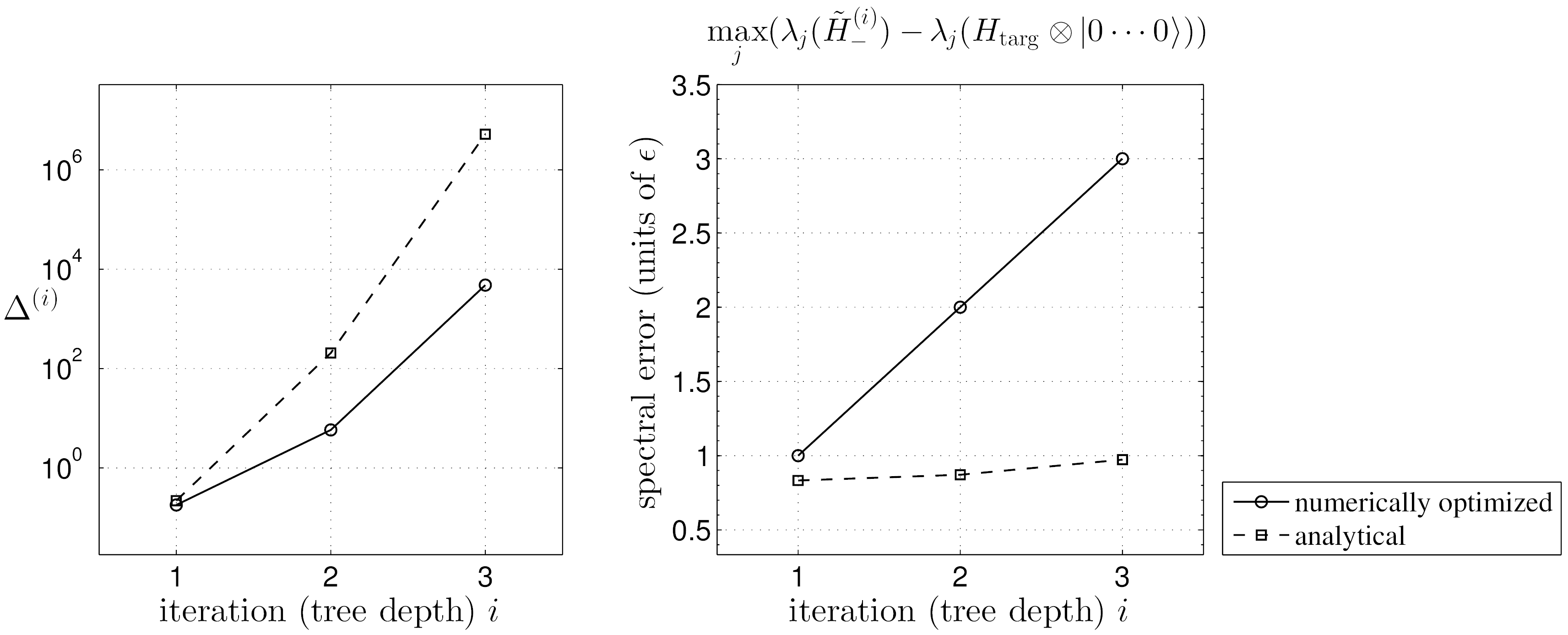}
\centerline{(b)\makebox[5.4cm][l]{ }(c)}
\caption{\normalsize (a) Reduction tree diagram for reducing a 7-body term to 3-body using parallel subdivision gadgets. Each $S_i$ is a single-qubit Pauli operator acting on qubit $i$. The vertical lines $|$ show where the subdivisions are made at each iteration to each term. (b) An example where we consider the target Hamiltonian $H_\text{targ}=\alpha S_1S_2S_3S_4S_5S_6S_7$ with $\alpha=5\times 10^{-3}$, $S_i=X_i$, $\forall i\in\{1,2,\cdots,7\}$, and reduce it to 3-body according to (a) up to error $\epsilon=5\times 10^{-4}$. This plot shows the energy gap applied onto the ancilla qubits introduced at each iteration. (c) The spectral error between the gadget Hamiltonian at each iteration $\tilde{H}^{(i)}$ and the target Hamiltonian $H_\text{targ}$. For both (b)(c) the data labelled as ``numerical'' correspond to the case where during each iteration $\Delta^{(i)}$ is optimized such that the maximum spectral difference between $\Pi_-^{(i)}\tilde{H}^{(i)}\Pi_-^{(i)}$ and $\tilde{H}^{(i-1)}\otimes P^{(i)}_-$ is $\epsilon$. For definitions of $\Delta^{(i)}$, $\tilde{H}^{(i)}$, $\Pi^{(i)}_-$ and $P^{(i)}_-$, see Eq.\ \ref{eq:k3_algo}. Those labelled as `analytical' correspond to cases where each iteration uses the gap bound derived in  Eq.\ \ref{eq:D_par_sub}.}
\label{fig:par_sub}
\end{figure}

We could show that after $s$ iterations, the maximum spectral error between $\Pi^{(s)}_-\tilde{H}^{(s)}\Pi^{(s)}_-$ and $\tilde{H}^{(0)}\bigotimes_{i=1}^s P^{(s)}_-$ is guaranteed to be within $s\epsilon$. Suppose we would like to make target Hamiltonian $\tilde{H}_0$, we construct a gadget $\tilde{H}=H^{(1)}+V^{(1)}$ according to algorithm \eqref{eq:k3_algo}, such that $|\lambda(\tilde{H}^{(1)})-\lambda(\tilde{H}^{(0)})|\le\epsilon$ for low-lying eigenvalues $\lambda(\cdot)$. Note that in a precise sense we should write $|\lambda(\Pi_-^{(1)}\tilde{H}^{(1)}\Pi_-^{(1)})-\lambda(\tilde{H}^{(0)}\otimes P_-^{(0)})|$. Since the projectors $\Pi_-^{(i)}$ and $P_-^{(i)}$ do not affect the low-lying spectrum of $\tilde{H}^{(i)}$ and $\tilde{H}^{(i-1)}$, for simplicity and clarity we write only $\tilde{H}^{(i-1)}$ and $\tilde{H}^{(i)}$. After $\tilde{H}^{(1)}$ is introduced, according to algorithm \eqref{eq:k3_algo} the second gadget $\tilde{H}^{(2)}$ is then constructed by considering the \emph{entire} $\tilde{H}^{(1)}$ as the new target Hamiltonian and introducing ancilla particles with unperturbed Hamiltonian $H^{(2)}$ and perturbation $V^{(2)}$ such that the low-energy spectrum of $\tilde{H}^{(2)}$ approximates the spectrum of $\tilde{H}^{(1)}$ up to error $\epsilon$. In other words $|\lambda(\tilde{H}^{(1)})-\lambda(\tilde{H}^{(2)})|\le\epsilon$. With the serial application of gadgets we have produced a sequence of Hamiltonians $\tilde{H}^{(0)}\rightarrow\tilde{H}^{(1)}\rightarrow\tilde{H}^{(2)}\rightarrow\cdots\rightarrow\tilde{H}^{(k)}$ where $\tilde{H}^{(0)}$ is the target Hamiltonian and each subsequent gadget Hamiltonian $\tilde{H}^{(i)}$ captures the \emph{entire} previous gadget $\tilde{H}^{(i-1)}$ in its low-energy sector with $|\lambda(\tilde{H}^{(i)})-\lambda(\tilde{H}^{(i-1)})|\le\epsilon$. Hence to bound the spectral error between the last gadget $\tilde{H}^{(k)}$ and the target Hamiltonian $\tilde{H}^{(0)}$ we could use triangle inequality: $|\lambda(\tilde{H}^{(s)})-\lambda(\tilde{H}^{(0)})|\le|\lambda(\tilde{H}^{(s)})-\lambda(\tilde{H}^{(s-1)})|+\cdots+|\lambda(\tilde{H}^{(1)})-\lambda(\tilde{H}^{(0)})|\le s\epsilon$.

%\noindent{A}fter $s$ iterations, the maximum spectral error between $\Pi^{(s)}_-\tilde{H}^{(s)}\Pi^{(s)}_-$ and $\tilde{H}^{(0)}\bigotimes_{i=1}^s P^{(s)}_-$ is guaranteed to be within $s\epsilon$. This is due to repeated use of the triangle inequality. Let $\lambda_j^{(0)}$ be the $j$-th lowest eigenvalue of $\tilde{H}^{(0)}$ and similarly define $\lambda_j^{(1)}$ and $\lambda_j^{(2)}$ for $\tilde{H}^{(1)}$ and $\tilde{H}^{(2)}$. The construction of the gadgets during each reduction step combined with Theorem \ref{th:perturbation} guarantees that $|\lambda_j^{(1)}-\lambda_j^{(0)}|\le\epsilon$ and $|\lambda_j^{(2)}-\lambda_j^{(1)}|\le\epsilon$. Hence $|\lambda_j^{(2)}-\lambda_j^{(0)}|\le |\lambda_j^{(1)}-\lambda_j^{(0)}|+|\lambda_j^{(2)}-\lambda_j^{(1)}|\le 2\epsilon$. Repeating the same idea $s$ times we have $|\lambda_j^{(s)}-\lambda_j^{(0)}|\le s\epsilon$ for all $j$ indexing the spectrum of the target Hamiltonian $\tilde{H^{(0)}}$.
$\quad$\\
$\quad$\\
\noindent{\emph{Total number of iterations for a $k$- to 3-body reduction.}} In general, given a $k$-body Hamiltonian, we apply the following parallel reduction scheme at each iteration until every term is 3-body: if $k$ is even, this reduces it to two $(k/2+1)$-body terms; if $k$ is odd, this reduces it to a $(\frac{k+1}{2}+1)$- and a $(\frac{k-1}{2}+1)$-body term. Define a function $f$ such that a $k$-body term needs $f(k)$ iterations to be reduced to 3-body. Then we have the recurrence
\begin{equation}
f(k)=\left\{
\begin{array}{cr}
\displaystyle f\left(\frac{k}{2}+1\right)+1 & \text{$k$ even} \\[0.1in]
\displaystyle f\left(\frac{k+1}{2}+1\right)+1 & \text{$k$ odd}
\end{array}\right.
\end{equation}
\noindent{}with $f(3)=0$ and $f(4)=1$. One can check that $f(k)=\lceil\log_2(k-2)\rceil$, $k\ge 4$ satisfies this recurrence. Therefore, using subdivision gadgets, one can reduce a $k$-body interaction to $3$-body in $s=\lceil\log_2(k-2)\rceil$ iterations and the spectral error between $\tilde{H}^{(s)}$ and $\tilde{H}^{(0)}$ is within $\lceil\log_2(k-2)\rceil\epsilon$.
$\quad$\\
$\quad$\\
\noindent{\emph{Gap scaling.}} From the iterative scheme shown previously one can conclude that $\Delta^{(i+1)}=\Theta(\epsilon^{-1}(\Delta^{(i)})^{3/2})$ for the $(i+1)^\text{th}$ iteration, which implies that for a total of $s$ iterations,
\begin{equation}
\Delta^{(s)}=\Theta\left(\epsilon^{-2[(3/2)^{s-1}-1]}(\Delta^{(1)})^{(3/2)^{s-1}}\right).
\end{equation}
\noindent{}Since $s=\lceil\log_2(k-2)\rceil$ and $\Delta^{(1)}=\Theta(\epsilon^{-1})$ we have
\begin{equation}
\Delta^{(s)}=\Theta\left(\epsilon^{-3(\frac{1}{2}\lceil k-2\rceil)^{\log_2(3/2)}-2}\right)=\Theta\left(\epsilon^{-\text{poly}(k)}\right)
\end{equation}
\noindent{}accumulating exponentially as a function of $k$. The exponential nature of the scaling with respect to $k$ agrees with results by Bravyi et al.\ \cite{BDLT08}. However, in our construction, due to the improvement of gap scaling in a single subdivision gadget from $\Delta=\Theta(\epsilon^{-2})$ to $\Theta(\epsilon^{-1})$, the scaling exponents in $\Delta^{(i+1)}=\Theta(\epsilon^{-1}(\Delta^{(i)})^{3/2})$ are also improved quadratically over those in \cite{BDLT08}, which is $\Delta^{(i+1)}=\Theta(\epsilon^{-2}(\Delta^{(i)})^{3})$.
$\quad$\\
$\quad$\\
\noindent{\emph{Qubit cost.}} Based on the reduction scheme described in Eq.\ \ref{eq:k3_algo} (illustrated in Fig.\ \ref{fig:par_sub}a for 7-body), the number of ancilla qubits needed for reducing a $k$-body term to 3-body is $k-3$. Suppose we are given a $k$-body target term $S_1S_2\cdots S_k$ (where all of the operators $S_i$ act on separate spaces) and we would like to reduce it to 3-body using the iterative scheme Eq.\ \ref{eq:k3_algo}. At each iteration, if we describe every individual subdivision gadget by a vertical line $|$ at the location where the partition is made, for example $S_1S_2S_3S_4|S_5S_6S_7$ in the case of the first iteration in Fig.\ \ref{fig:par_sub}a, then after $\lceil\log_2(k-2)\rceil$ iterations all the partitions made to the $k$-body term can be described as $S_1S_2|S_3|S_4|\cdots|S_{k-2}|S_{k-1}S_k$. Note that there are $k-3$ vertical lines in total, each corresponding to an ancilla qubit needed for a subdivision gadget. Therefore in total $k-3$ ancilla qubits are needed for reducing a $k$-body term to 3-body.
$\quad$\\
$\quad$\\
\noindent{\emph{Example: Reducing 7-body to 3-body.}} We have used numerics to test the reduction algorithm Eq.\ \ref{eq:k3_algo} on a target Hamiltonian $H_\text{targ}=\alpha S_1S_2S_3S_4S_5S_6S_7$. Here we let $S_i=X_i$, $\forall i\in\{1,2,\cdots,7\}$, $\epsilon=5\times 10^{-4}$ and $\alpha=5\times 10^{-3}$. During each iteration the values of $\Delta^{(i)}$ are assigned according to the lower bound in Eq.\ \ref{eq:D_par_sub}. From Fig.\ \ref{fig:par_sub}c we can see that the lower bounds are sufficient for keeping the total spectral error between $\tilde{H}_-^{(3)}$ and $\tilde{H}^{(0)}\bigotimes_{i=1}^3 P^{(i)}_-$ within $3\epsilon$. Furthermore, numerical search is also used at each iteration to find the minimum value of $\Delta^{(i)}$ so that the spectral error between $\Pi_-^{(i)}\tilde{H}^{(i)}\Pi_-^{(i)}$ and $\tilde{H}^{(i-1)}\bigotimes_{j=1}^i P^{(j)}_-$ is $\epsilon$.  The numerically found gaps $\Delta^{(i)}$ are much smaller than their analytical counterparts at each iteration (Fig.\ \ref{fig:par_sub}b), at the price that the error is larger (Fig.\ \ref{fig:par_sub}c). In both the numerical and the analytical cases, the error appears to accumulate linearly as the iteration proceeds.

\section{Improved Oliveira and Terhal 3- to 2-body gadget}\label{sec:3body}

\noindent{\bf Summary}. Subdivision gadgets cannot be used for reducing from 3- to 2-body; accordingly, the final reduction requires a different type of gadget \cite{KKR06,OT06,BDLT08}. Consider 3-body target Hamiltonian of the form $H_\text{targ}=H_\text{else}+\alpha A\otimes B\otimes C$. Here $A$, $B$ and $C$ are unit-norm Hermitian operators acting on separate spaces $\mathcal{A}$, $\mathcal{B}$ and $\mathcal{C}$. Here we focus on the gadget construction  introduced in Oliveira and Terhal \cite{OT06} and also used in Bravyi, DiVincenzo, Loss and Terhal \cite{BDLT08}. To accomplish the 3- to 2-body reduction, we introduce an ancilla spin $w$ and apply a penalty Hamiltonian $H=\Delta|1\rangle\langle{1}|_w$. We then add a perturbation $V$ of form,
\begin{equation}\label{eq:V3}
V=H_\text{else}+ \mu C \otimes|1\rangle\langle{1}|_w+(\kappa A +\lambda B)\otimes X_w+V_1 + V_2
\end{equation}
where $V_1$ and $V_2$ are 2-local compensation terms (details presented later in this section):
%\begin{eqnarray}\label{eq:3body_V12}
%V_1 & = & \displaystyle \frac{1}{\Delta}({\kappa}^2A^2+{\lambda}^2B^2)|0\rangle\langle{0}|_w+\frac{2{\kappa}{\lambda}}{\Delta}{A}\otimes{B} - \frac{1}{\Delta^2}({\kappa}^2A^2+{\lambda}^2B^2){\mu}{C}\otimes|0\rangle\langle{0}|_w  \\
%V_2 & = & \displaystyle -\frac{2{\kappa}{\lambda}}{\Delta^3}\text{sgn}(\alpha)\bigg[({\kappa}^2A^2+{\lambda}^2B^2)|0\rangle\langle{0}|_w+2{\kappa}{\lambda}{A}\otimes{B}\bigg].
%\end{eqnarray}
\begin{equation}\label{eq:3body_V12}
\begin{array}{ccl}
V_1 & = & \displaystyle \frac{1}{\Delta}(\kappa^2+\lambda^2)|0\rangle\langle{0}|_w+\frac{2\kappa\lambda}{\Delta}{A}\otimes{B}-\frac{1}{\Delta^2}(\kappa^2+\lambda^2)\mu{C}\otimes|0\rangle\langle{0}|_w   \\[0.1in]
V_2 & = & \displaystyle -\frac{2\kappa\lambda}{\Delta^3}\text{sgn}(\alpha)\bigg[(\kappa^2+\lambda^2)|0\rangle\langle{0}|_w+2\kappa\lambda{A}\otimes{B}\bigg].
\end{array}
\end{equation}
Here we let $\kappa=\text{sgn}(\alpha)\left({\alpha}/{2}\right)^{1/3}\Delta^{3/4}$, $\lambda=\left({\alpha}/{2}\right)^{1/3}\Delta^{3/4}$ and $\mu=\left({\alpha}/{2}\right)^{1/3}\Delta^{1/2}$.

\begin{figure*}
\begin{minipage}{0.4\textwidth}
\setlength{\unitlength}{1cm}
\begin{picture}(10,7.5)
\put(-4.8,0.44){\includegraphics[scale=0.4]{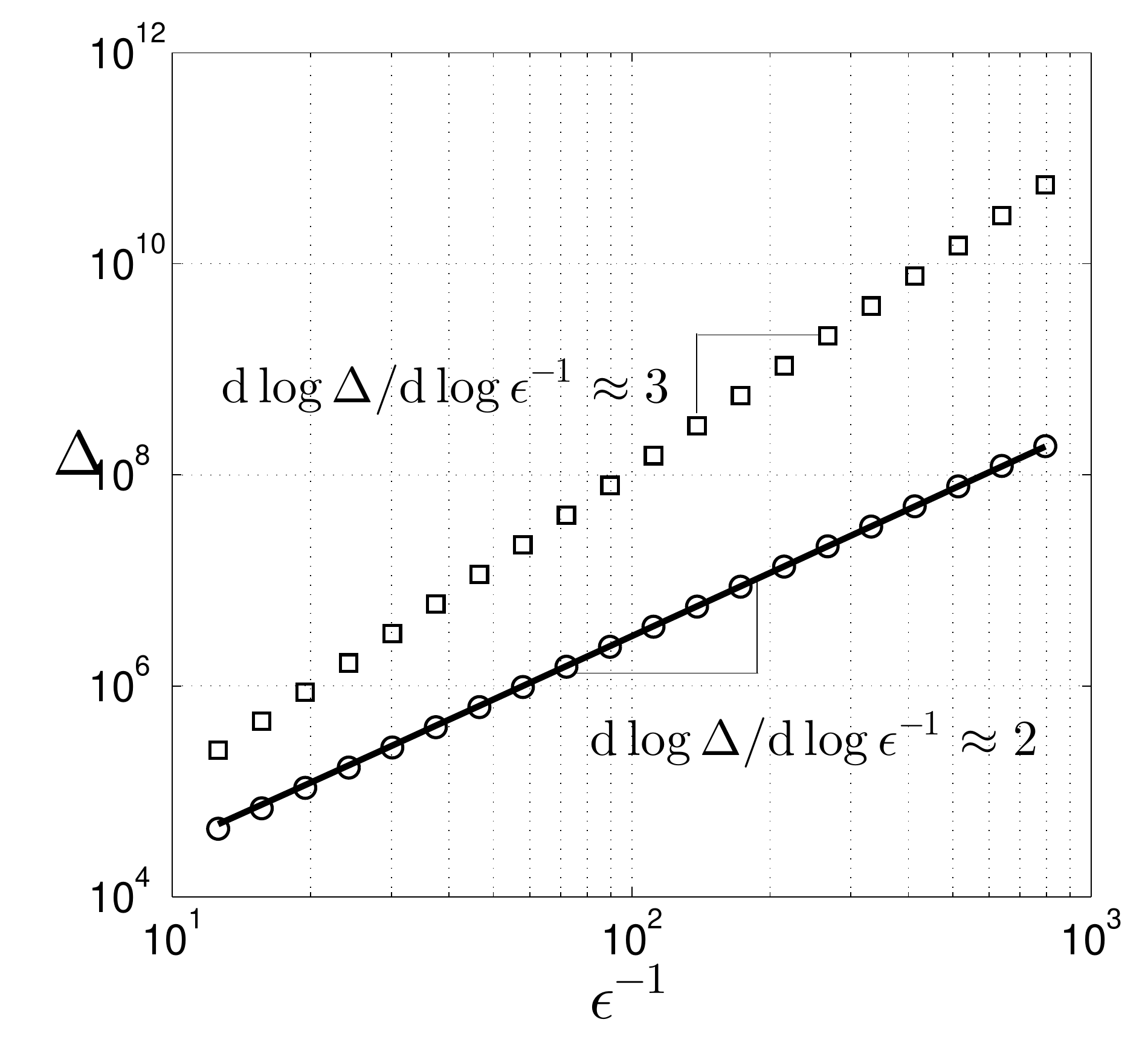}}
\put(2.8,0.5){\includegraphics[scale=0.76]{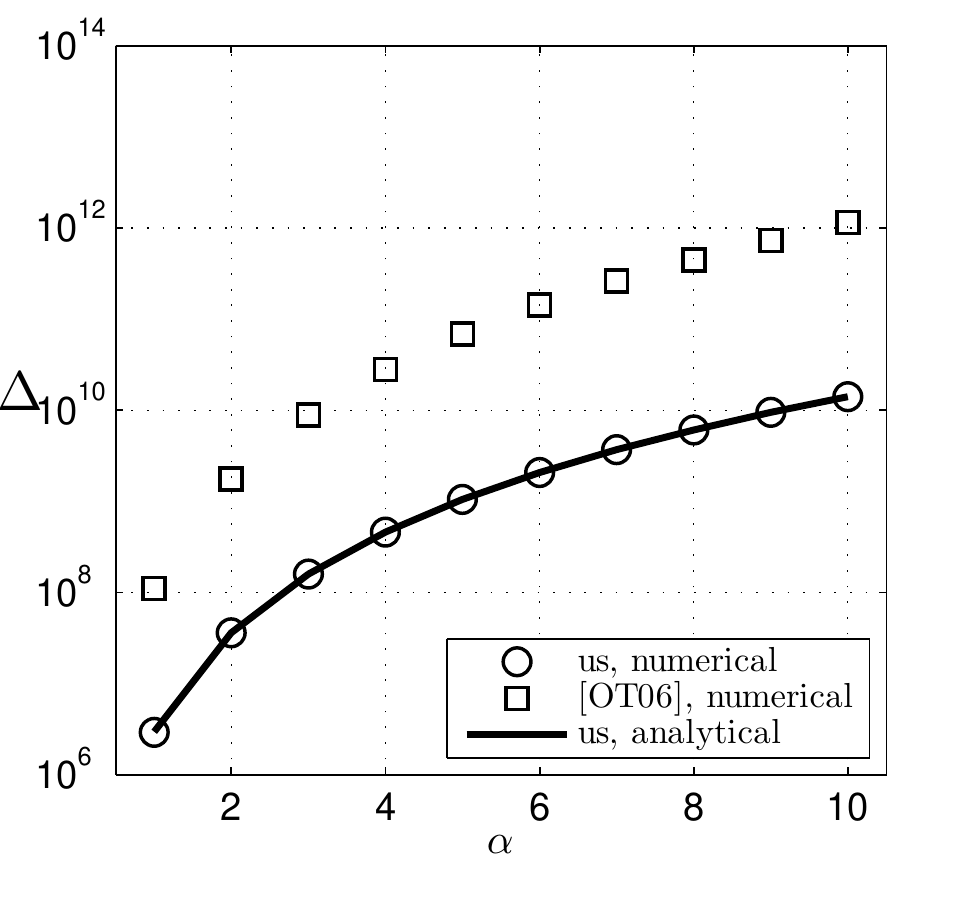}}
%\put(5.3,0.34){\includegraphics[scale=0.625]{sup_fig5b.png}}
\put(-0.9,0){\text{(a)}}
\put(6.5,0){\text{(b)}}
%\put(8.2,0){\text{(c)}}
\end{picture}
\end{minipage}
\caption{Comparison between our 3- to 2-body gadget with that of Oliveira and Terhal \cite{OT06}. As $\Delta$ is not explicitly assigned as a function of $\alpha$, $\|H_\text{else}\|$ and $\epsilon$ in \cite{OT06}, we numerically find the optimal $\Delta$ values for their constructions (marked as ``[OT06]"). (a) shows the scaling of the gap $\Delta$ as a function of error tolerance $\epsilon$. (b) shows the gap $\Delta$ as a function of the desired coupling $\alpha$. For the meanings of the labels in the legend, see Fig.\ \ref{fig:delta_compare_sub}. The fixed parameters in each subplots are: (a) $\|H_\text{else}\|=0$, $\alpha=1$. (b) $\epsilon=0.01$, $\|H_\text{else}\|=0$.  Note that our constructions have improved the $\Delta$ scaling for the ranges of $\alpha$ and $\epsilon$ considered. }
\label{fig:delta_compare_32}
\end{figure*}

For sufficiently large $\Delta$, the low-lying spectrum of the gadget Hamiltonian $\widetilde{H}$ captures the entire spectrum of $H_\text{targ}$ up to arbitrary error $\epsilon$. In the construction of \cite{BDLT08} it is shown that $\Delta=\Theta(\epsilon^{-3})$ is sufficient. In \cite{KKR06}, $\Delta=\Theta(\epsilon^{-3})$ is also assumed, though the construction of $V$ is slightly different from Eq.\ \ref{eq:V3}. By adding terms in $V$ to compensate for the perturbative error due to the modification, we find that $\Delta=\Theta(\epsilon^{-2})$ is sufficient for accomplishing the 3- to 2-body reduction:
\begin{equation}\label{eq:lower_bound_Delta3}
\Delta\ge\frac{1}{4}({-b+\sqrt{b^2-4c}})^2
\end{equation}
\noindent{}
where $b$ and $c$ are defined as
\begin{equation}
\begin{array}{ccl}
b & = & \displaystyle -\left[\xi+\frac{2^{4/3}\alpha^{2/3}}{\epsilon}(\max{z}+\eta+\xi^2)\right] \\[0.1in]
c & = & \displaystyle -\left(1+\frac{2^{4/3}\alpha^{2/3}}{\epsilon}\xi\right)(\max{z}+\eta)
\end{array}
\end{equation}
with $\max z=\|H_\text{else}\|+|\alpha|+\epsilon$, $\eta = \|H_\text{else}\|+2^{2/3}\alpha^{4/3}$ and $\xi=2^{-1/3}\alpha^{1/3}+2^{1/3}\alpha^{2/3}$. From Eq.\ \ref{eq:lower_bound_Delta3} we can see the lower bound to $\Delta$ is $\Theta({\epsilon^{-2}})$. Our improvement results in a power of $\epsilon^{-1}$ reduction in the gap. For the dependence of $\Delta$ on $\|H_\text{else}\|$, $\alpha$ and $\epsilon^{-1}$ for both the original \cite{OT06} and the optimized case, see Fig.\ \ref{fig:delta_compare_32}. Results show that the bound in Eq.\ \ref{eq:lower_bound_Delta3} is tight with respect to the minimum $\Delta$ numerically found that yields the spectral error between $\tilde{H}_-$ and $H_\text{targ}\otimes|0\rangle\langle{0}|_w$ to be $\epsilon$.
$\quad$\\
$\quad$\\
\noindent{\bf Analysis.} We will proceed by first presenting the improved construction of the 3- to 2-body gadget and then show that $\Delta=\Theta(\epsilon^{-2})$ is sufficient for the spectral error to be $\le\epsilon$.  Then we present the construction in the literature \cite{OT06,BDLT08} and argue that $\Delta=\Theta(\epsilon^{-3})$ is required for yielding a spectral error between $\tilde{H}$ and $H_\text{eff}$ within $\epsilon$ using this construction. 

\begin{figure}
\begin{tabular}{cc}

\setlength{\unitlength}{0.8cm}
\begin{picture}(8,5)
\put(1,2){\vector(1,0){6}}
\put(2,0){\vector(0,1){5}}
\put(7.2,1.8){\makebox{$r$}}
\put(1.7,1.6){\makebox{0}}
\put(1,5.1){\makebox{$f(r)$}}
\dashline{0.1}(2,4)(6,0)
\put(6.2,-0.3){\makebox{$1-2r$}}
\put(1.7,3.8){\makebox{1}}
\put(3.5,1.6){\makebox{$\frac{1}{2}$}}
\dashline{0.1}(4,0)(6,4)
\put(2.9,-0.3){\makebox{$4r-3$}}
\put(6,2){\circle*{0.15}}
\dashline{0.1}(4.666,0)(6,4)
\put(6,3){\makebox{$6r-5$}}
\put(6,1.6){\makebox{$1$}}
\put(4,2){\circle*{0.15}}
\put(6,4){\circle*{0.15}}
\put(4.6666666,1.3333333){\circle*{0.15}}
\put(4.5,2.3){\makebox{$\frac{2}{3}$}}
\put(4,2){\line(1,-1){.666}}
\put(6,4){\line(-1,-2){1.333}}
\dottedline{0.03}(4.6666666,1.3333333)(4.666,2)
\dottedline{0.03}(4.6666666,1.3333333)(2,1.333)
\put(1.4,1){\makebox{$-\frac{1}{3}$}}
\end{picture}
&
\setlength{\unitlength}{0.8cm}
\begin{picture}(8,5)
\put(1,2){\vector(1,0){6}}
\put(2,0){\vector(0,1){5}}
\put(7.2,1.8){\makebox{$r$}}
\put(1.7,1.6){\makebox{0}}
\put(1,5.1){\makebox{$f(r)$}}
\dashline{0.1}(2,4)(6,0)
\put(6.2,-0.3){\makebox{$1-2r$}}
\put(1.7,3.8){\makebox{1}}
\put(3.5,1.6){\makebox{$\frac{1}{2}$}}
%\dashline{0.1}(4,0)(6,4)
\put(6,2){\circle*{0.15}}
\dashline{0.1}(4.666,0)(6,4)
\put(6,3){\makebox{$6r-5$}}
\put(6,1.6){\makebox{$1$}}
\put(4,2){\circle*{0.15}}
\put(6,4){\circle*{0.15}}
\put(5,1){\circle*{0.15}}
\put(4.8,2.3){\makebox{$\frac{3}{4}$}}
\put(4,2){\line(1,-1){1}}
\put(6,4){\line(-1,-3){1}}
\dottedline{0.03}(5,1)(5,2)
\dottedline{0.03}(5,1)(2,1)
\put(1.4,0.8){\makebox{$-\frac{1}{2}$}} 
\end{picture}
\\[0.1in]
(a) & (b)
\end{tabular}
\caption{\normalsize The function $f(r)$ shows the dominant power of $\Delta$ in the error terms in the perturbative expansion. (a) When the error term $E_4$ in Eq.\ \ref{eq:Sz_3body}, which contributes to the $4r-3$ component of $f(r)$ in Eq.\ \ref{eq:fr}, is not compensated in the original construction by Oliveira and Terhal, the dominant power of $\Delta$ in the error term $f(r)$ takes minimum value of $-1/3$, indicating that $\Delta=\Theta(\epsilon^{-3})$ is required. (b) In the improved construction, $\min_{r\in(1/2,1)}f(r)=-1/2$ indicating that $\Delta=\Theta(\epsilon^{-2})$.}
\label{fig:3body_fg}
\end{figure}
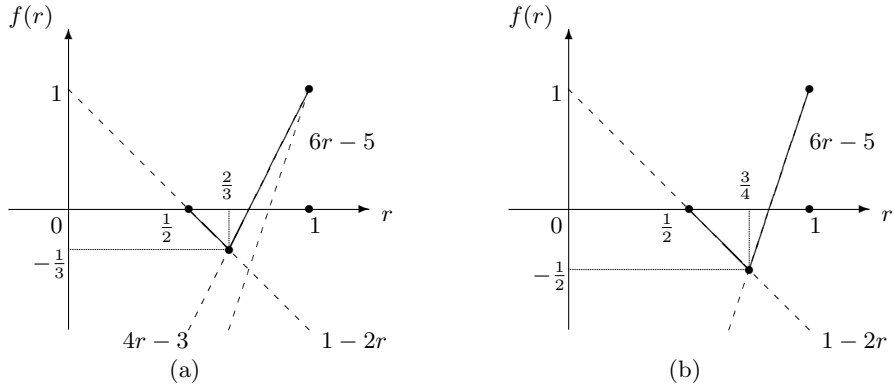

In the improved construction we define the perturbation $V$ as in Eq.\ \ref{eq:V3}. Here the coefficients are chosen to be $\kappa=\Theta(\Delta^{3/4})$, $\lambda=\Theta(\Delta^{3/4})$ and $\mu=\Theta(\Delta^{1/2})$. In order to show that the assigned powers of $\Delta$ in the coefficients are optimal, we introduce a parameter $r$ such that
\begin{equation}\label{eq:J_3body}
{\kappa}=\text{sgn}(\alpha)\left(\frac{\alpha}{2}\right)^{1/3}\Delta^r, 
\qquad {\lambda}=\left(\frac{\alpha}{2}\right)^{1/3}\Delta^r,
\qquad {\mu}=\left(\frac{\alpha}{2}\right)^{1/3}\Delta^{2-2r}.
\end{equation}
\noindent{}It is required that $\|V\|\le\Delta/2$ (Theorem \ref{th:perturbation}) for the convergence of the perturbative series. Therefore let $r<1$ and $2-2r<1$, which gives $1/2<r<1$. With the definitions $\mathcal{L}_-$ and $\mathcal{L}_+$ being the ground and excited state subspaces respectively, $V_-$, $V_+$, $V_{-+}$, $V_{+-}$ can be calculated as the following: 
\begin{equation}
\begin{array}{ccl}
V_- & = & \displaystyle \left[H_\text{else}+\frac{1}{\Delta}(\kappa A+\lambda B)^2-\frac{1}{\Delta}(\kappa^2+\lambda^2)\mu C - \frac{2\kappa\lambda}{\Delta^3}\text{sgn}(\alpha)(\kappa A+\lambda B)^2\right]\otimes|0\rangle\langle{0}|_w \\[0.1in]
V_+ & = & \displaystyle \left[H_\text{else}+\mu C + \frac{2\kappa\lambda}{\Delta}A\otimes B-\frac{4\kappa^2\lambda^2}{\Delta^3}\text{sgn}(\alpha)A\otimes B\right]\otimes|1\rangle\langle{1}|_w \\[0.1in]
V_{-+} & = & \displaystyle (\kappa A+\lambda B)\otimes |0\rangle\langle{1}|_w \\[0.1in]
V_{+-} & = & \displaystyle (\kappa A+\lambda B)\otimes |1\rangle\langle{0}|_w.
\end{array}
\end{equation}
The self-energy expansion, referring to Eq.\ \ref{eq:selfenergy}, becomes
\begin{equation}\label{eq:sigmaz_breakdown}
\begin{array}{ccl}
\Sigma_-(z) & = & \displaystyle V_-+\frac{1}{z-\Delta}V_{-+}V_{+-}+\frac{1}{(z-\Delta)^2}V_{-+}V_+V_{+-}+\sum_{k=2}^\infty\frac{V_{-+}V_+^kV_{+-}}{(z-\Delta)^{k+1}} \\[0.1in]
& = & \displaystyle 
\underbrace{H_\text{else}}_{(a)}
+
\underbrace{\frac{1}{\Delta}(\kappa A+\lambda B)^2}_{(b)}
\underbrace{-\frac{1}{\Delta}(\kappa^2+\lambda^2)\mu C}_{(c)} 
\underbrace{- \frac{2\kappa\lambda}{\Delta^3}\text{sgn}(\alpha)(\kappa A+\lambda B)^2}_{(d)} 
+ 
\underbrace{\frac{1}{z-\Delta}(\kappa A+\lambda B)^2}_{(e)} \\[0.1in]
& + & \displaystyle \frac{1}{(z-\Delta)^2}(\kappa A+\lambda B)\left[
\underbrace{H_\text{else}}_{(f)}
+
\underbrace{\mu C}_{(g)} 
+ 
\underbrace{\frac{2\kappa\lambda}{\Delta}A\otimes B}_{(h)}
\underbrace{-\frac{4\kappa^2\lambda^2}{\Delta^3}\text{sgn}(\alpha)A\otimes B}_{(i)}
\right](\kappa A+\lambda B) \\[0.1in]
& + & \displaystyle \underbrace{\sum_{k=2}^\infty\frac{V_{-+}V_+^kV_{+-}}{(z-\Delta)^{k+1}}}_{(j)}.
\end{array}
\end{equation}
Now we rearrange the terms in the self energy expansion so that the target Hamiltonian arising from the leading order terms can be separated from the rest, whcih are error terms. Observe that term $(g)$ combined with the factors outside the bracket could give rise to a 3-body $A\otimes B\otimes C$ term:
\begin{equation}
\begin{array}{ccl}
\displaystyle \frac{1}{(z-\Delta)^2}(\kappa A+\lambda B)^2\mu C & = & \displaystyle
\underbrace{\frac{2\kappa\lambda\mu}{\Delta^2}A\otimes B\otimes C}_{(g_1)} + 
\underbrace{\left(\frac{1}{(z-\Delta)^2}-\frac{1}{\Delta^2}\right)2{\kappa}{\lambda}{\mu}{A}\otimes{B}\otimes{C}}_{(g_2)} \\[0.1in]
& + & \displaystyle 
\underbrace{\frac{1}{(z-\Delta)^2}(\kappa^2+\lambda^2)\mu C}_{(g_3)}.
\end{array}
\end{equation}
Here $(g_1)$ combined with term $(a)$ in \eqref{eq:sigmaz_breakdown} gives $H_\text{targ}$. $(g_2)$ and $(g_3)$ are error terms. Now we further rearrange the error terms as the following. We combine term $(b)$ and $(e)$ to form $E_1$, term $(c)$ and $(g_3)$ to form $E_2$, term $(f)$ and the factors outside the bracket to be $E_3$. Rename $(g_2)$ to be $E_4$. Using the identity $({\kappa}{A}+{\lambda}{B})({A}\otimes{B})({\kappa}{A}+{\lambda}{B})=\text{sgn}(\alpha)({\kappa}{A}+{\lambda}{B})^2$ we combine term $(d)$ and $(h)$ along with the factors outside the bracket to be $E_5$. Rename $(i)$ to be $E_6$ and $(j)$ to be $E_7$. The rearranged self-energy expanision reads
\begin{equation}\label{eq:Sz_3body_new}
\begin{array}{ccl}
\Sigma_-(z) & = & \displaystyle \bigg[\underbrace{H_\text{else}+\frac{2{\kappa}{\lambda}{\mu}}{\Delta^2}{A}\otimes{B}\otimes{C}}_{H_\text{targ}}+\underbrace{\left(\frac{1}{\Delta}+\frac{1}{z-\Delta}\right)({\kappa}{A}+{\lambda}{B})^2}_{E_1} \\[0.1in]
& + & \displaystyle \underbrace{\left(\frac{1}{(z-\Delta)^2}-\frac{1}{\Delta^2}\right)({\kappa}^2+{\lambda}^2){\mu}{C}}_{E_2}+\underbrace{\frac{1}{(z-\Delta)^2}({\kappa}{A}+{\lambda}{B})H_\text{else}({\kappa}{A}+{\lambda}{B})}_{E_3} \\[0.1in]
& + & \displaystyle \underbrace{\left(\frac{1}{(z-\Delta)^2}-\frac{1}{\Delta^2}\right)2{\kappa}{\lambda}{\mu}{A}\otimes{B}\otimes{C}}_{E_4}+\underbrace{\left(\frac{1}{(z-\Delta)^2}-\frac{1}{\Delta^2}\right)\frac{2{\kappa}{\lambda}}{\Delta}\text{sgn}(\alpha)({\kappa}{A}+{\lambda}{B})^2}_{E_5} \\[0.1in]
& - & \underbrace{\frac{1}{(z-\Delta)^2}\cdot\frac{4{\kappa}^2{\lambda}^2}{\Delta^3}({\kappa}{A}+{\lambda}{B})^2}_{E_6}\bigg]\otimes|0\rangle\langle{0}|_w+ \underbrace{\sum_{k=2}^\infty\frac{V_{-+}V_+^kV_{+-}}{(z-\Delta)^{k+1}}}_{E_7}.
\end{array}
\end{equation}
\noindent{}We bound the norm of each error term in the self energy expansion Eq.\ \ref{eq:Sz_3body_new} by substituting the definitions of $\kappa$, $\lambda$ and $\mu$ in Eq.\ \ref{eq:J_3body} and letting $z$ be the maximum value permitted by Theorem \ref{th:perturbation} which is $\max z=|\alpha|+\epsilon+\|H_\text{else}\|$:
\begin{equation}\label{eq:E_1E_2}
\|E_1\|\le\displaystyle\frac{\max{z}{\cdot}2^{4/3}\alpha^{2/3}\Delta^{2r-1}}{\Delta-\max{z}}=\Theta(\Delta^{2r-2}),\qquad
\|E_2\| \le \displaystyle\frac{(2\Delta-\max{z})\max{z}}{(\Delta-\max{z})^2}\cdot\alpha=\Theta(\Delta^{-1})
\end{equation}
\begin{equation}\label{eq:E_3E_4}
\|E_3\| \le  \displaystyle\frac{2^{4/3}\alpha^{2/3}\Delta^{2r}\|H_\text{else}\|}{(\Delta-\max{z})^2}=\Theta(\Delta^{2r-2}),\qquad
\|E_4\| \le \displaystyle\frac{(2\Delta-\max{z})\max{z}}{(\Delta-\max{z})^2}\cdot\alpha=\Theta(\Delta^{-1})
\end{equation}
\begin{equation}\label{eq:E_5E_6}
\|E_5\| \le \displaystyle\frac{(2\Delta-\max{z})\max{z}}{(\Delta-\max{z})^2}\cdot{2^{5/3}}\alpha^{4/3}\Delta^{4r-3}=\Theta(\Delta^{4r-4}),\qquad
\|E_6\| \le \displaystyle \frac{4\alpha^2\Delta^{6r-3}}{(\Delta-\max{z})^2}=\Theta(\Delta^{6r-5})
\end{equation}
\begin{equation}\label{eq:E_7}
\begin{array}{ccl}
\|E_7\| & \le & \displaystyle \sum_{k=2}^\infty\left\|\frac{({\kappa}A+{\lambda}B)\left(H_\text{else}+{\mu}C+\frac{2{\kappa}{\lambda}}{\Delta}\left(1+\frac{2{\kappa}{\lambda}}{\Delta^2}\right)A\otimes B\right)^k({\kappa}A+{\lambda}B)}{(\Delta-\max{z})^{k+1}}\right\| \\[0.1in]
& \le & \displaystyle \frac{2^{4/3}\alpha^{2/3}\Delta^{2r}}{(\Delta-\max{z})}\sum_{k=2}^\infty\frac{\left(\|H_\text{else}\|+2^{-1/3}\alpha^{1/3}\Delta^{2-2r}+2^{1/3}\alpha^{2/3}\Delta^{2r-1}+2^{2/3}\alpha^{4/3}\Delta^{4r-3}\right)^k}{(\Delta-\max{z})^k} \\[0.1in]
& = & \displaystyle \Theta(\Delta^{\max\{1-2r,6r-5,10r-9\}}).
\end{array}
\end{equation}
\noindent{}Now the self energy expansion can be written as
\[\Sigma_-(z)=H_\text{targ}\otimes|0\rangle\langle{0}|_w+\Theta(\Delta^{f(r)})\]where the function $f(r)<0$ determines the dominant power in $\Delta$ from $\|E_1\|$ through $\|E_6\|$:
\begin{equation}\label{eq:fr_new}
f(r)=\max\{1-2r,6r-5\},\quad\frac{1}{2}<r<1.
\end{equation}
In order to keep the error $O(\epsilon)$, it is required that $\Delta=\Theta(\epsilon^{1/f(r)})$. To optimize the gap scaling as a function of $\epsilon$, $f(r)$ must take the minimum value. As is shown in Fig.\ \ref{fig:3body_fg}b, when $r=3/4$, the minimum value $f(r)=-1/2$ is obtained, which corresponds to $\Delta=\Theta(\epsilon^{-2})$. We have hence shown that the powers of $\Delta$ in the assignments of $\kappa$, $\lambda$ and $\mu$ in Eq.\ \ref{eq:J_3body} are optimal for the improved gadget construction. The optimal scaling of $\Theta(\epsilon^{-2})$ is also numerically confirmed in Fig.\ \ref{fig:delta_compare_32}a. As one can see, the optimized slope $\text{d}\log\Delta/\text{d}\log\epsilon^{-1}$ is approximately 2 for small $\epsilon$.

One natural question to ask next is whether it is possible to further improve the gap scaling as a function of $\epsilon$. This turns out to be difficult. Observe that the $6r-5$ component of $f(r)$ in Eq.\ \ref{eq:fr_new} comes from $E_6$ and $E_7$ in Eq.\ \ref{eq:Sz_3body_new}. In $E_7$, the $\Theta(\Delta^{6r-5})$ contribution is attributed to the term $\frac{1}{\Delta}({\kappa}{A}+{\lambda}{B})^2$ in $V_1$ of Eq.\ \ref{eq:3body_V12}, which is intended for compensating the $2^\text{nd}$ order perturbative term and therefore cannot be removed from the construction.

We now let $r=3/4$ be a fixed constant and derive the lower bound for $\Delta$ such that for given $\alpha$, $H_\text{else}$ and $\epsilon$, the spectral error between the effective Hamiltonian $H_\text{eff}=H_\text{targ}\otimes|0\rangle\langle{0}|_w$ and $\tilde{H}_-$ is within $\epsilon$. This amounts to satisfying the condition of Theorem \ref{th:perturbation}: 
\begin{equation}\label{eq:th_cond}
\|\Sigma_-(z)-H_\text{eff}\|\le\epsilon.
\end{equation} 
\noindent{}Define the total error $E=\Sigma_-(z)-H_\text{eff}=E_1+\cdots+E_7$. For convenience we also define $\eta=\|H_\text{else}\|+2^{2/3}\alpha^{4/3}$ and $\xi=2^{-1/3}\alpha^{1/3}+2^{1/3}\alpha^{2/3}$. Then
\begin{equation}\label{eq:E_7}
\begin{array}{ccl}
\|E_7\| & \le & \displaystyle \frac{2^{4/3}\alpha^{2/3}\Delta^{3/2}}{\Delta-\max{z}}\sum_{k=2}^\infty\frac{(\eta+\xi\Delta^{1/2})^k}{(\Delta-\max{z})^k}=\frac{2^{4/3}\alpha^{2/3}\Delta^{3/2}}{\Delta-\max{z}-(\eta+\xi\Delta^{1/2})}\left(\frac{\eta+\xi\Delta^{1/2}}{\Delta-\max{z}}\right)^2.
\end{array}
\end{equation} 
\noindent{}The upper bound for $\|E\|$ is then found by summing over Eq.\ \ref{eq:E_1E_2},  \ref{eq:E_3E_4}, \ref{eq:E_5E_6} and \ref{eq:E_7}:

\begin{equation}\label{eq:E_norm}
\begin{array}{ccl}
\|E\| & \le & \displaystyle \frac{\max{z}{\cdot}2^{4/3}\alpha^{2/3}\Delta^{1/2}}{\Delta-\max{z}}+\frac{(2\Delta-\max{z})\max{z}}{(\Delta-\max{z})^2}\cdot{2}^{4/3}\alpha^{3/2}\xi+\frac{2^{4/3}\alpha^{2/3}\Delta^{3/2}\eta}{(\Delta-\max{z})^2} \\[0.2in]
& + & \displaystyle  \frac{2^{4/3}\alpha^{2/3}\Delta^{3/2}}{\Delta-\max{z}-(\eta+\xi\Delta^{1/2})}\left(\frac{\eta+\xi\Delta^{1/2}}{\Delta-\max{z}}\right)^2. \\[0.2in]
\end{array}
\end{equation}
\noindent{}By rearranging the terms in Eq.\ \ref{eq:E_norm} we arrive at a simplified expression for the upper bound presented below. Requiring the upper bound of $\|E\|$ to be within $\epsilon$ gives
\begin{equation}\label{eq:E_delta_bound}
\begin{array}{ccl}
\|E\| & \le & \displaystyle 2^{4/3}\alpha^{2/3}\frac{(\max{z}+\eta+\xi^2)\Delta^{1/2}+\xi(\max{z}+\eta)}{\Delta-\xi\Delta^{1/2}-(\max{z}+\eta)}\le \epsilon.
\end{array}
\end{equation}
\noindent{}Eq.\ \ref{eq:E_delta_bound} is a quadratic constraint with respect to $\Delta^{1/2}$. Solving the inequality gives the lower bound of $\Delta$ given in Eq.\ \ref{eq:lower_bound_Delta3}. Note here that $\Delta=\Theta(\epsilon^{-2})$, which improves over the previously assumed $\Delta=\Theta(\epsilon^{-3})$ in the literature \cite{OT06,KKR06,BDLT08}. This bound is shown in Fig.\ \ref{fig:delta_compare_32}b as the ``analytical lower bound". Comparison between the analytical lower bound and the numerically optimized gap in Fig.\ \ref{fig:delta_compare_32}b indicates that the lower bound is relatively tight when $\|H_\text{else}\|=0$.
$\quad$\\
$\quad$\\
\noindent{\emph{Comparison with Oliveira and Terhal \cite{OT06}.}} Given operators $Q$, $R$ and $T$ acting on separate spaces $\mathcal{A}$, $\mathcal{B}$ and $\mathcal{C}$ respectively, the 3- to 2-body construction in \cite{OT06,KKR06} approximates the target Hamiltonian $H_\text{targ}=H_\text{else}+Q\otimes R\otimes T$. In order to compare with their construction, however, we let $\alpha=\|Q\|\cdot\|R\|\cdot\|T\|$ and define $Q=\alpha^{1/3}A$, $R=\alpha^{1/3}B$ and $T=\alpha^{1/3}C$. Hence the target Hamiltonian $H_\text{targ}=H_\text{else}+\alpha A\otimes B\otimes C$ with $A$, $B$ and $C$ being unit-norm Hermitian operators. Introduce an ancilla qubit $w$ and apply the penalty Hamiltonian $H=\Delta|1\rangle\langle{1}|_w$. In the construction by Oliveira and Terhal \cite{OT06}, the perturbation $V$ is defined as
\begin{equation}\label{eq:OT_V}
V=H_\text{else}\otimes\openone_w+{\mu}C\otimes|1\rangle\langle{1}|_w+({\kappa} A+{\lambda} B)\otimes X_w+V'_1
\end{equation}
\noindent{}where the compensation term $V'_1$ is
\begin{equation}\label{eq:V1_old}
\displaystyle
V'_1=\frac{1}{\Delta}({\kappa}A+{\lambda}B)^2-\frac{1}{\Delta^2}(\kappa^2A^2+\lambda^2B^2){\mu}C.
\end{equation}
\noindent{}Comparing Eq.\ \ref{eq:V1_old} with the expression for $V_1$ in Eq.\ \ref{eq:3body_V12}, one observes that $V_1$ slightly improves over $V'_1$ by projecting 1-local terms to $\mathcal{L}_-$ so that $V$ will have less contribution to $V_+$, which reduces the high order error terms in the perturbative expansion. However, this modification comes at a cost of requiring more 2-local terms in the perturbation $V$. 

From the gadget construction shown in \cite[Eq.\ 26]{OT06}, the equivalent choices of the coefficients ${\kappa}$, ${\lambda}$ and ${\mu}$ are
\begin{equation}\label{eq:J_3body_old}
{\kappa}=-\left(\frac{\alpha}{2}\right)^{1/3}\frac{1}{\sqrt{2}}\Delta^r,\quad{\lambda}=\left(\frac{\alpha}{2}\right)^{1/3}\frac{1}{\sqrt{2}}\Delta^r,\quad
{\mu}=-\left(\frac{\alpha}{2}\right)^{1/3}\Delta^{2-2r}
\end{equation}
\noindent{}where $r=2/3$ in the constructions used in \cite{OT06,BDLT08}. In fact this value of $r$ is optimal for the construction in the sense that it leads to the optimal gap scaling $\Delta=\Theta(\epsilon^{-3})$. Expanding the self-energy to $3^\text{rd}$ order, following a similar procedure as in \eqref{eq:sigmaz_breakdown}, we have

\begin{equation}\label{eq:Sz_3body}
\begin{array}{ccl}
\Sigma_-(z) & = & \displaystyle \bigg[
\underbrace{H_\text{else}+\frac{2{\kappa}{\lambda}{\mu}}{\Delta^2}{A}\otimes{B}\otimes{C}}_{H_\text{targ}}+
\underbrace{\left(\frac{1}{\Delta}+\frac{1}{z-\Delta}\right)({\kappa}{A}+{\lambda}{B})^2}_{E_1} \\[0.1in]
& + & \displaystyle 
\underbrace{\left(\frac{1}{(z-\Delta)^2}-\frac{1}{\Delta^2}\right)({\kappa}^2A^2+{\lambda}^2B^2){\mu}{C}}_{E_2}+
\underbrace{\frac{1}{(z-\Delta)^2}({\kappa}{A}+{\lambda}{B})H_\text{else}({\kappa}{A}+{\lambda}{B})}_{E_3} \\[0.1in]
& + & \displaystyle 
\underbrace{\frac{1}{(z-\Delta)^2}\cdot\frac{1}{\Delta}(\kappa A+\lambda B)^4}_{E_4}-
\underbrace{\frac{1}{(z-\Delta)^2}\cdot\frac{1}{\Delta^2}(\kappa^2A^2+\lambda^2B^2)\mu(\kappa A+\lambda B)^2\otimes C}_{E_5}\bigg]\otimes|0\rangle\langle{0}|_w \\[0.1in]
& + & \displaystyle \underbrace{\sum_{k=2}^\infty\frac{V_{-+}V_+^kV_{+-}}{(z-\Delta)^{k+1}}}_{E_6}.
\end{array}
\end{equation}
Similar to the derivation of Eq.\ \ref{eq:E_1E_2}, \ref{eq:E_3E_4}, and \ref{eq:E_5E_6} by letting $z\mapsto\max z$, where $\max z=|\alpha|+\epsilon+\|H_\text{else}\|$ is the largest value of $z$ permitted by the Theorem \ref{th:perturbation}, and using the triangle inequality to bound the norm, we can bound the norm of the error terms $E_1$ through $E_6$. For example, \[\|E_1\|\le\left(\frac{1}{\Delta-\max z}-\frac{1}{\Delta}\right)2^2\cdot\left(\frac{\alpha}{2}\right)^{2/3}\Delta^{2r}=\Theta(\Delta^{2r-2}).\] Applying the same calculation to $E_2,E_3,\cdots$ we find that $\|E_2\|=\Theta(\Delta^{-1})$, $\|E_3\|=\Theta(\Delta^{2r-2})$, $\|E_4\|=\Theta(\Delta^{4r-3})$, $\|E_5\|=\Theta(\Delta^{4r-4})$. The norm of the high order terms $E_6$ can be bounded as
\begin{equation}
\begin{array}{ccl}
\|E_6\| & \le & \displaystyle \sum_{k=2}^\infty\frac{\|V_{-+}\|\cdot\|V_+\|^k\cdot\|V_{+-}\|}{(\Delta-\max(z))^{k+1}}\le\frac{4\left(\frac{\alpha}{2}\right)^{1/3}\Delta^{2r}}{\Delta-\max(z)}\sum_{k=2}^\infty\left(\frac{\rho}{\Delta-\max(z)}\right)^k \\[0.1in]
& = & \displaystyle \frac{2^{4/3}\alpha^{2/3}\Delta^{2r}}{\Delta-\max(z)-\rho}\left(\frac{\rho}{\Delta-\max(z)}\right)^2=\Theta(\Delta^{2r-1+2\max\{1-2r,2r-2\}})=\Theta(\Delta^{\max\{1-2r,6r-5\}})
\end{array}
\end{equation}
\noindent{}where $\rho=\|H_\text{else}\|+2^{-1/3}\alpha^{1/3}\Delta^{2-2r}+2^{1/3}\alpha^{2/3}\Delta^{2r-1}$. If we again write the self energy expansion Eq.\ \ref{eq:Sz_3body} as \[\Sigma_-(z)=H_\text{targ}\otimes|0\rangle\langle{0}|_w+\Theta(\Delta^{f(r)}),\]the function $f(r)<0$, which determines the dominant power in $\Delta$ among $E_1$ through $E_6$, can be found as 
\begin{equation}\label{eq:fr}
f(r)=\max\{1-2r,2r-2,4r-3,6r-5\},\quad\frac{1}{2}<r<1.
\end{equation}
\noindent{}Similar to the discussion after Eq.\ \ref{eq:fr_new}, the optimal scaling of $\Delta=\Theta(\epsilon^{1/f(r)})$ gives $r=\text{argmin}f(r)=2/3$, when $f(r)=-1/3$ and $\Delta=\Theta(\epsilon^{-3})$, as is shown in Fig.\ \ref{fig:3body_fg}a. Note that the $4r-3$ component in $f(r)$, Eq.\ \ref{eq:fr}, comes from the error term $E_4$ in Eq.\ \ref{eq:Sz_3body}. The idea for improving the gadget construction comes from the observation in Fig.\ \ref{fig:3body_fg}a that when we add a term in $V$ to compensate for $E_4$, the dominant power of $\Delta$ in the perturbation series, $f(r)$, could admit a lower minimum as shown in Fig.\ \ref{fig:3body_fg}b. In the previous calculation we have shown that this is indeed the case and the minimum value of $f(r)$ becomes $-1/2$ in the improved case, indicating that $\Delta=\Theta(\epsilon^{-2})$ is sufficient for keeping the error terms $O(\epsilon)$.

\section{Creating 3-body gadget from local X}\label{sec:5th_32}

\noindent{\bf Summary}. In general, terms in perturbative gadgets involve mixed couplings (e.g. $X_i Z_j$). Although such couplings can be realized by certain gadget constructions \cite{BL07},
%\yudong{Here ``creation gadgets" is changed to ``certain gadget constructions".} 
physical couplings of this type are difficult to realize in an experimental setting. However, there has been significant progress towards experimentally implementing Ising models with transverse fields of the type \cite{2006cond.mat..8253H}:
\begin{equation}\label{eq:dwave}
H_{ZZ}=\sum_i\delta_iX_i+\sum_ih_iZ_i+\sum_{i,j}J_{ij} Z_iZ_j.
\end{equation}
Accordingly, an interesting question is whether we can approximate 3-body terms such as $\alpha \cdot Z_i\otimes Z_j\otimes Z_k$ using a Hamiltonian of this form. This turns out to be possible by employing a perturbative calculation which considers terms up to $5^\text{th}$ order. 

Similar to the 3- to 2-body reduction discussed previously, we introduce an ancilla $w$ and apply the Hamiltonian $H=\Delta|1\rangle\langle{1}|_w$. We apply the perturbation
\begin{equation}\label{eq:V5}
V = H_\text{else}+\mu(Z_i+Z_j+Z_k)\otimes|1\rangle\langle{1}|_w  + \mu\openone\otimes X_w+V_\textrm{comp}
\end{equation}
where $\mu = \left(\alpha \Delta^4 / 6\right)^{1/5}$ and $V_\textrm{comp}$ is
\begin{equation}
\begin{array}{ccl}
V_\textrm{comp} & = & \displaystyle \frac{\mu^2}{\Delta} |0\rangle\langle{0}|_w-\left(\frac{\mu^3}{\Delta^2}+ 7 \frac{\mu^5}{\Delta^4}\right)\left(Z_i+Z_j+Z_k \right)\otimes|0\rangle\langle{0}|_w+ \frac{\mu^4}{\Delta^3}\left(3 \openone +2 Z_i Z_j+2 Z_i Z_k +2 Z_j Z_k\right).
\end{array}
\end{equation}

To illustrate the basic idea of the $5^\text{th}$ order gadget, define subspaces $\mathcal{L}_-$ and $\mathcal{L}_+$ in the usual way and define $P_-$ and $P_+$ as projectors into these respective subspaces. Then the second term in Eq.\ \ref{eq:V5} with $\otimes|1\rangle\langle{1}|_w$ contributes a linear combination $\mu Z_i+\mu Z_j+ \mu Z_k$ to $V_+=P_+VP_+$. The third term in Eq.\ \ref{eq:V5} induces a transition between $\mathcal{L}_-$ and $\mathcal{L}_+$ yet since it operates trivially on qubits 1-3, it only contributes a constant $\mu$ to the projections $V_{-+}=P_-VP_+$ and $V_{+-}=P_+VP_-$. In the perturbative expansion, the $5^\text{th}$ order contains a term
\begin{equation}\label{eq:V55}
\frac{V_{-+}V_+V_+V_+V_{+-}}{(z-\Delta)^4}=\frac{\mu^5 (Z_i+Z_j+Z_k)^3}{(z-\Delta)^4}
\end{equation}
due to the combined the contribution of the second and third term in Eq.\ \ref{eq:V5}.
\noindent{}This yields a term proportional to $\alpha\cdot Z_i\otimes Z_j \otimes Z_k$ along with some 2-local error terms. These error terms, combined with the unwanted terms that arise at $1^\text{st}$ through $4^\text{th}$ order perturbation, are compensated by $V_\text{comp}$. Note that terms at 6$^\textrm{th}$ order and higher are $\Theta(\Delta^{-1/5})$. This means in order to satisfy the gadget theorem of Kempe \emph{et al.} (\cite[Theorem 3]{KKR06}, or Theorem I.1) $\Delta$ needs to be $\Theta(\epsilon^{-5})$. This is the first perturbative gadget that simulates a 3-body target Hamiltonian using the Hamiltonian Eq.\ \ref{eq:dwave}. By rotating the ancilla space, subdivision gadgets can also be implemented using this Hamiltonian: in the $X$ basis, $Z$ terms will induce a transition between the two energy levels of $X$. Therefore $Z_i Z_j$ coupling could be used for a perturbation of the form in Eq.\ \ref{eq:2body_V} in the rotated basis. In principle using {the transverse Ising model in Eq.\ \ref{eq:dwave},} one can reduce some {diagonal} $k$-body Hamiltonian to 3-body by iteratively applying the subdivision gadget and then to 2-body by using the 3-body reduction gadget.
$\quad$\\
$\quad$\\
\noindent{\bf Analysis.} Similar to the gadgets we have presented so far, we introduce an ancilla spin $w$. Applying an energy gap $\Delta$ on the ancilla spin gives the unperturbed Hamiltonian $H=\Delta|1\rangle\langle{1}|_w$. We then perturb the Hamiltonian $H$ using a perturbation $V$ described in \eqref{eq:V5}. Using the same definitions of subspaces $\mathcal{L}_+$ and $\mathcal{L}_-$ as the previous 3-body gadget, the projections of $V$ into these subspaces can be written as
\begin{equation}\label{eq:V_proj_fifth}
\begin{array}{ccl}
V_+ & = & \displaystyle \bigg\{H_\text{else} + \mu(Z_1+Z_2+Z_3) + \frac{{\mu}^4}{\Delta^3}\big[3{\openone}+ 2(Z_1Z_2+Z_1Z_3+Z_2Z_3)\big]\bigg\}\otimes|1\rangle\langle{1}|_w \\[0.1in]
%& + & \displaystyle 2({\kappa}{\lambda}Z_1Z_2+{\kappa}{\mu}Z_1Z_3+{\lambda}{\mu}Z_2Z_3)\big]\bigg\}\otimes|1\rangle\langle{1}|_w \\[0.1in]
V_- & = & \displaystyle \bigg\{H_\text{else}+\frac{{\mu}^2}{\Delta}{\openone}-\frac{{\mu}^3}{\Delta^2}(Z_1+Z_2+Z_3){\openone}+\frac{{\mu}^4}{\Delta^3}\big[3\openone+2(Z_1Z_2+Z_1Z_3+Z_2Z_3)\big] \\[0.1in]
%& + & \displaystyle 2({\kappa}{\lambda}Z_1Z_2+{\kappa}{\mu}Z_1Z_3+{\lambda}{\mu}Z_2Z_3)\big] \\[0.1in]
& & \displaystyle -\frac{7{\mu}^5}{\Delta^4}\big(Z_1+Z_2+Z_3\big)\bigg\}\otimes|0\rangle\langle{0}|_w \\[0.1in]
%& & \displaystyle \qquad +(3{\kappa}^2+3{\lambda}^2+{\mu}^2){\mu}Z_3\big]\otimes|0\rangle\langle{0}|_w \\[0.1in]
V_{-+} & = & {\mu}{\openone}\otimes|0\rangle\langle{1}|_w,\quad V_{+-}= {\mu}{\openone}\otimes|1\rangle\langle{0}|_w. \\[0.1in]
\end{array}
\end{equation}
\noindent{}The low-lying spectrum of $\tilde{H}$ is approximated by the self energy expansion $\Sigma_-(z)$ below with $z\in[-\max{z},\max{z}]$ where $\max{z}=\|H_\text{else}\|+|\alpha|+\epsilon$.
With the choice of $\mu$ above the expression of $V_+$ in Eq.\ \ref{eq:V_proj_fifth} can be written as
\begin{equation}\label{eq:Vp_simple}
V_+=\left(H_\text{else}+{\mu}(Z_1+Z_2+Z_3)+O(\Delta^{1/5})\right)\otimes|1\rangle\langle{1}|_w.
\end{equation}
\noindent{}Because we are looking for the $5^\text{th}$ order term in the perturbation expansion that gives a term proportional to $Z_1Z_2Z_3$, expand the self energy in Eq.\ \ref{eq:selfenergy} up to $5^\text{th}$ order:
\begin{equation}\label{eq:self_energy_fifth}
\begin{array}{ccl}
\Sigma_-(z) & = & \displaystyle V_-\otimes|0\rangle\langle{0}|_w+\frac{V_{-+}V_{+-}}{z-\Delta}\otimes|0\rangle\langle{0}|_w+\frac{V_{-+}V_+V_{+-}}{(z-\Delta)^2}\otimes|0\rangle\langle{0}|_w+\frac{V_{-+}V_+V_+V_{+-}}{(z-\Delta)^3}\otimes|0\rangle\langle{0}|_w \\[0.1in]
& + & \displaystyle \frac{V_{-+}V_+V_+V_+V_{+-}}{(z-\Delta)^4}\otimes|0\rangle\langle{0}|_w+\sum_{k=4}^\infty\frac{V_{-+}V_+^kV_{+-}}{(z-\Delta)^{k+1}}\otimes|0\rangle\langle{0}|_w.
\end{array}
\end{equation}
\noindent{}Using this simplification as well as the expressions for $V_-$, $V_{-+}$ and $V_{+-}$ in Eq.\ \ref{eq:V_proj_fifth}, the self energy expansion Eq.\ \ref{eq:self_energy_fifth} up to $5^\text{th}$ order becomes
\begin{equation}\label{eq:self_energy_fifth2}
\begin{array}{ccl}
\Sigma_-(z) & = & \displaystyle 
\underbrace{\left(H_\text{else}+\frac{6\mu^5}{\Delta^4}Z_1Z_2Z_3\right)\otimes|0\rangle\langle{0}|_w}_\text{$H_\text{eff}$}+\underbrace{\left(\frac{1}{\Delta}+\frac{1}{z-\Delta}\right){\mu}^2{\openone}\otimes|0\rangle\langle{0}|_w}_\text{$E_1$} \\[0.1in]
& + & \displaystyle\underbrace{\left(\frac{1}{(z-\Delta)^2}-\frac{1}{\Delta^2}\right)\mu^3(Z_1+Z_2+Z_3)\otimes|0\rangle\langle{0}|_w}_\text{$E_2$}+\underbrace{\left(\frac{1}{\Delta^3}+\frac{1}{(z-\Delta)^3}\right)\cdot \mu^4\cdot(Z_1+Z_2+Z_3)^2\otimes|0\rangle\langle{0}|_w}_\text{$E_3$} \\[0.1in]
& + & \displaystyle \underbrace{\left(\frac{1}{(z-\Delta)^4}-\frac{1}{\Delta^4}\right)7{\mu}^5(Z_1+Z_2+Z_3)\otimes|0\rangle\langle{0}|_w}_\text{$E_4$}+\underbrace{\frac{{\mu}^2}{(z-\Delta)^2}\cdot\frac{{\mu}^4}{\Delta^3}(Z_1+Z_2+Z_3)^2\otimes|0\rangle\langle{0}|_w}_\text{$E_6$} \\[0.1in]
& + & O(\Delta^{-2/5})+O(\|H_\text{else}\|\Delta^{-2/5})+O(\|H_\text{else}\|^2\Delta^{-7/5})+O(\|H_\text{else}\|^3\Delta^{-12/5})+\underbrace{\sum_{k=4}^\infty\frac{V_{-+}V_+^kV_{+-}}{(z-\Delta)^{k+1}}\otimes|0\rangle\langle{0}|_w}_\text{$E_7$}. \\[0.1in]
\end{array}
\end{equation}
\noindent{}Similar to what we have done in the previous sections, the norm of the error terms $E_1$ through $E_7$ can be bounded from above by letting $z\mapsto\max{z}$. Then we find that
\begin{equation}\label{eq:error_total_fifth}
\begin{array}{ccl}
\|\Sigma_-(z)-H_\text{targ}\otimes|0\rangle\langle{0}|_w\| & \le & \Theta(\Delta^{-1/5})
\end{array}
\end{equation}
\noindent{}if we only consider the dominant dependence on $\Delta$ and regard $\|H_\text{else}\|$ as a given constant. To guarantee that $\|\Sigma_-(z)-H_\text{targ}\otimes|0\rangle\langle{0}|_w\|\le\epsilon$, we let the right hand side of Eq.\ \ref{eq:error_total_fifth} to be $\le\epsilon$, which translates to $\Delta=\Theta(\epsilon^{-5})$. 

\begin{figure}
\includegraphics[scale=0.15]{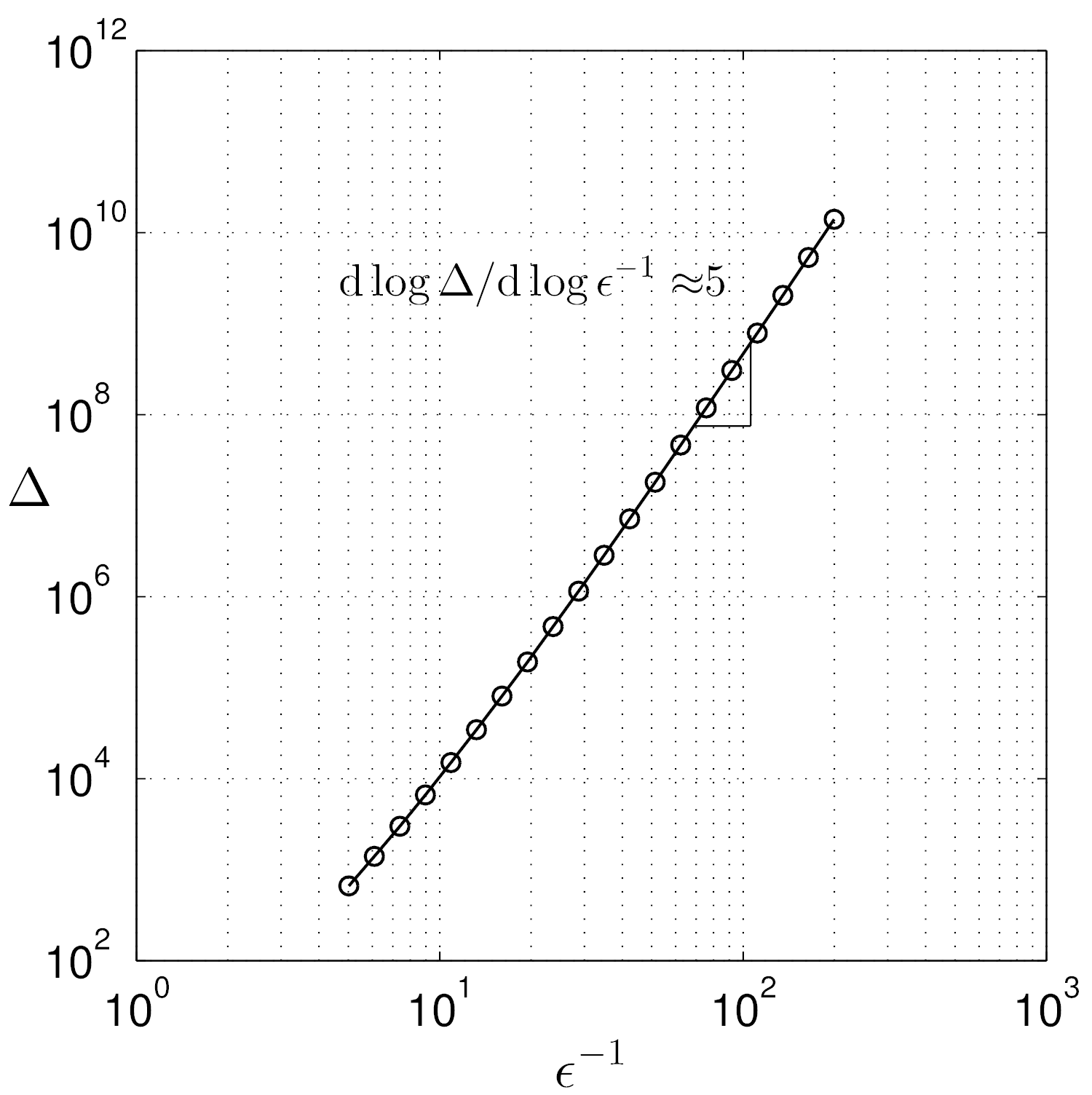}
\includegraphics[scale=0.15]{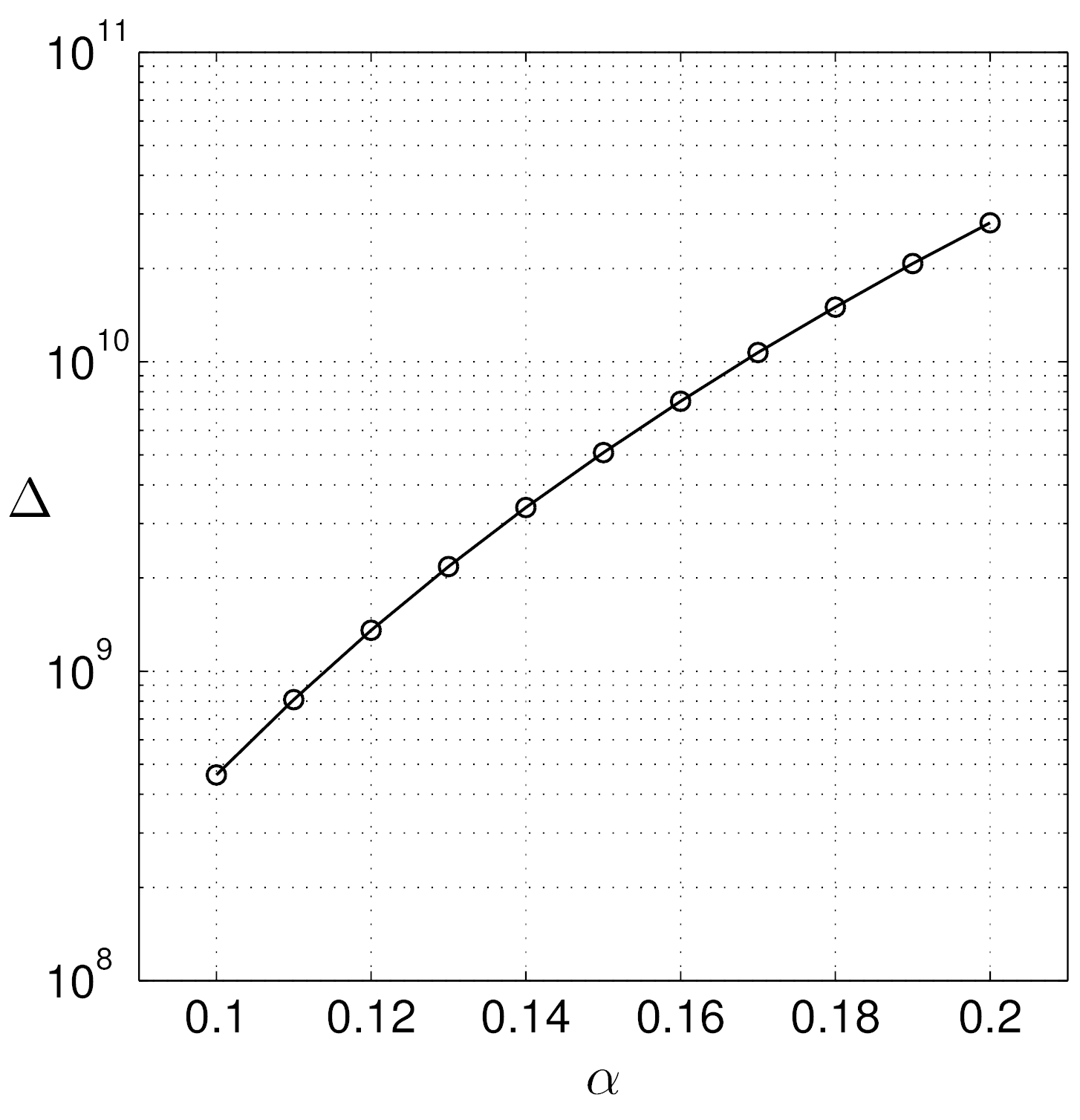}
\makebox[0.8cm]{}(a)\makebox[8cm]{}(b)
\caption{\normalsize (a) The scaling of minimum $\Delta$ needed to ensure $\|\Sigma_-(z)-H_\text{eff}\|\le\epsilon$ as a function of $\epsilon^{-1}$. Here we choose $\|H_\text{else}\|=0$, $\alpha=0.1$ and $\epsilon$ ranging from $10^{-0.7}$ to $10^{-2.3}$. The values of minimum $\Delta$ are numerically optimized \cite{footnote:num_op}. The slope of the line at large $\epsilon^{-1}$ is $4.97\approx{5}$, which provides evidence that with the assignments of ${\mu}=(\alpha\Delta^4/6)^{1/5}$, the optimal scaling of $\Delta$ is $\Theta(\epsilon^{-5})$. (b) The numerically optimized \cite{footnote:num_op} gap versus the desired coupling $\alpha$ in the target Hamiltonian. Here $\epsilon=0.01$ and $\|H_\text{else}\|=0$.}
\label{fig:ZZZ_fifth_Delta_eps}
\end{figure}

This $\Theta(\epsilon^{-5})$ scaling is numerically illustrated (Fig.\ \ref{fig:ZZZ_fifth_Delta_eps}a). Although in principle the $5^\text{th}$ order gadget can be implemented on a Hamiltonian of form Eq.\ \ref{eq:dwave}, for a small range of $\alpha$, the minimum $\Delta$ needed is already large (Fig.\ \ref{fig:ZZZ_fifth_Delta_eps}b), rendering it challenging to demonstrate the gadget experimentally with current resources. However, this is the only currently known gadget realizable with a transverse Ising model that is able to address the case where $H_\text{else}$ is not necessarily diagonal. 
%Further investigations are invited on what other gadgets might be feasible on TIM Hamiltonians. It has been shown in \cite{CM13} that the complexity class \textsc{NP}$\subseteq$\textsc{TIM}. The inclusion is simple to show since any \textsc{TIM} Hamiltonian with zero transverse field contains only ZZ terms, whose ground state can be \textsc{NP-hard} to find in the worst. This gadget shows an alternative reduction from many-body diagonal Hamiltonian (whose ground state problem is polynomially equivalent to Hamiltonians of only 2-body diagonal terms) to \textsc{TIM} Hamiltonian, which is consistent with the fact that \textsc{NP}$\subseteq$\textsc{TIM}.

\section{YY gadget}\label{sec:yy}
\noindent{\bf Summary}. The gadgets which we have presented so far are intended to reduce the locality of the target Hamiltonian. Here we present another type of gadget, called ``creation'' gadgets \cite{BL07}, which simulate the type of effective couplings that are not present in the gadget Hamiltonian. Many creation gadgets proposed so far are modifications of existing reduction gadgets. For example, the ZZXX gadget in \cite{BL07}, which is intended to simulate $Z_iX_j$ terms using Hamiltonians of the form
\begin{equation}\label{eq:dwave2}
\begin{array}{ccl}
H_{ZZXX} & = & \displaystyle \sum_i\Delta_iX_i+\sum_ih_iZ_i+\sum_{i,j}J_{ij} Z_iZ_j+\sum_{i,j}K_{ij}X_i X_j,
\end{array}
\end{equation}
is essentially a 3- to 2-body gadget with the target term $A\otimes B\otimes C$ being such that the operators $A$, $B$ and $C$ are $X$, $Z$ and identity respectively. Therefore the analyses on 3- to 2- body reduction gadgets that we have presented for finding the lower bound for the gap $\Delta$ are also applicable to this ZZXX creation gadget.

Note that YY terms can be easily realized via bases rotation if single-qubit Y terms are present in the Hamiltonian in Eq.\ \ref{eq:dwave2}. Otherwise it is not \emph{a priori} clear how to realize YY terms using $H_{ZZXX}$ in Eq.\ \ref{eq:dwave2}. We will now present the first YY gadget which starts with a universal Hamiltonian of the form Eq.\ \ref{eq:dwave2} and simulates the target Hamiltonian $H_\text{targ}=H_\text{else}+\alpha Y_i Y_j$. The basic idea is to use the identity $X_iZ_i=\iota Y_i$ where $\iota=\sqrt{-1}$ and induce a term of the form $X_iZ_iZ_jX_j=Y_iY_j$ at the $4^\text{th}$ order. Introduce ancilla qubit $w$ and apply a penalty $H=\Delta|1\rangle\langle{1}|_w$. With a perturbation $V$ we could perform the same perturbative expansion as previously. Given that the $4^\text{th}$ order perturbation is $V_{-+}V_+V_+V_{+-}$ up to a scaling constant. we could let single $X_i$ and $X_j$ be coupled with $X_w$, which causes both $X_i$ and $X_j$ to appear in $V_{-+}$ and $V_{+-}$. Furthermore, we couple single $Z_i$ and $Z_j$ terms with $Z_w$. Then $\frac{1}{2}(\openone +Z_w)$ projects single $Z_i$ and $Z_j$ onto the $+$ subspace and causes them to appear in $V_+$. For $H_\text{targ}=H_\text{else}+\alpha Y_1Y_2$, the full expressions for the gadget Hamiltonian is the following: the penalty Hamiltonian $H=\Delta|1\rangle\langle{1}|_w$ acts on the ancilla qubit. The perturbation $V=V_0+V_1+V_2$ where $V_0$, $V_1$, and $V_2$ are defined as
\begin{equation}\label{eq:yy_V1}
\begin{array}{ccl}
V_0 & = & \displaystyle H_\text{else}+\mu({ Z_1+ Z_2})\otimes{|1\rangle\langle{1}|_w}+\mu( X_1-\text{sgn}(\alpha) X_2)\otimes X_w \\[0.05in]
V_1  & = & \displaystyle \frac{2\mu^2}{\Delta}(\openone\otimes|0\rangle\langle{0}|_w+X_1X_2) \\[0.05in]
V_2  & = & \displaystyle -\frac{2\mu^4}{\Delta^{3}}Z_1Z_2.
\end{array}
\end{equation}
with $\mu = (|\alpha|\Delta^3/4)^{1/4}$. For a specified error tolerance $\epsilon$, we have constructed a YY gadget Hamiltonian of gap scaling $\Delta=O(\epsilon^{-4})$ and the low-lying spectrum of the gadget Hamiltonian captures the spectrum of $H_\text{targ}\otimes|0\rangle\langle{0}|_w$ up to error $\epsilon$.

The YY gadget implies that a wider class of Hamiltonians such as 
\begin{equation}
\begin{array}{ccl}
H_{ZZYY} & = & \displaystyle \sum_i h_iX_i + \sum_i \Delta_iZ_i + \sum_{i,j}J_{ij}Z_iZ_j+ \sum_{i,j}K_{ij}Y_iY_j
\end{array}
\end{equation}
and
\begin{equation}
\begin{array}{ccl}
H_{XXYY} & = & \displaystyle \sum_i h_iX_i + \sum_i \Delta_iZ_i + \sum_{i,j}J_{ij}X_iX_j+ \sum_{i,j}K_{ij}Y_iY_j
\end{array}
\end{equation}
can be simulated using the Hamiltonian of the form in Eq.\ \ref{eq:dwave2}. Therefore using the Hamiltonian in Eq.\ \ref{eq:dwave2} one can in principle simulate any finite-norm real valued Hamiltonian on qubits. Although by the QMA-completeness of $H_{ZZXX}$ one could already simulate such Hamiltonian via suitable embedding, our YY gadget provides a more direct alternative for the simulation.
%This gadget opens the door to simulate molecular Hamiltonians in quantum chemistry \cite{2014arXiv1401.3186V} and simulating Heisenberg coupling $\sum_{i,j}X_iX_j+Y_iY_j+Z_iZ_j$ in condensed matter physics. One could also use the YY gadget for simulating electronic structure on experimental systems that implement Hamiltonians of the form Eq.\ \ref{eq:dwave2}. For example, mapping from physical Hamiltonians to spin operators can be done using the Bravyi-Kitaev transformation \cite{BK02}, giving rise to log-local many-body Pauli Hamiltonians representing electronic structure problems \cite{SRL12}.   

$\quad$\\
\noindent{\bf Analysis. } The results in \cite{BL07} shows that Hamiltonians of the form in Eq.\ \ref{eq:dwave2} supports universal adiabatic quantum computation and finding the ground state of such a Hamiltonian is \textsc{QMA-complete}. This form of Hamiltonian is also interesting because of its relevance to experimental implementation \cite{2006cond.mat..8253H}. Here we show that with a Hamiltonian of the form in Eq.\ \ref{eq:dwave2} we could simulate a target Hamiltonian $H_\text{targ}=H_\text{else}+\alpha Y_1Y_2$. Introduce an ancilla $w$ and define the penalty Hamiltonian as $H=\Delta|1\rangle\langle{1}|_w$. Let the perturbation $V=V_0+V_1+V_2$ be
\begin{equation}\label{eq:yy_V}
\begin{array}{ccl}
V_0 & = & H_\text{else}+\kappa({Z_1+ Z_2})\otimes{|1\rangle\langle{1}|_w}+\kappa(X_1-\text{sgn}(\alpha) X_2)\otimes X_w \\[0.05in]
V_1  & = & 2\kappa^2\Delta^{-1}[|0\rangle\langle{0}|_w-\text{sgn}(\alpha) X_1X_2] \\[0.05in]
V_2  & = & -4\kappa^4\Delta^{-3}Z_1Z_2.
\end{array}
\end{equation}
Then the gadget Hamiltonian $\tilde{H}=H+V$ is of the form in Eq.\ \ref{eq:dwave2}. Here we choose the parameter $\kappa=(|\alpha|\Delta^3/4)^{1/4}$. In order to show that the low lying spectrum of $\tilde{H}$ captures that of the target Hamiltonian, define $\mathcal{L}_-=\text{span}\{|\psi\rangle\text{ such that }\tilde{H}|\psi\rangle=\lambda|\psi\rangle,\lambda<\Delta/2\}$ as the low energy subspace of $\tilde{H}$ and $\mathcal{L}_+=\openone-\mathcal{L}_-$. Define $\Pi_-$ and $\Pi_+$ as the projectors onto $\mathcal{L}_-$ and $\mathcal{L}_+$ respectively. 

With these notations in place, here we show that the spectrum of $\tilde{H}_-=\Pi_-\tilde{H}\Pi_-$ approximates the spectrum of $H_\text{targ}\otimes|0\rangle\langle{0}|_w$ with error $\epsilon$. To begin with, the projections of $V$ into the subspaces $\mathcal{L}_-$ and $\mathcal{L}_+$ can be written as
\begin{equation}\label{eq:yy_Vproj}
\begin{array}{ccl}
V_- & = & \displaystyle\bigg(H_\text{else}+\underbrace{\frac{\kappa^2}{\Delta}(X_1-\text{sgn}(\alpha) X_2)^2}_{(a)}\underbrace{-\frac{4\kappa^4}{\Delta^3}Z_1Z_2}_{(b)}\bigg)\otimes|0\rangle\langle{0}|_w \\[0.05in]
V_+ & = & \displaystyle\left(H_\text{else}+\kappa(Z_1+Z_2)-\frac{2\kappa^2}{\Delta}\text{sgn}(\alpha) X_1X_2 -\frac{4\kappa^4}{\Delta^3}Z_1Z_2\right)\otimes|1\rangle\langle{1}|_w \\[0.05in]
V_{-+} & = & \kappa(X_1 - \text{sgn}(\alpha) X_2)\otimes|0\rangle\langle{1}|_w \\[0.05in]
V_{+-} & = & \kappa(X_1 - \text{sgn}(\alpha) X_2)\otimes|1\rangle\langle{0}|_w
\end{array}
\end{equation}
Given the penalty Hamiltonian $H$, we have the operator valued resolvent $G(z)=(z\openone-H)^{-1}$ that satisfies $G_+(z)=\Pi_+G(z)\Pi_+=(z-\Delta)^{-1}|1\rangle\langle{1}|_w$. Then the low lying sector of the gadget Hamiltonian $\tilde H$ can be approximated by the perturbative expansion Eq.\ \ref{eq:selfenergy}. For our purposes we will consider terms up to the $4^\text{th}$ order:
\begin{equation}
\Sigma_-(z)=V_-+\frac{1}{z-\Delta}V_{-+}V_{+-}+\frac{1}{(z-\Delta)^2}V_{-+}V_+V_{+-}+\frac{1}{(z-\Delta)^3}V_{-+}V_+V_+V_{+-}+\sum_{k=3}^\infty\frac{V_{-+}V_+^kV_{+-}}{(z-\Delta)^{k+1}}.
\end{equation}
Now we explain the perturbative terms that arise at each order. The $1^\text{st}$ order is the same as $V_-$ in Eq.\ \ref{eq:yy_Vproj}. The $2^\text{nd}$ order term gives
\begin{equation}\label{eq:yy_2nd}
\frac{1}{z-\Delta}V_{-+}V_{+-}=\underbrace{\frac{1}{z-\Delta}\cdot\kappa^2(X_1-\text{agn}(\alpha) X_2)^2}_{(c)}\otimes|0\rangle\langle{0}|_w.
\end{equation}
At the $3^\text{rd}$ order, we have
\begin{equation}\label{eq:yy_3rd}
\begin{array}{ccl}
\displaystyle \frac{1}{(z-\Delta)^2}V_{-+}V_+V_{+-} & = & \displaystyle \bigg(\frac{1}{(z-\Delta)^2}\cdot\kappa^2(X_1-\text{agn}(\alpha) X_2)H_\text{else}( X_1-\text{sgn}(\alpha) X_2) \\[0.05in]
& + & \displaystyle \underbrace{\frac{1}{(z-\Delta)^2}\frac{4\kappa^4}{\Delta}(X_1X_2-\text{sgn}(\alpha)\openone)}_{(d)}\bigg)\otimes|0\rangle\langle{0}|_w+ O(\Delta^{-1/4}).
\end{array}
\end{equation}
The $4^\text{th}$ order contains the desired YY term:
\begin{equation}\label{eq:yy_4th}
\begin{array}{ccl}
\displaystyle \frac{1}{(z-\Delta)^3}V_{-+}V_+V_+V_{+-} & = & \displaystyle \bigg(\underbrace{\frac{1}{(z-\Delta)^3}\cdot 2\kappa^4(X_1-\text{sgn}(\alpha) X_2)^2}_{(e)}-\underbrace{\frac{1}{(z-\Delta)^3}4\kappa^4Z_1Z_2}_{(f)} \\[0.05in]
& + & \displaystyle \frac{4\kappa^4\text{sgn}(\alpha)}{(z-\Delta)^3}Y_1Y_2\bigg)\otimes|0\rangle\langle{0}|_w+O(\|H_\text{else}\|\cdot\Delta^{-3/4})+O(\|H_\text{else}\|^2\cdot\Delta^{-1/2})
\end{array}
\end{equation}
Note that with the choice of $\kappa=(|\alpha|\Delta^3/4)^{1/4}$, all terms of $5^\text{th}$ order and higher are of norm $O(\Delta^{-1/4})$. In the $1^\text{st}$ order through $4^\text{th}$ order perturbations the unwanted terms are labelled as $(a)$ through $(f)$ in Eqs.\ \ref{eq:yy_Vproj}, \ref{eq:yy_2nd}, \ref{eq:yy_3rd}, and \ref{eq:yy_4th}. Note how they compensate in pairs: the sum of $(a)$ and $(c)$ is $O(\Delta^{-1/4})$. The same holds for $(d)$ and $(e)$, $(b)$ and $(f)$. Then the self energy is then
\begin{equation}\label{eq:yy_selfenergy}
\Sigma_-(z)=(H_\text{else}+\alpha Y_1Y_2)\otimes|0\rangle\langle{0}|_w+O(\Delta^{-1/4}).
\end{equation}
Let $\Delta=\Theta(\epsilon^{-4})$, then by the Gadget Theorem (\ref{th:perturbation}), the low-lying sector of the gadget Hamiltonian $\tilde{H}_-$ captures the spectrum of $H_\text{targ}\otimes|0\rangle\langle{0}|_w$ up to error $\epsilon$. 

The fact that the gadget relies on $4^\text{th}$ order perturbation renders the gap scaling relatively larger than it is in the case of subdivision or 3- to 2-body reduction gadgets. However, this does not diminish its usefulness in various  applications. 

\section*{Conclusion}
We have presented improved constructions for the most commonly used gadgets, which in turn implies a reduction in the resources for the many works which employ these current constructions.  We presented the first comparison between the known gadget constructions and the first numerical optimizations of gadget parameters.  Our analytical results are found to agree with the optimised solutions.  The introduction of our gadget which simulates YY-interactions opens many prospects for universal adiabatic quantum computation, particularly the simulation of physics feasible on currently realizable Hamiltonians. 

\section*{Acknowledgements}
We thank Andrew Landahl for helpful comments. JDB and YC completed parts of this study while visiting the Institute for Quantum Computing at the University of Waterloo. RB was supported by the United States Department of Defense. The views and conclusions contained in this document are those of the authors and should not be interpreted as representing the official policies, either expressed or implied, of the U.S. Government. JDB completed parts of this study while visiting the Qatar Energy and Environment Research Institute and would like to acknowledge the Foundational Questions Institute (under grant FQXi-RFP3-1322) for financial support.

\appendix

\section{Parallel 3- to 2-body gadget}\label{sec:3body_par}

\begin{figure}
\begin{center}
% First subfigure.
\begin{subfigure}[b]{.6\textwidth}
\resizebox{\linewidth}{!}{
    \begin{tikzpicture}[
      scale=0.5,
      level/.style={thick},
      trans/.style={->, thick},
      plus/.style={->, bend right, thick},
    ]
    % Draw the energy levels.
 \draw[thick, dotted] (-7.5,1) node[below] {${\cal L_-}$}  node[above] {${\cal L_+}$} -- (5,1);
  \draw (-8,4) node {$\displaystyle G_+\left(z\right) = \frac{1}{z- \Delta}$};
    \draw (-7,7) node {$\displaystyle G_+\left(z\right) = \frac{1}{z- 2\Delta}$};
    \draw[level] (-2,0) -- (2,0) node[midway, below] {\scriptsize $|0\,.\,.\,.\!\!\!\underbrace{0}_{i}\!\!\!.\,.\,.\!\!\!\underbrace{0}_{j}\!\!\!.\,.\,.\,0\rangle$};
    \draw[level] (-5,4) -- (-1,4) node[midway, above] {\scriptsize $|0\ldots1\ldots0\ldots0\rangle$};
    \draw[level] (-2,7) -- (2,7) node[midway, above] {\scriptsize $|0\ldots1\ldots1\ldots0\rangle$};
    \draw[level] (1,4) -- (5,4) node[midway, above] {\scriptsize $|0\ldots0\ldots1\ldots0\rangle$};
    % Draw the transitions.
    \draw[trans] (-1.5,0.25) to node[left] {$\small V_{-+}$} (-3.5,3.75);
        \draw[trans] (-2.5,3.75) to node[right] {$\small V_{+-}$} (-.5,0.25);
            \draw[trans] (2.5,3.75) to node[left] {$\small V_{+-}$} (.5,.25);
        \draw[trans] (1.5,.25) to node[right] {$\small \,V_{-+}$} (3.5,3.75);
            \draw[trans] (-3.5,5.) to node[left] {$\small V_f\,$} (-1.5,6.75);
            \draw[trans] (-.5,6.75) to node[right] {$\small \,\,V_f$} (-2.5,5.);
                \draw[trans] (3.5,5.) to node[right] {$\small V_f\,$} (1.5,6.75);
            \draw[trans] (.5,6.75) to node[left] {$\small \,\,V_f$} (2.5,5.);
    \end{tikzpicture}}
\centerline{(a)}
\end{subfigure}
% Second subfigure
\begin{subfigure}[b]{.6\textwidth}
\resizebox{\linewidth}{!}{
    \begin{tikzpicture}[
      scale=0.5,
      level/.style={thick},
      trans/.style={->, thick},
      plus/.style={->, bend right, thick},
    ]
    % Draw the energy levels.
 \draw[thick, dotted] (-7.5,1) node[below] {${\cal L_-}$}  node[above] {${\cal L_+}$} -- (5,1);
  \draw (-8,4) node {$\displaystyle G_+\left(z\right) = \frac{1}{z- \Delta}$};
    \draw[level] (-2,0) -- (2,0) node[midway, below] {\scriptsize $|0\,.\,.\,.\!\!\!\underbrace{0}_{i}\!\!\!.\,.\,.\!\!\!\underbrace{0}_{j}\!\!\!.\,.\,.\,0\rangle$};
    \draw[level] (-5,4) -- (-1,4) node[midway, above] {\scriptsize $|0\ldots1\ldots0\ldots0\rangle$};
    \draw[level] (1,4) -- (5,4) node[midway, above] {\scriptsize $|0\ldots0\ldots1\ldots0\rangle$};
    % Draw the transitions.
    \draw[trans] (-1.5,0.25) to node[left] {$\small V_{-+}$} (-3.5,3.75);
        \draw[trans] (-2.5,3.75) to node[right] {$\small V_{+-}$} (-.5,0.25);
            \draw[trans] (2.5,3.75) to node[left] {$\small V_{+-}$} (.5,.25);
        \draw[trans] (1.5,.25) to node[right] {$\small \,V_{-+}$} (3.5,3.75);
         \draw[->, thick, loop above, looseness=2] (-5,5) to node[above] {$\small V_s$}  (-3.25,5);
                  \draw[->, thick, loop above, looseness=2] (-2.75,5) to node[above] {$\small V_s$}  (-1.,5);
         \draw[->, thick, loop above, looseness=2] (5,5) to node[above] {$\small V_s$}  (3.25,5);
                  \draw[->, thick, loop above, looseness=2] (2.75,5) to node[above] {$\small V_s$}  (1.,5);
    \end{tikzpicture}}
\end{subfigure}
\centerline{(b)}
\end{center}
\caption{Diagrams illustrating the transitions that occur at 4th order. The two diagrams each represent a type of transition that occurs at 4th order. Each diagram is divided by a horizontal line where below the line is $\mathcal{L}_-$ space and above is $\mathcal{L}_+$ subspace. Each diagram deals with a fixed pair of ancilla qubits labelled $i$ and $j$. The diagram (a) has three horizontal layers connected with vertically going arrows. $V_f$ and $V_s$ are both components of $V_+$. In fact $V_+ = V_f + V_s$ where $V_f$ is responsible for the flipping and $V_s$ contributes when the transition does not have flipping. At the left of each horizontal layer lies the expression for $G_+(z)$, which is different for states in $\mathcal{L}_+$ with different Hamming weights. The diagram (b) is constructed in a similar fashion except that we are dealing with the type of 4th order transition where the state stays the same for two transitions in $\mathcal{L}_+$, hence the $V_s$ symbols and the arrows going from one state to itself. The diagram (a) reflects the type of 4th order transition that induces cross-gadget contribution and given our gadget parameter setting, this contribution could be $O(1)$ when otherwise compensated. The diagram (b) shows two paths that don not interfere with each other and thus having no cross-gadget contributions.}
\label{fig:diagram}
\end{figure}
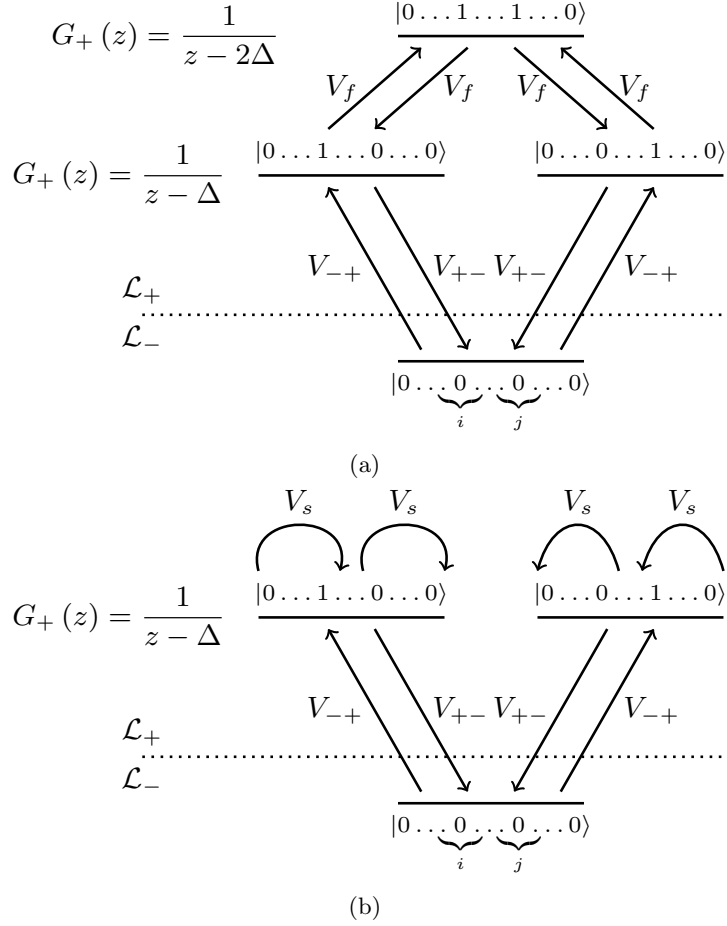

\noindent{\bf Summary.} In Sec.\ \ref{sec:par_sub} we have shown that by using parallel subdivision gadgets iteratively, one can reduce a $k$-body target term to $3$-body. We now turn our attention to considering $H_\text{targ}=H_\text{else}+\sum_{i=1}^m\alpha_i A_i\otimes B_i\otimes C_i$, which is a sum of $m$ 3-body terms. A straightforward approach to the reduction is to deal with the 3-body terms in series \emph{i.e.} one at a time: apply a 3-body gadget on one term, and include the entire gadget in the $H_\text{else}$ of the target Hamiltonian in reducing the next 3-body term. In this construction, $\Delta$ scales exponentially as a function of $m$. In order to avoid that overhead, we apply all gadgets in parallel, which means introducing $m$ ancilla spins, one for each 3-body term and applying the same $\Delta$ onto it. This poses additional challenges as the operator valued resolvent $G(z)$ now has multiple poles. Enumerating high order terms in the perturbation series requires consideration of the combinatorial properties of the bit flipping processes (Fig.\ \ref{fig:diagram}). 

If we apply the current construction \cite{OT06,BDLT08} of 3-body gadgets in parallel, which requires $\Delta=\Theta(\epsilon^{-3})$, it can be shown \cite{BDLT08} that the cross-gadget contribution is $O(\epsilon)$. However, if we apply our improved construction of the 3- to 2-body gadget in parallel, the perturbation expansion will contain $\Theta(1)$ cross-gadget terms that are dependent on the commutation relations between $A_i$, $B_i$ and $A_j$, $B_j$. Compensation terms are designed to ensure that these error terms are suppressed in the perturbative expansion. With our improved parallel 3-body construction, $\Delta=\Theta({\epsilon^{-2}}\text{poly}(m))$ is sufficient.

The combination of parallel subdivision with the parallel 3- to 2-body reduction allows us to reduce an arbitrary $k$-body target Hamiltonian $H_\text{targ}=H_\text{else}+\alpha\sigma_1\sigma_2\cdots\sigma_k$ to 2-body \cite{BDLT08}. In this paper we have improved both parallel 2-body and 3- to 2-body gadgets. When numerically optimized at each iteration, our construction requires a smaller gap than the original construction \cite{BDLT08} for the range of $k$ concerned. 

$\quad$\\
\noindent{\bf Analysis.} In Sec.\ \ref{sec:par_sub} we have shown that with subdivision gadgets one can reduce a $k$-body interaction term down to 3-body. To complete the discussion on reducing a $k$-body term to $2$-body, now we deal with reducing a 3-body target Hamiltonian of form
\[
H_\text{targ}=H_\text{else}+\sum_{i=1}^m\alpha_i{A_i}\otimes{B_i}\otimes{C_i}
\]
\noindent{}where $H_\text{else}$ is a finite-norm Hamiltonian and all of $A_i$, $B_i$, $C_i$ are single-qubit Pauli operators acting on one of the $n$ qubits that $H_\text{targ}$ acts on. Here without loss of generality, we assume $A_i$, $B_i$ and $C_i$ are single-qubit Pauli operators as our construction depends on the commutation relationships among these operators. The Pauli operator assumption ensures that the commutative relationship can be determined efficiently a priori. 

We label the $n$ qubits by integers from 1 to $n$. We assume that in each 3-body term of the target Hamiltonian, ${A_i}$, ${B_i}$ and ${C_i}$ act on three different qubits whose labels are in increasing order i.e.\ if we label the qubits with integers from 1 to $n$, ${A_i}$ acts on qubit $a_i$, ${B_i}$ acts on $b_i$, ${C_i}$ on $c_i$, we assume that $1\le a_i<b_i<c_i\le n$ must hold for all values of $i$ from 1 to $m$. 

One important feature of this gadget is that the gap $\Delta$ scales as $\Theta(\epsilon^{-2})$ instead of the common $\Theta(\epsilon^{-3})$ scaling assumed by the other 3-body constructions in the literature \cite{KKR06,OT06,BDLT08}.

To reduce the $H_\text{targ}$ to 2-body, introduce $m$ qubits labelled as $u_1$, $u_2$, $\cdots$, $u_m$ and apply an energy penalty $\Delta$ onto the excited subspace of each qubit, as in the case of parallel subdivision gadgets presented previously. Then we have
\begin{equation}\label{eq:H_par3}
H=\sum_{i=1}^m\Delta|1\rangle\langle{1}|_{u_i}=\sum_{x\in\{0,1\}^m}h(x)\Delta|x\rangle\langle x|.
\end{equation}
\noindent{}where $h(x)$ is the Hamming weight of the $m$-bit string $x$.
%For the perturbation $V$, in the literature \cite{BDLT08} the individual 3- to 2-body gadgets are designed to work independently. Therefore the construction of $V$ follows from Eq.\ \ref{eq:OT_V} in a straightforward manner:
%\begin{equation}\label{eq:par3_V_1}
%V=H_\text{else}+\sum_{i=1}^m{\mu_i}{C_i}\otimes|1\rangle\langle{1}|_{u_i}+\sum_{i=1}^m({\kappa_i}{A_i}+{\lambda_i}{B_i})\otimes X_{u_i}+V_1
%\end{equation}
%\noindent{}where the compensation term $V_1$ is naturally following from Eq.\ \ref{eq:V1_old}:
%\begin{equation}\label{eq:par3_V1}
%V_1=\frac{1}{\Delta}\sum_{i=1}^m({\kappa_i}{A_i}+{\lambda_i}{B_i})^2-\frac{1}{\Delta^2}\sum_{i=1}^m({\kappa_i^2}+{\lambda_i^2}){\mu_i}{C_i}.
%\end{equation}
%\noindent{}With ${\kappa_i}$, ${\lambda_i}$ and ${\mu_i}$ being $\Theta(\Delta^{2/3})$, it is shown in \cite{BDLT08} that $\Delta=\Theta(\epsilon^{-3})$ suffices to render the spectral error between the low-lying sector of $\tilde{H}=H+V$ and the entire spectrum of the target Hamiltonian $O(\epsilon)$. In this case the cross-gadget contribution in the perturbative expansion can be shown to scale as $\Theta(\Delta^{-1/3})$ and is therefore $O(\epsilon)$. For the purpose of brevity we omit the calculation. For details one can refer to \cite[Sec. II]{BDLT08}. Following the argument in Sec.\ \ref{sec:3body}, we can reduce the gap scaling to $\Theta(\epsilon^{-2})$. 
In this new construction the perturbation $V$ is defined as
\begin{equation}\label{eq:V_par3}
\begin{array}{ccl}
V & = & \displaystyle H_\text{else}+\sum_{i=1}^m{\mu_i}{C_i}\otimes|1\rangle\langle{1}|_{u_i}+\sum_{i=1}^m({\kappa_i}{A_i}+{\lambda_i}{B_i})\otimes X_{u_i}+V_1+V_2+V_3
\end{array}
\end{equation}
\noindent{}where $V_1$ is defined as
\begin{equation}\label{eq:par3_V1}
V_1=\frac{1}{\Delta}\sum_{i=1}^m({\kappa_i}{A_i}+{\lambda_i}{B_i})^2-\frac{1}{\Delta^2}\sum_{i=1}^m({\kappa_i^2}+{\lambda_i^2}){\mu_i}{C_i}
\end{equation} 
and $V_2$ is defined as
\begin{equation}
V_2=-\frac{1}{\Delta^3}\sum_{i=1}^m({\kappa_i}{A_i}+{\lambda_i}{B_i})^4.
\end{equation}
\noindent{}$V_3$ will be explained later. Following the discussion in Sec.\ \ref{sec:3body}, the coefficients ${\kappa_i}$, ${\lambda_i}$ and ${\mu_i}$ are defined as
\begin{equation}\label{eq:J_3par}
{\kappa_i}=\text{sgn}(\alpha_i)\left(\frac{|\alpha_i|}{2}\right)^{\frac{1}{3}}\Delta^{\frac{3}{4}},\quad {\lambda_i}=\left(\frac{|\alpha_i|}{2}\right)^{\frac{1}{3}}\Delta^{\frac{3}{4}},\quad {\mu_i}=\left(\frac{|\alpha_i|}{2}\right)^{\frac{1}{3}}\Delta^{\frac{1}{2}}.
\end{equation}
\noindent{}However, as we will show in detail later in this section, a close examination of the perturbation expansion based on the $V$ in Eq.\ \ref{eq:V_par3} shows that with assignments of ${\kappa_i}$, ${\lambda_i}$ and ${\mu_i}$ in Eq.\ \ref{eq:J_3par} if $V$ has only $V_1$ and $V_2$ as compensation terms, the cross-gadget contribution in the expansion causes $\Theta(1)$ error terms to arise. In order to compensate for the $\Theta(1)$ error terms, we introduce the compensation \[V_3=\sum_{i=1}^m\sum_{j=1,j\neq i}^m\bar{V}_{ij}\] into $V$ and $\bar{V}_{ij}$ is the compensation term for cross-gadget contribution \cite{footnote:cross}. Before presenting the detailed form of $\bar{V}_{ij}$, let $s_1^{(i,j)}=s_{11}^{(i,j)}+s_{12}^{(i,j)}$ where
\begin{equation}\label{eq:s11}
s_{11}^{(i,j)}=\left\{
\begin{array}{cl}
1 & \text{if  }\left\{\begin{tabular}{c}$[{A_i},{A_j}]\neq 0$ \\ $[{B_i},{B_j}]=0$\end{tabular}\right. \text{or } \left\{\begin{tabular}{c}$[{B_i},{B_j}]\neq 0$ \\ $[{A_i},{A_j}]=0$\end{tabular}\right. \\
0 & \text{otherwise}
\end{array}\right.
\end{equation}
\begin{equation}\label{eq:s12}
s_{12}^{(i,j)}=\left\{
\begin{array}{cl}
1 & \text{if $[{A_i},{B_j}]\neq 0$ or  $[{B_i},{A_j}]\neq 0$} \\[0.1in]
0 & \text{otherwise}
\end{array}
\right.
\end{equation}
and further define $s_2^{(i,j)}$ as
\begin{equation}\label{eq:s2}
s_2^{(i,j)}=\left\{
\begin{array}{cl}
1 & \text{if $[{A_i},{A_j}]\neq 0$ and $[{B_i},{B_j}]\neq 0$} \\[0.1in]
0 & \text{otherwise.}
\end{array}\right.
\end{equation}
Then we define $\bar{V}_{ij}$ as
\begin{equation}\label{eq:Vij}
\begin{array}{ccl}
\bar{V}_{ij} & = & \displaystyle -s_1^{(i,j)}\cdot\frac{1}{\Delta^3}({\kappa_i}{\kappa_j})^2{\openone}-s_2^{(i,j)}\bigg(\frac{2}{\Delta^3}({\kappa_i}{\kappa_j})^2{\openone}-\frac{2}{\Delta^3}{\kappa_i}{\kappa_j}{\lambda_i}{\lambda_j}{A_i}{A_j}{B_i}{B_j}\bigg) 
\end{array}
\end{equation}
\noindent{}where $s_1^{(i,j)}$ and $s_2^{(i,j)}$ are coefficients that depend on the commuting relations between the operators in the $i$-th term and the $j$-th term. {Note that in Eq.\ \ref{eq:Vij}, although the term $A_iA_jB_iB_j$ is 4-local, it arises only in cases where $s_2^{(i,j)}=1$. In this case, an additional gadget with a new ancilla $u_{ij}$ can be introduced to generate the 4-local term. For succinctness we present the details of this construction in Appendix \ref{appendix:4local}.}
\noindent{With} the penalty Hamiltonian $H$ defined in Eq.\ \ref{eq:H_par3}, the operator-valued resolvent (or the Green's function) can be written as
\begin{equation}
G(z)=\sum_{x\in\{0,1\}^m}\frac{1}{z-h(x)\Delta}|x\rangle\langle{x}|.
\end{equation}
\noindent{}Define subspaces of the ancilla register $\mathcal{L}_-=\text{span}\{|00\cdots 0\rangle\}$ and $\mathcal{L}_+=\text{span}\{|x\rangle|x\neq 00\cdots 0\}$. Define ${P_-}$ and ${P_+}$ as the projectors onto $\mathcal{L}_-$ and $\mathcal{L}_+$. Then the projections of $V$ onto the subspaces can be written as
\begin{equation}\label{eq:V_3par}
\begin{array}{ccl}
V_+ & = & \displaystyle \bigg(H_\text{else}+\frac{1}{\Delta}\sum_{i=1}^m({\kappa_i}{A_i}+{\lambda_i}{B_i})^2-\frac{1}{\Delta^2}\sum_{i=1}^m({\kappa_i^2}+{\lambda_i^2}){\mu_i}{C_i}-\frac{1}{\Delta^3}\sum_{i=1}^m({\kappa_i}{A_i}+{\lambda_i}{B_i})^4+\sum_{i=1}^m\sum_{j=1,j\neq i}^m\bar{V}_{ij}\bigg)\otimes{P_+} \\[0.1in] 
& + & \displaystyle \sum_{i=1}^m{\mu_i}{C_i}\otimes{P_+}|1\rangle\langle{1}|_{u_i}{P_+}+
\underbrace{\sum_{i=1}^m({\kappa_i}{A_i}+{\lambda_i}{B_i})\otimes{P_+}X_{u_i}{P_+}}_{V_f}=V_s+V_f \\[0.1in]
V_{-+} & = & \displaystyle \sum_{i=1}^m({\kappa_i}{A_i}+{\lambda_i}{B_i})\otimes{P_-}X_{u_i}{P_+},\quad V_{+-}=
\sum_{i=1}^m({\kappa_i}{A_i}+{\lambda_i}{B_i})\otimes{P_+}X_{u_i}{P_-} \\[0.1in]
V_- & = & \displaystyle \bigg(H_\text{else}+\frac{1}{\Delta}\sum_{i=1}^m({\kappa_i}{A_i}+{\lambda_i}{B_i})^2-\frac{1}{\Delta^2}\sum_{i=1}^m({\kappa_i^2}+{\lambda_i^2}){\mu_i}{C_i}- \frac{1}{\Delta^3}\sum_{i=1}^m({\kappa_i}{A_i}+{\lambda_i}{B_i})^4+\sum_{i=1}^m\sum_{j=1,j\neq i}^m\bar{V}_{ij}\bigg)\otimes{P_-}.
\end{array}
\end{equation} 
\noindent{}Here the $V_+$ projection is intentionally divided up into $V_f$ and $V_s$ components. $V_f$ is the component of $V_+$ that contributes to the perturbative expansion only when the perturbative term corresponds to flipping processes in the $\mathcal{L}_+$ subspace. $V_s$ is the component that contributes only when the perturbative term corresponds to transitions that involve the state of the $m$-qubit ancilla register staying the same. 

The projection of the Green's function $G(z)$ onto $\mathcal{L}_+$ can be written as
\begin{equation}
G_+(z)=\sum_{x\neq 0\cdots 00}\frac{1}{z-h(x)\Delta}|x\rangle\langle{x}|.
\end{equation}
We now explain the self energy expansion
\begin{equation}\label{eq:Sz_par_3body}
\Sigma_-(z)=V_-+V_{-+}G_+V_{+-}+V_{-+}G_+V_+G_+V_{+-}+V_{-+}(G_+V_+)^2G_+V_{+-}+V_{-+}(G_+V_+)^3G_+V_{+-}+\cdots
\end{equation}
\noindent{}in detail term by term. The $1^\text{st}$ order term is simply $V_-$ from Equation Eq.\ \ref{eq:V_3par}. The $2^\text{nd}$ order term corresponds to processes of starting from an all-zero state of the $m$ ancilla qubits, flipping one qubit and then flipping it back:
\begin{equation}
\begin{array}{ccl}
V_{-+}G_+V_{+-}& = & \displaystyle \frac{1}{z-\Delta}\sum_{i=1}^m({\kappa_i}{A_i}+{\lambda_i}{B_i})^2 \\[0.1in]
\end{array}
\end{equation}
\noindent{}The $3^\text{rd}$ order term corresponds to processes of starting from an all-zero state of the ancilla register, flipping one qubit, staying at the same state for $V_+$ and then flipping the same qubit back. Therefore only the $V_f$ component in $V_+$ in Equation Eq.\ \ref{eq:V_3par} will contribute to the perturbative expansion:

\begin{equation}
\begin{array}{ccl}
V_{-+}G_+V_+G_+V_{+-} & = & \displaystyle \frac{1}{(z-\Delta)^2}\sum_{i=1}^m({\kappa_i}{A_i}+{\lambda_i}{B_i})\bigg[H_\text{else}+{\mu_i}{C_i}+\frac{1}{\Delta}\sum_{j=1}^m({\kappa_j}{A_j}+{\lambda_j}{B_j})^2 \\[0.1in]
& + & \displaystyle \frac{1}{\Delta^2}\sum_{j=1}^m\bigg[(\kappa_j^2+\lambda_j^2)\mu_j{C_j}-\frac{1}{\Delta^3}\sum_{j=1}^m({\kappa_j}{A_j}+{\lambda_j}{B_j})^4+\sum_{j=1}^m\sum_{l=1,l\neq j}^m\bar{V}_{jl}\bigg] \\[0.1in]
&  & \displaystyle ({\kappa_i}{A_i}+{\lambda_i}{B_i}). \\[0.1in] 
\end{array}
\end{equation}
\noindent{}The $4^\text{th}$ order term is more involved. Here we consider two types of transition processes (for diagrammatic illustration refer to Fig.\ \ref{fig:diagram}): 
\begin{enumerate}
\item Starting from the all-zero state, flipping one of the qubits, flipping another qubit, then using the remaining $V_+$ and $V_{+-}$ to flip both qubits back one after the other (there are 2 different possible sequences, see Fig.\ \ref{fig:diagram}a). 
\item Starting from the all-zero state of the ancilla register, flipping one of the qubits, staying twice for the two $V_+$ components and finally flipping back the qubit during $V_{+-}$ (Fig.\ \ref{fig:diagram}b).
\end{enumerate}

Therefore in the transition processes of type (1), $V_+$ will only contribute its $V_f$ component and the detailed form of its contribution depends on which qubit in the ancilla register is flipped. The two possibilities of flipping the two qubits back explains why the second term in Eq.\ \ref{eq:V_3par_4} takes the form of a summation of two components. Because two qubits are flipped during the transition, $G_+$ will contribute a $\frac{1}{z-2\Delta}$ factor and two $\frac{1}{z-\Delta}$ factors to the perturbative term.

In the transition processes of type (2), $V_+$ will only contribute its $V_s$ component to the $4^\text{th}$ order term since the states stay the same during both $V_+$ operators in the perturbative term. $G_+$ will only contribute a factor of $\frac{1}{z-\Delta}$ because the Hamming weight of the bit string represented by the state of the ancilla register is always 1. This explains the form of the first term in Eq.\ \ref{eq:V_3par_4}.

\begin{equation}\label{eq:V_3par_4}
\begin{array}{ccl}
V_{-+}(G_+V_+)^2G_+V_{+-} & = & \displaystyle \frac{1}{(z-\Delta)^3}\sum_{i=1}^m({\kappa_i}{A_i}+{\lambda_i}{B_i})\bigg[H_\text{else}+{\mu_i}{C_i}+\frac{1}{\Delta}\sum_{j=1}^m({\kappa_j}{A_j}+{\lambda_j}{B_j})^2 \\[0.1in]
& - & \displaystyle \frac{1}{\Delta^2}\sum_{j=1}^m(\kappa_j^2+\lambda_j^2)\mu_j{C_j}-\frac{1}{\Delta^3}\sum_{j=1}^m({\kappa_j}{A_j}+{\lambda_j}{B_j})^4+\sum_{j=1}^m\sum_{l=1,l\neq j}^m\bar{V}_{jl}\bigg]^2 \\[0.1in]
& & ({\kappa_i}{A_i}+{\lambda_i}{B_i}) \\[0.1in]
& + & \displaystyle \frac{1}{(z-\Delta)^2(z-2\Delta)}\sum_{i=1}^m\sum_{j=1,j\neq i}^m\bigg[({\kappa_i}{A_i}+{\lambda_i}{B_i})({\kappa_j}{A_j}+{\lambda_j}{B_j}) \\[0.1in]
& & \makebox[4.65cm]{}{}({\kappa_i}{A_i}+{\lambda_i}{B_i})({\kappa_j}{A_j}+{\lambda_j}{B_j}) \\[0.1in]
& + & \displaystyle ({\kappa_i}{A_i}+{\lambda_i}{B_i})({\kappa_j}{A_j}+{\lambda_j}{B_j})({\kappa_j}{A_j}+{\lambda_j}{B_j})({\kappa_i}{A_i}+{\lambda_i}{B_i}) \bigg].  \\[0.1in]
\end{array}
\end{equation}
\noindent{}Although the $4^\text{th}$ order does not contain terms that are useful for simulating the 3-body target Hamiltonian, our assignments of ${\kappa_i}$, ${\lambda_i}$ and ${\mu_i}$ values in Eq.\ \ref{eq:J_3par} imply that some of the terms at this order can be $\Theta(1)$. Indeed, the entire second term in Eq.\ \ref{eq:V_3par_4} is of order $\Theta(1)$ based on Eq.\ \ref{eq:J_3par}. Therefore it is necessary to study in detail what error terms arise at this order and how to compensate for them in the perturbation $V$. A detailed analysis on how to compensate the $\Theta(1)$ errors is presented in the Appendix \ref{appendix:4local}. The $5^\text{th}$ order and higher terms are errors that can be reduced by increasing $\Delta$:
\begin{equation}\label{eq:high}
\begin{array}{ccl}
&  & \displaystyle \sum_{k=3}^\infty V_{-+}(G_+V_+)^kG_+V_{+-}.
\end{array}
\end{equation}
\noindent{}At first glance, with assignments of ${\kappa_i}$, ${\lambda_i}$ and ${\mu_i}$ in Eq.\ \ref{eq:J_3par}, it would appear that this error term is $\Theta(\Delta^{-1/4})$ since $\|V_{-+}\|=\Theta(\Delta^{3/4})$, $\|V_{+-}\|=\Theta(\Delta^{3/4})$, $\|V_+\|=\Theta(\Delta^{3/4})$ and $\|G_+\|=\Theta(\Delta^{-1})$,
\begin{equation}\label{eq:crude}
\begin{array}{ccl}
\displaystyle \sum_{k=3}^\infty V_{-+}(G_+V_+)^kG_+V_{+-} & \le & \displaystyle \sum_{k=3}^\infty \|V_{-+}\|\cdot\|G_+V_+\|^k\|G_+\|\cdot\|V_{+-}\| \\[0.1in]
& = & \displaystyle \|V_{-+}(G_+V_+)^3G_+V_{+-}\|\sum_{k=0}^\infty\|G_+V_+\|^k \\[0.1in]
& = & \displaystyle O(\Delta^{-1/4})
\end{array}
\end{equation}
\noindent{}as $\sum_{k=0}^\infty\|G_+V_+\|^k=O(1)$. However, here we show that in fact this term in Eq.\ \ref{eq:high} is $\Theta(\Delta^{-1/2})$. Note that the entire term Eq.\ \ref{eq:high} consists of contributions from the transition processes where one starts with a transition from the all-zero state to a state $|x\rangle$ with $x\in\{0,1\}^m$ and $h(x)=1$. If we focus on the perturbative term of order $k+2$: \[V_{-+}(G_+V_+)^kG_+V_{+-},\]after $k$ steps. During every step one can choose to either flip one of the ancilla qubits or stay in the same state of the ancilla register, the state of the ancilla register will go back to a state $|y\rangle$ with $y\in\{0,1\}^m$ and $h(y)=1$. Finally the $|1\rangle$ qubit in $|y\rangle$ is flipped back to $|0\rangle$ and we are back to the all-zero state which spans the ground state subspace $\mathcal{L}_-$. Define the total number of flipping steps to be $k_f$. Then for a given $k$, $k_f$ takes only values from
\begin{equation}
K(k)=\left\{
\begin{array}{cl}
\{k,k-2,\cdots,2\} & \text{if $k$ is even} \\[0.1in]
\{k-1,k-3,\cdots,2\} & \text{if $k$ is odd}.
\end{array}
\right.
\end{equation}

\noindent{}For the term of order $k+2$, all the transition processes that contribute non-trivially to the term can be categorized into two types:
\begin{enumerate}
\item If $x=y$, the minimum number of flipping steps is 0. The contribution of all such processes to the $(k+2)$-th order perturbative term is bounded by \cite{footnote:comb}
\begin{equation}
\begin{array}{cl}
 \le & \displaystyle m^{k_f}\cdot \binom{k}{k_f}\cdot \|V_f\|^{k_f}\cdot \|V_s\|^{k-k_f}\cdot \frac{\|V_{-+}\|\cdot\|V_{+-}\|}{(\Delta-\max(z))^{k+1}}
\end{array}
\end{equation}
\noindent{}where the factor $m^{k_f}$ is the number of all possible ways of flipping $k_f$ times, each time one of the $m$ ancilla qubits. This serves as an upper bound for the number of transition processes that contribute non-trivially to the perturbative term. The factor $\binom{k}{k_f}$ describes the number of possible ways to choose which $(k-k_f)$ steps among the total $k$ steps involve the state of the ancilla register staying the same. $\|G_+\|\le\frac{1}{\Delta-\max(z)}$ is used in the upper bound.

\item If $x\neq y$, the minimum number of flipping steps is 2. The contribution of all such processes to the $(k+2)$-th order perturbative term is bounded by
\begin{equation}
\begin{array}{cl}
 \le & \displaystyle \binom{k}{k_f}\cdot\binom{k_f}{2}\cdot 2!\cdot\|V_f\|^{k_f}\|V_s\|^{k-k_f}\cdot m^{k_f-2}\cdot \frac{\|V_{-+}\|\cdot\|V_{+-}\|}{(\Delta-\max(z))^{k+1}}
\end{array}
\end{equation}
\noindent{}where the factor $\binom{k}{k_f}$ is the number of all possible ways to choose which $(k-k_f)$ steps among the $k$ steps should the state remain the same. $\binom{k_f}{2}$ is the number of possible ways to choose from the $k_f$ flipping steps the 2 minimum flips. $2!$ is for taking into account the ordering of the 2 flipping steps. $\|G_+\|\le\frac{1}{\Delta-\max(z)}$ is used in the upper bound.
\end{enumerate}

\begin{figure}
\includegraphics[scale=0.15]{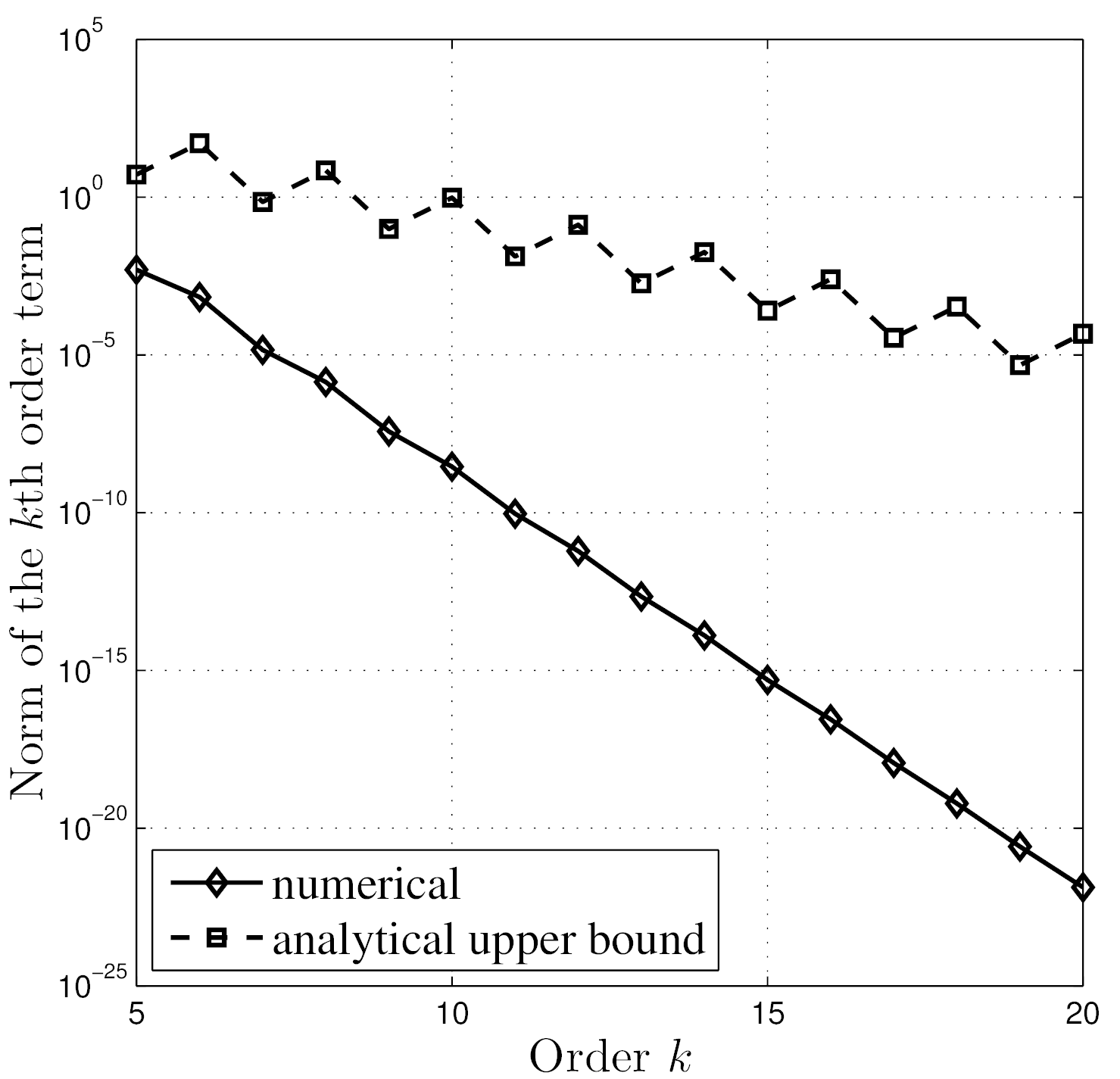}
\caption{\normalsize Numerical verification for the upper bound to the norm of the $(k+2)$-th order perturbative term in Eq.\ \ref{eq:high_sum}. Here we use the parallel 3-body gadget for reducing $H_\text{targ}=0.1X_1Z_2Z_3-0.2X_1X_2Z_3$ up to error $\epsilon=0.01$. The gap in the gadget construction is numerically optimized \cite{footnote:num_op}. Here the calculation of the analytical upper bound uses the result in Eq.\ \ref{eq:high_sum}. The calculation is then compared with the norm of the corresponding perturbative term numerically calculated according to the self-energy expansion.}
\label{fig:par3_high_bound}
\end{figure}

For a general $m$-qubit ancilla register, there are in total $m$ different cases of the first type of transition processes and $\binom{m}{2}$ different cases of the second type of transition processes. Therefore we have the upper bound to the norm of the $(k+2)$-th term (Fig.\ \ref{fig:par3_high_bound})
\begin{equation}\label{eq:high_sum}
\begin{array}{ccl}
\|V_{-+}(G_+V_+)^kG_+V_{+-}\| & \le & \displaystyle  m\sum_{k_f\in K(k)}m^{k_f}\binom{k}{k_f}\cdot\|V_f\|^{k_f}\cdot\|V_s\|^{k-k_f}\frac{\|V_{-+}\|\cdot\|V_{+-}\|}{(\Delta-\max(z))^{k+1}} \\[0.1in]
& & \displaystyle +\binom{m}{2} \sum_{k=3}^\infty \binom{k}{k_f}\cdot\binom{k_f}{2}\cdot 2!\cdot\|V_f\|^{k_f}\|V_s\|^{k-k_f}\cdot m^{k_f-2}\cdot \frac{\|V_{-+}\|\cdot\|V_{+-}\|}{(\Delta-\max(z))^{k+1}} \\[0.1in]
& & \displaystyle = \sum_{k_f\in K(k)}\left(m+\frac{m-1}{m}\right)2^k\cdot \frac{\|V_{-+}\|\cdot(m\|V_f\|)^{k_f}\cdot\|V_s\|^{k-k_f}\cdot\|V_{+-}\|}{(\Delta-\max(z))^{k+1}} \\[0.1in]
& & \displaystyle \le \frac{\|V_{-+}\|\cdot\|V_{+-}\|}{\Delta-\max(z)}(m+1)\sum_{k=3}^\infty \left(\frac{\|V_s\|}{\Delta-\max(z)}\right)^k\sum_{k_f\in K(k)}\left(m\frac{\|V_f\|}{\|V_s\|}\right)^{k_f}.
\end{array}
\end{equation} 
\noindent{}Since $\|\sum_{i=1}^m\sum_{j=1,j\neq i}^m\bar{V}_{ij}\|$ is bounded by $\frac{1}{\Delta^3}\sum_{i=1}^m\sum_{j=1,j\neq i}^m 8({\kappa_i}{\kappa_j})^2{\openone}$, from Eq.\ \ref{eq:V_3par} we see that
\begin{equation}\label{eq:Vbound}
\begin{array}{ccl}
\|V_s\| & \le & \displaystyle \|H_\text{else}\|+2^{-1/3}\Delta^{1/2}\sum_{i=1}^m|\alpha_i|^{1/3}+2^{4/3}\Delta^{1/2}\sum_{i=1}^m|\alpha_i|^{2/3}+ \sum_{i=1}^m|\alpha_i| \\[0.1in]
& & \displaystyle +2^{8/3}\sum_{i=1}^m|\alpha_i|^{4/3}+\sum_{i=1}^m\sum_{j=1,j\neq i}^m 8\cdot 2^{-4/3}|\alpha_i|^{2/3}|\alpha_j|^{2/3} \equiv v_s \\[0.1in]
\|V_f\| & \le & \displaystyle 2^{2/3}\Delta^{3/4}\sum_{i=1}^m|\alpha_i|^{1/3}\equiv v_f.
\end{array}
\end{equation}
\noindent{}With bounds of $\|V_s\|$ and $\|V_f\|$ in Eq.\ \ref{eq:V_3par}, the summation in Equation Eq.\ \ref{eq:high_sum} can be written as
\begin{equation}\label{eq:high_simp}
\begin{array}{l}
\displaystyle \|\sum_{k=3}^\infty V_{-+}(G_+V_+)^kG_+V_{+-}\| \le \frac{\|V_{-+}\|\cdot\|V_{+-}\|}{\Delta-\max(z)}(m+1) \\[0.1in]
\displaystyle \bigg[\sum_{r=1}^\infty\left(\frac{2v_s}{\Delta-\max(z)}\right)^{2r+1}\sum_{s=1}^r\left(m\frac{v_f}{v_s}\right)^{2s}+\sum_{r=2}^\infty\left(\frac{2v_s}{\Delta-\max(z)}\right)^{2r}\sum_{s=1}^r\left(m\frac{v_f}{v_s}\right)^{2s}\bigg].
\end{array}
\end{equation}
\noindent{}To guarantee convergence of the summation in Eq.\ \ref{eq:high_simp} we require that $\Delta$ satisfies
\begin{eqnarray}
\frac{2mv_f}{\Delta-\max(z)}<1 \\ [0.1in]
m\left(\frac{v_f}{v_s}\right)>1,
\end{eqnarray}
\noindent{}both of which are in general satisfied. The summation in Eq.\ \ref{eq:high_simp} can then be written as
\begin{equation}\label{eq:high_bound}
\begin{array}{c}
\displaystyle \|\sum_{k=3}^\infty V_{-+}(G_+V_+)^kG_+V_{+-}\|\le\frac{\|V_{-+}\|\cdot\|V_{+-}\|}{\Delta-\max(z)}\cdot\frac{\left(m\frac{v_f}{v_s}\right)^2}{\left(m\frac{v_f}{v_s}\right)^2-1} \\[0.1in]
\displaystyle \frac{\left(\frac{2mv_f}{\Delta-\max(z)}\right)^2}{1-\left(\frac{2mv_f}{\Delta-\max(z)}\right)^2}(m+1)\left[\left(\frac{2mv_f}{\Delta-\max(z)}\right)^2+\frac{2v_s}{\Delta-\max(z)}\right]=\Theta(\Delta^{-1/2}),
\end{array}
\end{equation}
\noindent{}which shows that the high order terms are $\Theta(\Delta^{-1/2})$. This is tighter than the crude bound $\Theta(\Delta^{-1/4})$ shown in Eq.\ \ref{eq:crude}. The self-energy expansion Eq.\ \ref{eq:Sz_par_3body} then satisfies
\begin{equation}
\|\Sigma_-(z)-H_\text{targ}\otimes{P_-}\|\le \Theta(\Delta^{-1/2})
\end{equation}
\noindent{}which indicates that $\Delta=\Theta(\epsilon^{-2})$ is sufficient for the parallel 3-body gadget to capture the entire spectrum of $H_\text{targ}\otimes{P_-}$ up to error $\epsilon$.

\begin{figure}
\makebox[3cm]{}\includegraphics[scale=0.17]{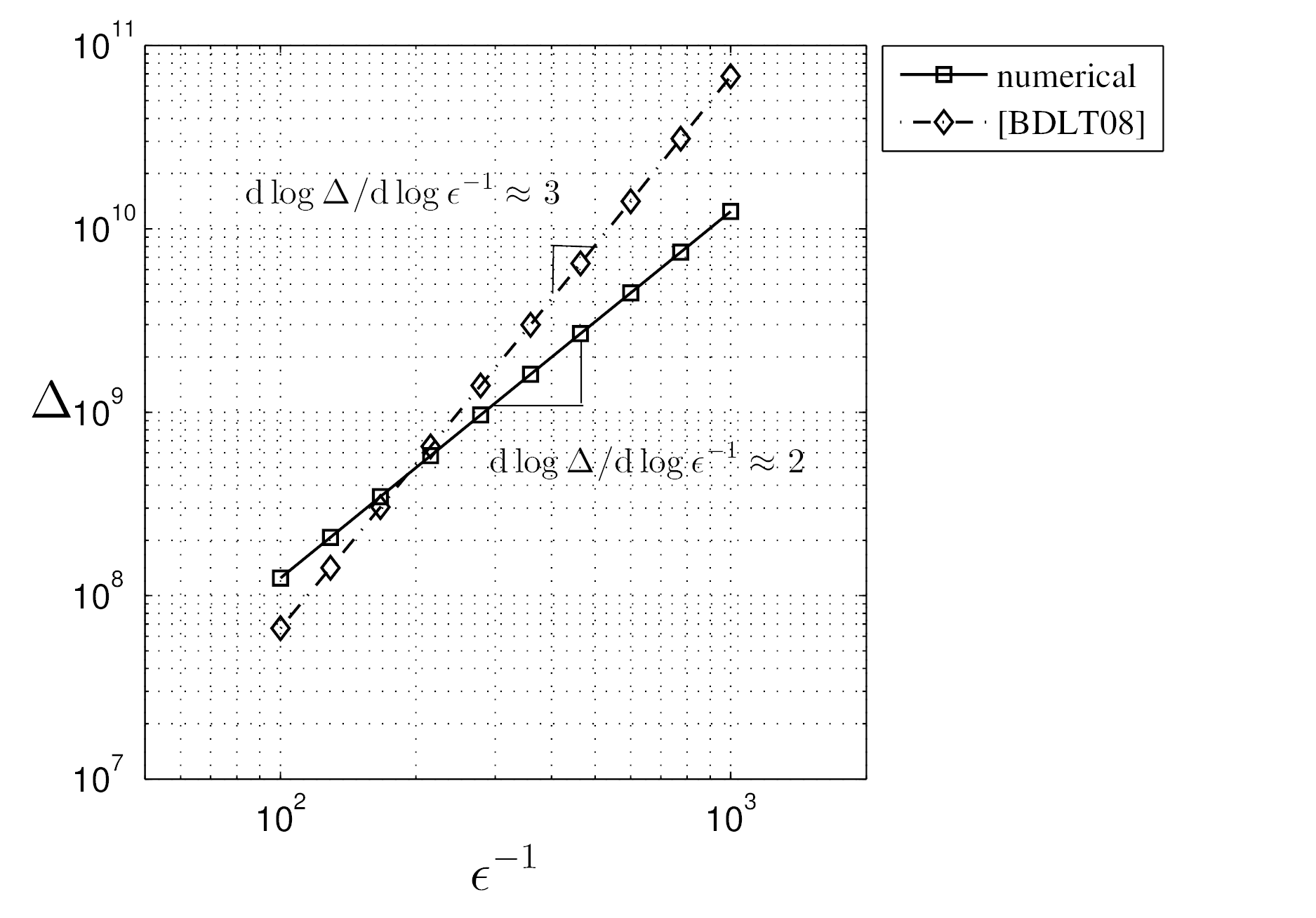}
\caption{\normalsize Scaling of the spectral gap $\Delta$ as a function of error $\epsilon$ for the parallel 3-body example that is intended to reduce the target Hamiltonian $H_\text{targ}=Z_1Z_2Z_3-X_1X_2X_3$ to 2-body. Here $\epsilon=0.01$. We show both numerically optimized values (``numerical'') in our construction and the construction in \cite{BDLT08}, which is referred to as ``[BDLT08]''. }
\label{fig:par3_Delta_eps}
\end{figure}

We have used numerics to verify the $\Theta(\epsilon^{-2})$ scaling, as shown in Fig.\ \ref{fig:par3_high_bound}. Furthermore, for a range of specified $\epsilon$, the minimum $\Delta$ needed for the spectral error between the gadget Hamiltonian and the target Hamiltonian is numerically found. In the optimized cases, the slope ${\rm d}\log\Delta/{\rm d}\log\epsilon^{-1}$ for the construction in \cite{BDLT08} is approximately 3, showing that $\Delta=\Theta(\epsilon^{-3})$ is the optimal scaling for the construction in \cite{BDLT08}. For our construction both the analytical bound and the optimized $\Delta$ scale as $\Theta(\epsilon^{-2})$ (see Fig.\ \ref{fig:par3_Delta_eps}).

\section{Compensation for the 4-local error terms in parallel 3- to 2-body gadget}\label{appendix:4local}

Continuing the discussion in Appendix \ref{sec:3body_par}, here we deal with $\Theta(1)$ error terms that arise in the $3^\text{rd}$ and $4^\text{th}$ order perturbative expansion when $V$ in Eq.\ \ref{eq:V_par3} is without $V_3$ and in so doing explain the construction of $\bar{V}_{ij}$ in Eq.\ \ref{eq:Vij}. From the previous description of the $3^\text{rd}$ and $4^\text{th}$ order terms, for each pair of terms $(i)$ and $(j)$ where $i$ and $j$ are integers between 1 and $m$, let \[
\begin{array}{ccl}
M_1 & = & ({\kappa_i}{A_i}+{\lambda_i}{B_i})({\kappa_j}{A_j}+{\lambda_j}{B_j}) \\[0.1in]
M_2 & = & ({\kappa_j}{A_j}+{\lambda_j}{B_j})({\kappa_i}{A_i}+{\lambda_i}{B_i})
\end{array}
\]
and then the $\Theta(1)$ error term arising from the $3^\text{rd}$ and $4^\text{th}$ order perturbative expansion can be written as
\begin{equation}
\frac{1}{(z-\Delta)^2}\bigg[\frac{1}{z-2\Delta}(M_1^2+M_2^2)+\left(\frac{1}{\Delta}+\frac{1}{z-2\Delta}\right)(M_1M_2+M_2M_1)\bigg].
\end{equation}
Based on the number of non-commuting pairs among ${A_i}$, ${A_j}$, ${B_i}$ and ${B_j}$, all possible cases can be enumerated as the following:
\begin{equation}\label{eq:cases}
\begin{array}{ccl}
\text{case 0:} & & [{A_i},{A_j}]=0,[{B_i},{B_j}]=0,[{A_i},{B_j}]=0,[{B_i},{A_j}]=0 \\[0.1in]
\text{case 1:} & 1.1: & [{A_i},{A_j}]=0,[{B_i},{B_j}]=0,[{A_j},{B_i}]\neq 0 \\[0.1in]
& 1.2: & [{A_i},{A_j}]=0,[{B_i},{B_j}]=0,[{A_i},{B_j}]\neq 0 \\[0.1in]
& 1.3: & [{A_i},{A_j}]=0,[{B_i},{B_j}]\neq 0 \\[0.1in]
& 1.4: & [{A_i},{A_j}]\neq 0,[{B_i},{B_j}]=0 \\[0.1in]
\text{case 2:} & & [{A_i},{A_j}]\neq 0,[{B_i},{B_j}]\neq 0.
\end{array}
\end{equation}
\noindent{}In case 0, clearly $M_1=M_2$. Then the $\Theta(1)$ error becomes \[\frac{1}{(z-\Delta)^2}\left(\frac{1}{\Delta}+\frac{2}{z-2\Delta}\right)\cdot 2M_1^2=\Theta(\Delta^{-1})\] which does not need any compensation.
In case 1, for example in the subcase 1.1, ${A_j}$ does not commute with ${B_i}$. Then $M_1$ and $M_2$ can be written as 
\[
\begin{array}{ccl}
M_1 & = & K+{\kappa_j}{\lambda_i}{B_i}{A_j} \\[0.1in]
M_2 & = & K+{\kappa_j}{\lambda_i}{A_j}{B_i}
\end{array}
\]
where $K$ contains the rest of the terms in $M_1$ and $M_2$. Furthermore,
\[
\begin{array}{c}
M_1^2+M_2^2= 2K^2-2({\kappa_j}{\lambda_i})^2{\openone} \\[0.1in]
M_1M_2+M_2M_1= 2K^2+2({\kappa_j}{\lambda_i})^2{\openone}.
\end{array}
\]
Hence the $\Theta(1)$ term in this case becomes
\begin{equation}
\frac{1}{(z-\Delta)^2}\bigg[\left(\frac{1}{\Delta}+\frac{2}{z-2\Delta}\right)2K^2+\frac{1}{\Delta}\cdot 2({\kappa_j}{\lambda_i})^2{\openone}\bigg]
\end{equation}
\noindent{}where the first term is $\Theta(\Delta^{-1})$ and the second term is $\Theta(1)$, which needs to be compensated. Similar calculations for cases 1.2, 1.3 and 1.4 will yield $\Theta(1)$ error with the same norm. In case 2, define $R={\kappa_i}{\lambda_j}{A_i}{B_j}+{\lambda_i}{\kappa_j}{B_i}{A_j}$ and $T={\kappa_i}{\kappa_j}{A_i}{A_i}+{\lambda_i}{\lambda_j}{B_i}{B_i}$. Then 
\[
\begin{array}{c}
M_1^2+M_2^2 = 2(R^2+T^2) \\[0.1in]
M_1M_2+M_2M_1 = 2(R^2-T^2).
\end{array}
\]
The $\Theta(1)$ error terms in the 3rd and 4th order perturbative expansion becomes
\begin{equation}
\frac{1}{(z-\Delta)^2}\bigg[\left(\frac{1}{\Delta}+\frac{2}{z-2\Delta}\right)\cdot 2R^2-\frac{1}{\Delta}\cdot 2T^2\bigg]
\end{equation}
\noindent{}where the first term is $\Theta(\Delta^{-1})$ and hence needs no compensation. The second term is $\Theta(1)$. 
Define
\begin{equation}
s_0^{(i,j)}=\left\{
\begin{array}{ccl}
1 & & \text{if case 0} \\[0.1in]
0 & & \text{Otherwise}
\end{array}
\right.
\end{equation}
\noindent{}With the definitions of $s_1^{(i,j)}$ and $s_2^{(i,j)}$ in Eq.\ \ref{eq:s11}, Eq.\ \ref{eq:s12} and Eq.\ \ref{eq:s2}, the contribution of the $i$-th and the $j$-th target terms to the $\Theta(1)$ error in the perturbative expansion $\Sigma_-(z)$ becomes
\begin{equation}\label{eq:O1}
\begin{array}{cl}
& \displaystyle s_0^{(i,j)}\cdot\frac{1}{(z-\Delta)^2}\left(\frac{1}{\Delta}+\frac{2}{z-2\Delta}\right)\cdot 2({\kappa_i}{A_i}+{\lambda_i}{B_i})^2({\kappa_j}{A_j}+{\lambda_j}{B_j})^2 \\[0.1in]
+ & \displaystyle s_1^{(i,j)}\cdot\frac{1}{(z-\Delta)^2}\bigg[\left(\frac{1}{\Delta}+\frac{2}{z-2\Delta}\right)\cdot 2K_{ij}^2+\frac{1}{\Delta}\cdot 2({\kappa_i}{\kappa_j})^2{\openone}\bigg] \\[0.1in]
+ & \displaystyle s_2^{(i,j)}\cdot\frac{1}{(z-\Delta)^2}\bigg[\left(\frac{1}{\Delta}+\frac{2}{z-2\Delta}\right)\cdot 2R_{ij}^2+\frac{1}{\Delta}\cdot 2\{[({\kappa_i}{\kappa_j})^2+({\lambda_i}{\lambda_j})^2]{\openone} \\[0.1in]
& \displaystyle -2{\kappa_i}{\kappa_j}{\lambda_i}{\lambda_j}{A_i}{A_j}{B_i}{B_j}\}\bigg].
\end{array}
\end{equation}
\noindent{}The term proportional to $s_0^{(i,j)}$ in Eq.\ \ref{eq:O1} does not need compensation since it is already $\Theta(\Delta^{-1})$. The term proportional to $s_1^{(i,j)}$ can be compensated by the corresponding term in $\bar{V}_{ij}$ in Eq.\ \ref{eq:Vij} that is proportional to $s_1^{(i,j)}$. Similarly, the $\Theta(1)$ error term proportional to $s_2^{(i,j)}$ can be compensated by the term in $\bar{V}_{ij}$ in Eq.\ \ref{eq:Vij} that is proportional to $s_2^{(i,j)}$.

{Now we deal with generating the 4-local term in $\bar{V}_{ij}$. Introduce an ancilla $u_{ij}$ and construct a gadget $\tilde{H}_{ij}=H_{ij}+V_{ij}$ such that $H_{ij}=\Delta|1\rangle\langle{1}|_{u_{ij}}$ and the perturbation $V_{ij}$ becomes
\begin{equation}
V_{ij}=(\kappa_i A_i+\lambda_j B_j)\otimes X_{u_{ij}}+(\kappa_j A_j+\lambda_i B_i)\otimes|1\rangle\langle{1}|_{u_{ij}}+V'_{ij}
\end{equation}
where $V'_{ij}$ is defined as
\begin{equation}
V'_{ij}=\frac{1}{\Delta}(\kappa_i A_i+\lambda_j B_j)^2+\frac{1}{\Delta^3}\left[(\kappa_j^2+\lambda_i^2)(\kappa_i A_i+\lambda_jB_j)^2-2\kappa_j\lambda_i(\kappa_j^2+\lambda_j^2)A_jB_i\right]
\end{equation}
The self-energy expansion $\Sigma_-(z)$ is now\[\Sigma_-(z)=\frac{1}{(z-\Delta)^3}4\kappa_i\kappa_j\lambda_i\lambda_jA_iA_jB_iB_j+O(\Delta^{-1/2})\]which is $O(\Delta^{-1/2})$ close to the 4-local compensation term in $\bar{V}_{ij}$. We apply the the gadget $\tilde{H}_{ij}$ for every pair of qubits with $s_2^{(i,j)}=1$. The cross-gadget contribution between the $\tilde{H}_{ij}$ gadgets as well as those cross-gadget contribution between $\tilde{H}_{ij}$ gadgets and gadgets based on ancilla qubits $u_1$ through $u_m$ both belong to the case 1 of the Eq.\ \ref{eq:cases} and hence are easy to deal with using 2-body terms.}

\bibliographystyle{unsrt}
\bibliography{ref}

\begin{thebibliography}{10}

\bibitem{2004quant.ph..5098A}
Dorit Aharonov, Julia Kempe, Seth Lloyd, Wim~Van Dam, Zeph Landau, and Oded
  Regev.
\newblock Adiabatic quantum computation is equivalent to standard quantum
  computation.
\newblock {\em SIAM Journal on Computing}, 37:166--194, 2007.
\newblock \href{http://arxiv.org/abs/quantph/0405098}{arXiv:quant-ph/0405098}.

\bibitem{2007PhRvL..99g0502M}
A.~{Mizel}, D.~A. {Lidar}, and M.~{Mitchell}.
\newblock Simple proof of equivalence between adiabatic quantum computation and
  the circuit model.
\newblock {\em Physical Review Letters}, 99(7):070502, August 2007.
\newblock \href{http://arxiv.org/abs/quant-ph/0609067}{arXiv:quant-ph/0609067}.

\bibitem{OT06}
R.~Oliveira and B.~Terhal.
\newblock The complexity of quantum spin systems on a two-dimensional square
  lattice.
\newblock {\em Quant. Inf. and Comp.}, 8(10):0900--0924, 2008.
\newblock \href{http://arxiv.org/abs/quant-ph/0504050}{arXiv:quant-ph/0504050}.

\bibitem{BL07}
J.~D. Biamonte and P.~J. Love.
\newblock Realizable hamiltonians for universal adiabatic quantum computers.
\newblock {\em Phys. Rev. A}, 8(1):012352, 2008.
\newblock \href{http://arxiv.org/abs/0704.1287}{arXiv:0704.1287}.

\bibitem{CL08}
B.~A. Chase and A.~J. Landahl.
\newblock {Universal quantum walks and adiabatic algorithms by 1D
  Hamiltonians}.
\newblock page arXiv:0802.1207, 2008.
\newblock \href{http://arxiv.org/abs/0802.1207}{arXiv:0802.1207v1}.

\bibitem{KSV02}
A.~Kitaev, A.~H. Shen, and M.~N. Vyalyi.
\newblock {\em Classical and Quantum Computation}.
\newblock AMS Graduate Studies in Mathematics, 2002.

\bibitem{KKR06}
J.~Kempe, A.~Kitaev, and O.~Regev.
\newblock The complexity of the local hamiltonian problem.
\newblock {\em SIAM J. Computing}, 35(5):1070--1097, 2006.
\newblock \href{http://arxiv.org/abs/quant-ph/0406180}{quant-ph/0406180}.

\bibitem{CM13}
Toby Cubitt and Ashley Montanaro.
\newblock Complexity classification of local hamiltonian problems.
\newblock 2013.
\newblock \href{http://arxiv.org/abs/1311.3161}{arXiv:1311.3161 [quant-ph]}.

\bibitem{sim11}
J.~D. {Biamonte}, V.~{Bergholm}, J.~D. {Whitfield}, J.~{Fitzsimons}, and
  A.~{Aspuru-Guzik}.
\newblock {Adiabatic quantum simulators}.
\newblock {\em AIP Advances}, 1(2):022126, 2011.
\newblock \href{http://arxiv.org/abs/1002.0368}{arXiv:1002.0368 [quant-ph]}.

\bibitem{2014arXiv1401.3186V}
L.~{Veis} and J.~{Pittner}.
\newblock {Adiabatic state preparation study of methylene}.
\newblock {\em ArXiv e-prints}, January 2014.
\newblock \href{http://arxiv.org/abs/1401.3186}{arXiv:1401.3186 [quant-ph]}.

\bibitem{BDLT08}
S.~Bravyi, D.~DiVincenzo, D.~Loss, and B.~Terhal.
\newblock Quantum simulation of many-body hamiltonians using perturbation
  theory with bounded-strength interactions.
\newblock {\em Phys. Rev. Lett.}, 101:070503, 2008.
\newblock \href{http://arxiv.org/abs/0803.2686}{arXiv:0803.2686v1}.

\bibitem{BDOT06}
S.~Bravyi, D.~DiVincenzo, R.~Oliveira, and B.~Terhal.
\newblock The complexity of stoquastic local hamiltonian problems.
\newblock {\em Quant. Inf. and Comp.}, 8(5), 2006.
\newblock \href{http://arxiv.org/abs/quant-ph/0606140}{quant-ph/0606140}.

\bibitem{Schuch09}
N.~Schuch and F.~Verstraete.
\newblock Computational complexity of interacting electrons and fundamental
  limitations of density functional theory.
\newblock {\em Nature Physics}, 5:732--735, 2009.
\newblock \href{http://arxiv.org/abs/0712.0483}{arXiv:0712.0483v2}.

\bibitem{Ganti2013}
A.~{Ganti}, U.~{Onunkwo}, and K.~{Young}.
\newblock {A family of [[6k, 2k, 2]] codes for practical, scalable adiabatic
  quantum computation}.
\newblock September 2013.
\newblock \href{http://arxiv.org/abs/1309.1674}{arXiv:1309.1674 [quant-ph]}.

\bibitem{JF08}
S.~P. Jordan and E.~Farhi.
\newblock Perturbative gadgets at arbitrary orders.
\newblock {\em Phys. Rev. A}, 062329, 2008.
\newblock \href{http://arxiv.org/abs/0802.1874}{arXiv:0802.1874v4}.

\bibitem{bloch58}
C.~Bloch.
\newblock {Sur la th\'{e}orie des perturbations des \'{e}tats li\'{e}s}.
\newblock {\em Nuclear Physics}, 6:329--347, 1958.

\bibitem{cory99}
M.~D. Price, S.~S. Somaroo, A.~E. Dunlop, T.~F. Havel, and D.~G. Cory.
\newblock {Generalized Methods for the Development of Quantum Logic Gates for
  an NMR Quantum Information Processor}.
\newblock {\em Phys. Rev. A}, 60:2777--2780, 1999.

\bibitem{cory00}
C.~H. Tseng, S.~S. Somaroo, Y.~S. Sharf, E.~Knill, R.~Laflamme, T.~F. Havel,
  and D.~G. Cory.
\newblock {Quantum Simulation of a three-body interaction Hamiltonian on an NMR
  Quantum Computer}.
\newblock {\em Phys. Rev. A}, 61:12302--12308, 2000.
\newblock \href{http://arxiv.org/abs/quant-ph/9908012}{arXiv:quant-ph/9908012}.

\bibitem{2006cond.mat..8253H}
R.~Harris et~al.
\newblock Sign and magnitude tunable coupler for superconducting flux qubits.
\newblock {\em Phys. Rev. Lett.}, 2007.
\newblock \href{http://arxiv.org/abs/cond-mat/0608253}{cond-mat/0608253}.

\bibitem{Boixo2012}
Sergio Boixo, Tameem Albash, Federico~M. Spedalieri, Nicholas Chancellor, and
  Daniel~A. Lidar.
\newblock {Experimental signature of programmable quantum annealing}.
\newblock {\em Nature Communications}, 4:2067, June 2012.
\newblock \href{http://arxiv.org/abs/1212.1739}{arXiv:1212.1739 [quant-ph]}.

\bibitem{BCM+13}
Zhengbing Bian, Fabian Chudak, William~G. Macready, Lane Clark, and Frank
  Gaitan.
\newblock Experimental determination of ramsey numbers.
\newblock {\em Phys. Rev. Lett.}, 111(130505), 2013.
\newblock \href{http://arxiv.org/abs/1201.1842}{arXiv:1201.1842 [quant-ph]}.

\bibitem{Lidar2014}
Kristen~L. Pudenz, Tameem Albash, and Daniel~A. Lidar.
\newblock {Error Corrected Quantum Annealing with Hundreds of Qubits}.
\newblock {\em Nature Communications}, 5:3243, 2014.
\newblock \href{http://arxiv.org/abs/1307.8190}{arXiv:1307.8190 [quant-ph]}.

\bibitem{B08}
J.~D. Biamonte.
\newblock Non-perturbative k-body to two-body commuting conversion hamiltonians
  and embedding problem instances into ising spins.
\newblock {\em Phys. Rev. A}, 77(5):052331, 2008.
\newblock \href{http://arxiv.org/abs/0801.3800}{arXiv:0801.3800}.

\bibitem{WFB12}
J.~D. Whitfield, M.~Faccin, and J.~D. Biamonte.
\newblock Ground state spin logic.
\newblock {\em Euro. Phys. Lett.}, 99(57004), 2012.
\newblock \href{http://arxiv.org/abs/1205.1742}{arXiv:1205.1742v1}.

\bibitem{BOA13}
R.~Babbush, B.~O'Gorman, and A.~Aspuru-Guzik.
\newblock Resource efficient gadgets for compiling adiabatic quantum
  optimization problems.
\newblock {\em Ann. Phys.}, 525(10-11):877--888, 2013.
\newblock \href{http://arxiv.org/abs/1307.8041}{arXiv:1307.8041 [quant-ph]}.

\bibitem{BDL11}
S.~Bravyi, D.~DiVincenzo, and D.~Loss.
\newblock Schrieffer-wolff transformation for quantum many-body systems.
\newblock {\em Ann. Phys.}, 326(10), 2011.
\newblock \href{http://arxiv.org/abs/1105.0675}{arXiv:1105.0675}.

\bibitem{footnote:num_op}
The notion of `optimized case' refers to the search for the gap $\Delta$ needed
  for yielding a spectral error of precisely $\epsilon$ between gadget and
  target Hamiltonian, which is described in Sec.\ II.

\bibitem{footnote:cross}
As is shown by \cite{BDLT08}, for the gadget construction with the assignments
  of ${\kappa_i}$, ${\lambda_i}$ and ${\mu_i}$ all being $O(\Delta^{2/3})$, the
  cross-gadget contribution can be reduced by increasing $\Delta$, thus no
  cross-gadget compensation is needed. However, with our assignments of
  ${\kappa_i}$, ${\lambda_i}$ and ${\mu_i}$ in \eqref{eq:J_3par} there are
  cross-gadget error terms in the perturbative expansion that are of order
  $O(1)$, which cannot be reduced by increasing $\Delta$. This is why we need
  $\bar{V}_{ij}$. Since the $O(1)$ error terms are dependent on the commuting
  relations between $A_i$, $B_i$, $A_j$ and $B_j$ of each pair of $i$-th and
  $j$-th terms in the target Hamiltonian, $\bar{V}_{ij}$ depends on their
  commutation relations too.

\bibitem{footnote:comb}
Here we use the notation ${\sf C}_m^n$ to represent the combinatorial number
  that is the number of ways to choose $n$ elements from a total of $m$ without
  distinguishing between different orderings.

\end{thebibliography}

\end{document}